\documentclass[journal]{IEEEtran}

\usepackage{cite}
\usepackage{amsmath,amsthm,amssymb,amsfonts}
\usepackage{algorithmic}
\usepackage{array,multirow}
\usepackage{graphicx,xcolor,slashed}
\usepackage{makecell,booktabs}

\usepackage[labelformat=simple]{subcaption}

\newtheorem{lemma}{Lemma}
\newtheorem{theorem}{Theorem}

\usepackage{stfloats}
\usepackage{url}

\newcommand{\hl}[1]{{{\color{black} #1}}}
\newcommand{\rev}[1]{{{\color{black} #1}}}

\begin{document}
\title{Shape of the Cloak: Formal Analysis of Clock Skew-Based Intrusion Detection System in Controller Area Networks}

\author{Xuhang~Ying, Sang~Uk~Sagong,~\IEEEmembership{Student Member,~IEEE,}
        Andrew~Clark,~\IEEEmembership{Member,~IEEE,}\\
        Linda~Bushnell,~\IEEEmembership{Fellow,~IEEE,}
        and~Radha~Poovendran,~\IEEEmembership{Fellow,~IEEE}% <-this % stops a space
\thanks{X. Ying, S. U. Sagong, L. Bushnell and R. Poovendran are with the Department of Electrical Engineering, University of Washington, Seattle, WA, 98195-2500.). {\tt \{xhying,sagong,lb2,rp3\}@uw.edu}}% <-this % stops a space
\thanks{A. Clark is with the Department of Electrical and Computer Engineering, Worcester Polytechnic Institute, Worcester, MA, 01609. {\tt aclark@wpi.edu}}% <-this % stops a space
%\thanks{Manuscript received XX, XX; revised XX, XX.}
%\thanks{This work was supported by NSF grants CNS-1446866 and CNS-1656981, ONR grants N00014-16-1-2710 and N00014-17-1-2946, and ARO grant W911NF-16-1-0485. Views and conclusions expressed are that of the authors and not be interpreted as that of the NSF, ONR or ARO.}
\thanks{Part of this work was presented at ACM/IEEE ICCPS 2018 \cite{sagong2018cloaking}.}
}

\markboth{IEEE TRANSACTIONS ON INFORMATION FORENSICS AND SECURITY, VOL. XX, NO. XX, XX 2018}%
{Ying \MakeLowercase{\textit{et al.}}: Shape of the Cloak: Formal Analysis of Clock Skew-Based Intrusion Detection System in Controller Area Networks}

% make the title area
\maketitle

\begin{abstract}
This paper presents a new masquerade attack called the cloaking attack and provides formal analyses for clock skew-based Intrusion Detection Systems (IDSs) that detect masquerade attacks in the Controller Area Network (CAN) in automobiles. 
In the cloaking attack, the adversary manipulates the message inter-transmission times of spoofed messages by adding delays so as to emulate a desired clock skew and avoid detection. 
In order to predict and characterize the impact of the cloaking attack in terms of the attack success probability on a given CAN bus and IDS, %without having to rely solely on experiments, 
we develop formal models for two clock skew-based IDSs, i.e., the state-of-the-art (SOTA) IDS and its adaptation to the widely used Network Time Protocol (NTP), using parameters of the attacker, the detector, and the hardware platform.
To the best of our knowledge, this is the first paper that provides formal analyses of clock skew-based IDSs in automotive CAN. 
We implement the cloaking attack on two hardware testbeds, a prototype and a real vehicle (the University of Washington (UW) EcoCAR), and demonstrate its effectiveness against both the SOTA and NTP-based IDSs. 
\hl{We validate our formal analyses through extensive experiments for different messages, IDS settings, and vehicles.}
By comparing each predicted attack success probability curve against its experimental curve, we find that the average prediction error  is within $3.0\%$ for the SOTA IDS and $5.7\%$ for the NTP-based IDS.

\end{abstract}


\IEEEpeerreviewmaketitle

\section{Introduction}
\label{sec:intro}
Recent studies have identified security vulnerabilities in networked automobiles, in which attackers have compromised in-vehicle Electronic Control Units (ECUs), and disabled brakes \cite{Checkoway:2011:comprehensive}, remotely controlled steering \cite{Tesla:Hackers}, and disabled vehicles on a highway \cite{Wired:Hackers}.
Such exploits of ECUs are feasible because in-vehicle network protocols, such as the Controller Area Network (CAN) \cite{bosch1991can},
were designed for closed systems and do not have security mechanisms such as message authentication.
%were designed for closed systems with physical access controls to prevent intrusions, and they thus lack security mechanisms such as message authentication.
Networked automobiles, however, contain externally accessible ECUs that can be compromised by remote adversaries \cite{Koscher:2010:experimental,Checkoway:2011:comprehensive,miller2014survey}.
Since the CAN bus is a broadcast medium and there is no message authentication, a compromised ECU can be used to inject spoofed messages with faked message IDs and masquerade as a targeted ECU (masquerade attack) \cite{Checkoway:2011:comprehensive}.
%, due to the lack of message authentication and possibility of blocking message transmission of the targeted ECU \cite{hoppe2011security}.

Given that CAN has a preset %predetermined
tight bit budget for messages and resource-constrained ECUs have real-time requirements, it has not been a practical option to incorporate cryptographic primitives as in \cite{Lin:2012:cyber,nilsson2008efficient,Herrewege:2011:CANAuth} into CAN.
As an alternative, Intrusion Detection Systems (IDSs) have been proposed that exploit physical properties such as message periodicity and network entropy
%to address these security concerns by exploiting physical characteristics %invariants % to fingerprint the compromised ECUs 
without modifying the CAN protocol %\cite{Miller:2013:adventure,Muter:2011:entropy,Muter:2010:structured,Shin:2016:finger,cho2017viden}.
\cite{murvay2014source,Shin:2016:finger,cho2017viden,choi2018identifying}.

\begin{figure}[t!]
	\centering
	\begin{subfigure}[h]{0.48\columnwidth} % {0.48\columnwidth}
		\includegraphics[width=\columnwidth]{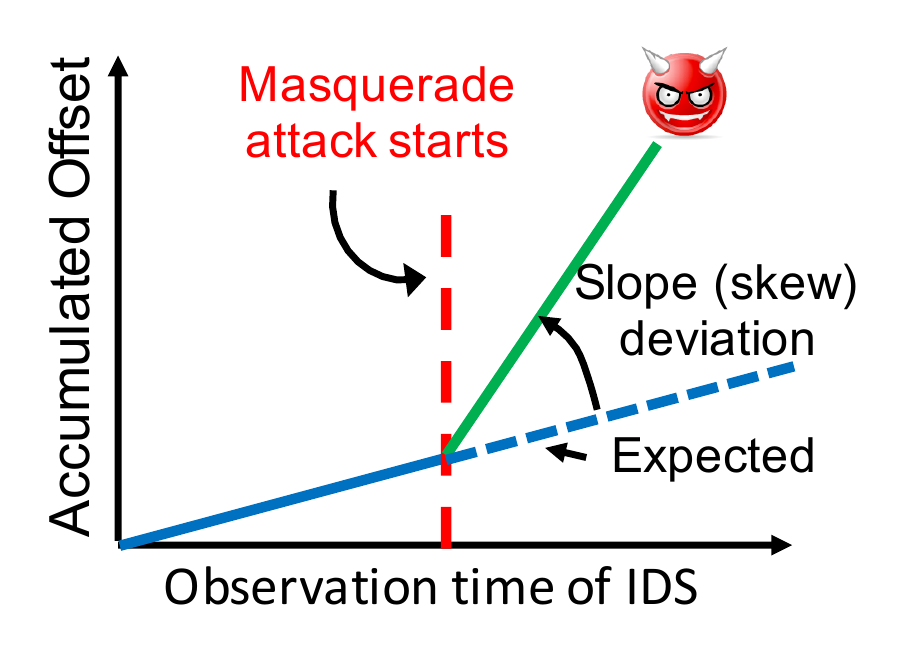}
		\caption{}
		\label{fig:cids_detects_masquerade_attack}
	\end{subfigure}
	\begin{subfigure}[h]{0.48\columnwidth} % {0.48\columnwidth}
		\includegraphics[width=\columnwidth]{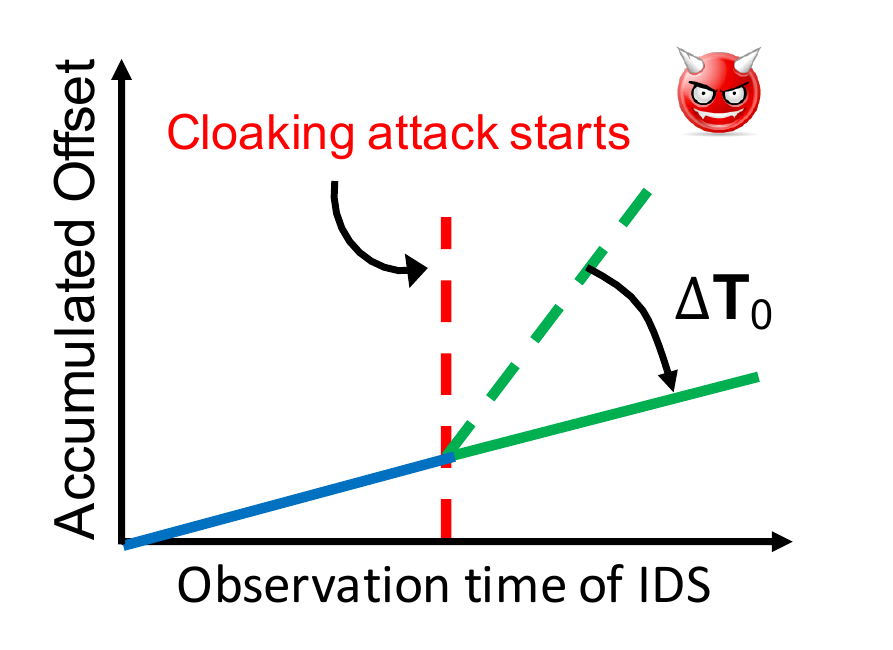}
		\caption{}
		\label{fig:cloaking_attack_bypasses_cids}
	\end{subfigure}
	\caption{Clock skew estimated by the IDS at the receiver. (a) An IDS tracks the clock skew of the transmitter and detects deviations due to  masquerade attacks. (b) A cloaking adversary adds a delay $\Delta T_0$ to the message inter-transmission times to emulate the targeted ECU's clock skew and bypass the IDS. }
	\label{fig:masquerading_vs_cloaking}
	\vspace{-0.3cm}
\end{figure}

One state-of-the-art (SOTA) IDS was proposed in USENIX 2016 \cite{Shin:2016:finger} based on two key observations: 1) almost all CAN messages are periodic, and 2) periodically received messages can be used to estimate the \textit{clock skew} of the transmitter, a unique physical invariant %property 
of each ECU due to variations in the clock's hardware crystal.
Therefore, a change in estimated clock skew at the receiver implies an anomaly in the transmitter's clock characteristics, which indicates the presence of a masquerade attack with high probability (Fig.~\ref{fig:cids_detects_masquerade_attack}).
The novelty of the SOTA IDS is the use of the clock skew for detecting a masquerade attack without requiring any synchronization and identifying the compromised ECU that mounts the attack.

%\begin{figure*}[t!]
%	\centering
%	\begin{subfigure}[h]{0.45\columnwidth}
%		\captionsetup{justification=centering}
%		\includegraphics[width=\columnwidth]{./Figures/example_exp_vs_fitted_sota_arduino}
%		\caption{CAN bus prototype\\SOTA IDS}
%	\end{subfigure}
%	\begin{subfigure}[h]{0.45\columnwidth}
%		\captionsetup{justification=centering}
%		\includegraphics[width=\columnwidth]{./Figures/example_exp_vs_fitted_sota_184}
%		\caption{Real vehicle testbed\\SOTA IDS}
%	\end{subfigure}
%	\begin{subfigure}[h]{0.45\columnwidth}
%		\captionsetup{justification=centering}
%		\includegraphics[width=\columnwidth]{./Figures/example_exp_vs_fitted_ntp_arduino}
%		\caption{CAN bus prototype\\~~~NTP-based IDS}
%	\end{subfigure}
%	\begin{subfigure}[h]{0.45\columnwidth}
%		\captionsetup{justification=centering}
%		\includegraphics[width=\columnwidth]{./Figures/example_exp_vs_fitted_ntp_184}
%		\caption{Real vehicle testbed\\~~~NTP-based IDS}
%	\end{subfigure}
%	\caption{Experimental attack success probability curves for the SOTA and NTP-based IDSs on two hardware platforms, the CAN bus prototype and the real vehicle testbed \cite{sagong2017cloaking}.
%	In each figure, the colored curves correspond to the cloaking attack of different durations in terms of the number of attack batches $n$ (each batch consisting of $20$ messages).
%	While the NTP-based IDS is more sensitive to $n$, the bell-shaped curve is consistently observed across different hardware platforms, and thus may be parameterized through a formal model.}
%	\label{fig:example_data_driven_fitting}
%\end{figure*}

In \hl{our preliminary work} \cite{sagong2018cloaking}, we \hl{investigated} IDSs that use the clock skew for \hl{detecting masquerade attacks}. 
Our key observation is that an adversary, who realizes that the IDS at the receiver ECU computes the clock skew using message inter-arrival times, can manipulate the inter-transmission times by adding delays to emulate the clock skew of the targeted ECU and avoid detection.
We refer to masquerade attacks of this kind as the \textit{cloaking attack} (Fig.~\ref{fig:cloaking_attack_bypasses_cids}).
We experimentally obtained the attack success probability curves (attack success probability as a function of the added inter-transmission delay) and noticed that they have a consistent bell-shaped structure across different hardware platforms, \hl{which may be captured by a formal model}.
In this paper, we provide such formal models \hl{that accurately predict and characterize the attack success probability curves for} the SOTA IDS and its adaptation to the Network Time Protocol (NTP), using parameters of the attacker, the detector, and the hardware platform.  %which would also provide insight into future designs of effective IDSs. 
\hl{Moreover, we collect additional 16+ hours of CAN data from the UW EcoCAR testbed for six representation messages with different periods, message ID levels, and transmitting ECUs for experimental evaluation.
We further demonstrate the applicability of our formal models for different IDS settings and vehicles. }
To the best of our knowledge, this is the first paper that provides formal analyses of clock skew-based IDSs in automotive CAN. 
%Such formal models would also provide insight into future designs of effective IDSs. 
Throughout this paper, we make the following specific contributions:
\begin{itemize}
	\item We propose the cloaking attack, in which an adversary adjusts message inter-transmission times and cloaks its clock to match the
	targeted ECU's clock skew and avoid detection.
	
	\item We analyze and formally model the attack success probability of the proposed attack on both the SOTA and NTP-based IDSs. 
	
%	\item For the SOTA IDS, %we observe from experiments that the attack is usually detected shortly after it begins or not detected at all. 
%	%Based on this insight, 
%	we model its behavior %the behavior of the IDS 
%	immediately after the attack and derive a closed-form approximation for the attack success probability.
%	For the NTP-based IDS, %we observe from experiments that the expected behavior of the IDS can be modeled using parameters including message period, mean inter-arrival time, clock skew, and inter-transmission delay added by the adversary.
%	%Based on this insight, 
%	we model its expected behavior %the behavior of the IDS 
%	during the attack and derive an approximation for the attack success probability. 
	
	\item We evaluate the proposed attack on hardware testbeds, including a CAN bus prototype and a real vehicle (the UW EcoCAR). Our results show that while the NTP-based IDS is more effective than the SOTA IDS in detecting masquerade attacks, the cloaking attack is successful against both IDSs during all hardware trials. 

	%\item We validate our formal analyses through extensive experiments using the data collected from our EcoCAR testbed. % and the Toyota dataset that was also used in \cite{Shin:2016:finger,sagong2018cloaking}.
	%\footnote{The vehicle used in our experiments and other research projects is anonymized to align with the double blind review policy.} \cite{ecocar2018anonymized}. %the University of Washington EcoCAR testbed \cite{ecocar}. 
	%Our results show that our formal models can accurately predict the attack success probability curves for messages with different periods, ID levels, and transmitting ECUs. 
	%We also validate the predictive capability and accuracy of our formal models across different IDS settings and vehicles.
	
	\item We validate our formal analyses using the data collected from the UW EcoCAR \hl{and the Toyota dataset that was also used in \cite{Shin:2016:finger}}.
	\hl{Our results show that our formal models provide accurate predictions of attack success probability curves for different messages, IDS settings, and vehicles.}
	We define a metric called the \textit{Area Deviation Error} (ADE) to measure the modeling accuracy, which is the ratio of the absolute difference of the areas under the predicted and experimental attack success probability curves to the area under the experimental curve. 
	%Since the metric amplifies the modeling error of a high-fidelity model, its value can exceed $1$. 
	Our results show that the average ADEs of the proposed formal models are within $3.0\%$ for the SOTA IDS and $5.7\%$ for the NTP-based IDS.
\end{itemize}

The remainder of this paper is organized as follows. 
Section \ref{sec:related} reviews the related work. Sections \ref{sec:model} presents our system and adversary models.
Section \ref{sec:ids} reviews the SOTA IDS and presents the proposed NTP-based IDS. 
The cloaking attack is proposed in Section~\ref{sec:cloaking_attack}. 
Section~\ref{sec:formal-analysis} presents formal models for the SOTA and NTP-based IDSs.
Section \ref{sec:evaluation} presents the experimental evaluation.  
Section \ref{sec:conclusion} concludes this paper.
\section{Related Work}
\label{sec:related}
Recent experimental studies have shown that automobiles are vulnerable to cyber attacks with potentially life-threatening consequences  such as disabling brakes or overriding steering 
\cite{Koscher:2010:experimental,Miller:2013:adventure,Miller:2015:remote,Checkoway:2011:comprehensive,miller2014survey,foster15fast}, most of which are caused by the lack of security protections in CAN \cite{Checkoway:2011:comprehensive,Lin:2012:cyber}.
%The lack of security mechanisms in CAN has been identified as a root cause behind many of these attacks \cite{Checkoway:2011:comprehensive,Lin:2012:cyber}.
Hence, there is an urgent need for securing CAN buses.

Security solutions for CAN can be broadly classified into schemes that add \rev{cryptographic measures to the CAN bus}~\cite{Herrewege:2011:CANAuth,nilsson2008efficient,Lin:2012:cyber,ying2019tacan} and anomaly-based IDSs that 1) analyze the traffic on the CAN bus including message contents \cite{kang2016intrusion,dario2017detecting,marchetti2017anomaly}, timing/frequency \cite{Miller:2013:adventure,song2016intrusion,moore2017modeling,taylor2015frequency,Hoppe:2008:security},  entropy \cite{Muter:2011:entropy}, \rev{and survival rates \cite{han2018anomaly}}, 2)  exploit the physical characteristics of ECUs extracted from in-vehicle sensing data \cite{ganesan2017exploiting,li2017poster,Muter:2010:structured} or measurements \cite{cho2017viden,cho2015cps,murvay2014source,choi2018identifying,choi2018voltageids}, \rev{and 3) exploit the characteristics of the CAN protocol, such as the remote frame \cite{lee2017otids}}.
%avatefipour2018linking
%and anomaly-based IDSs that detect attacks by analyzing CAN traffic~\cite{Muter:2011:entropy,Miller:2013:adventure,hoppe2011security} or  exploiting in-vehicle measurements \cite{cho2017viden,cho2015cps,murvay2014source,choi2016identifying,Muter:2010:structured}. 
Compared to the CAN traffic, it is more difficult for adversaries to imitate the physical characteristics of ECUs, % in the time and frequency domain, 
such as the mean squared error %and the root mean square amplitude \cite{choi2016identifying} 
of voltage measurements \cite{murvay2014source}. 
In \cite{cho2017viden}, Cho and Shin proposed an IDS called Viden that constructs voltage profiles to identify the attacker.
In \cite{choi2018voltageids}, Choi \textit{et al.} proposed VoltageIDS that leverages the time and frequency domain features of the electrical CAN signals to fingerprint ECUs. 
\rev{In \cite{kneib2018scission}, Kneib and Huth proposed Scission that exploits physical characteristics from analog values of CAN frames to determines if whether was transmitted by the legitimate ECU.}
However, real-time sensing/measurement and processing can be challenging for ECUs with limited resource, which may hinder the deployment of the existing schemes in practice. 
\rev{In addition, it has been shown in \cite{sagong2018exploring} that the extra wires required by voltage-based IDSs may introduce new attack surfaces for various voltage-based attacks.}

\begin{table}[t!]
	\footnotesize
	\centering
	\caption{Frequently used notations.}
	\begin{tabular}{|c|l|}
		\hline
		\textbf{Notation} & \textbf{Description} \\
		\hline
		$a_{k,i}$ & Arrival time of $i$-th message in $k$-th batch \\
		\hline
		$\eta_{k,i}$ & Noise in arrival time of $i$-th message in $k$-th batch \\
		\hline
		$\mu$ & Mean of all inter-arrival times before the attack \\
		\hline
		$\mu[k]$ & Mean of inter-arrival times in $k$-th batch \\
		\hline
		$\sigma$ & Standard deviation of all inter-arrival times \\
		\hline
		$\sigma_\eta$ & Standard deviation of noise in arrival times\\
		\hline
		$N$ & Batch size\\
		\hline
		$O$ & (Constant) clock offset in each period $T$ \\
		\hline
		$O_{avg}[k]$ & Average offset in $k$-th batch \\
		\hline
		$O_{acc}[k]$ & Accumulated offset up to $k$-th batch \\
		\hline
		$S[k]$ & Clock skew estimate in $k$-th batch \\
		\hline
		$t[k]$ & Elapsed time up to last message in $k$-th batch \\
		\hline
		$e[k]$ & (Unnormalized) identification error in $k$-th batch \\
		\hline
		$\mu_{\text{CUSUM}}$ & Mean of reference identification errors \\
		\hline
		$\sigma_{\text{CUSUM}}$ & Standard deviation of reference  identification errors \\
		\hline
		$e_n[k]$ & Normalized identification error in $k$-th batch \\
		\hline
		$e_{ref}[k]$ & Identification error used as reference in CUSUM\\
		\hline
		$L^{+}[k]$, $L^{-}[k]$ & Upper and lower control limits in $k$-th batch \\
		\hline
		$\Gamma$ & CUSUM detection threshold \\
		\hline
		$\gamma$ & CUSUM update threshold \\
		\hline
		$\kappa$ & CUSUM sensitivity parameter \\
		\hline
		\multirow{ 2}{*}{$\Delta T_0$} & Inter-transmission delay added by adversary that
		\\ & exactly achieves the targeted ECU's clock skew\\
		\hline
		$\Delta T$ & Difference between the total added delay and $\Delta T_0$\\
		\hline
		$P_s$ & Probability of a successful cloaking attack \\
		\hline
		\multirow{ 2}{*}{$\tau$} & Rate of decrease of normalized identification error  \\ & after an attack occurs (for the SOTA IDS) \\
		\hline
		$\hat{S}[k], \hat{t}[k]$  & Expected value of $S[k], t[k], O_{acc}[k], e[k]$ (for the\\
		$\hat{O}_{acc}[k], \hat{e}[k]$ &  NTP-based IDS) \\
		\hline
	\end{tabular}
	
	\label{table:notation}
	\normalsize
\end{table}

A novel IDS that uses the clock skew to fingerprint ECUs was proposed in \cite{Shin:2016:finger}. 
As a physical invariant, the clock skew can be estimated from the timestamps of periodically received CAN messages and used for detecting masquerade attacks. 
In this paper, we propose the cloaking attack, in which the adversary alters the message inter-transmission times to match the clock skew of the targeted ECU and evade detection with a high probability.
We further propose formal models that predict the attack success probability for a given CAN bus and IDS with high accuracy.

%It was later shown in \cite{sagong2018cloaking} that an intelligent adversary can evade detection with a high probability by altering message inter-transmission times to match the clock skew of the targeted ECU. The evaluations of \cite{Shin:2016:finger,sagong2018cloaking}, however, are experiment-driven and do not include a formal model for predicting the attack success probability for a given CAN bus and IDS. 
%To fill this gap, we provide a formal analysis for clock skew-based IDSs with high prediction accuracy.

\section{System Model}
\label{sec:model}
In this section, we provide brief background on the CAN protocol, review clock-related concepts as defined in NTP, and present our timing model for the CAN bus. 
A list of frequently used notations is provided in Table~\ref{table:notation}. 

\subsection{CAN Background}
The CAN protocol \cite{ISO:2015,Bosch:1991} is one of the most widely used in-vehicle network standards.
It allows in-vehicle ECUs to broadcast messages, and almost all CAN messages are periodic. % in automobiles.
%As shown in Fig.~\ref{fig:CAN_DataFrame}, 
In particular, CAN messages do not have transmit timestamps and do not support encryption or authentication.

%Each CAN message has a message ID, and a lower ID suggests a higher priority. If multiple ECUs are attempting transmissions at the same time, the one with the lowest ID will win out through a process known as \textit{arbitration}. Arbitration is built on the fact that the CAN bus acts as a logical AND gate. For instance, consider two ECUs that attempt to transmit messages with IDs 0x100 and 0x010 simultaneously. Both ECUs will transmit one bit at a time, starting from the most significant bit of its message ID. The ECU that transmits a higher message ID 0x100 will observe a 0 bit while it had transmitted a 1 bit. It immediately realizes that a higher priority message is being transmitted, and thus stops its transmission. 

\subsection{Clock-Related Concepts in NTP}
\label{sec:clock_related_concepts} 
Let $C_{A}(t)$ denote the time kept by clock $A$, and $C_{true}(t) = t$ be the true time. 
According to the NTP \cite{Moon:1998,Mills:1992:NTP}, %Paxson:1998:CMP:277858.277865
the \textit{clock offset} of clock A is given by  
\begin{equation}
\label{eq:NTP_def_offset}
O_{A}(t) = C_{A}(t) - C_{true}(t),
\end{equation}
which is the difference between the time reported by $C_A$ and the true time. 
%Let us first define ${C_{true}}$ as the ``true'' clock that  runs at a constant rate, i.e., ${C_{true}}(t)=t$.
%Let $C_A(t)$ denote the time kept by clock $A$. 
%The \textit{clock offset} of $C_A$, denoted as $O_A(t)$, is the difference between the time reported by $C_A$ and the ``true'' time, i.e.,
%\begin{equation}
%O_A(t) = C_A(t) - C_{true}(t).\label{eq:NTP_def_offset}
%\end{equation}
The \textit{frequency} of $C_A$ at time $t$, denoted $C'_A(t)$, is the first derivative of $C_{A}(t)$, while the \textit{clock skew} $S_{A}(t)$ is the first derivative of the clock offset $O_{A}(t)$. 
%The \textit{clock skew} of $C_A$, denoted as $S_A(t)$, is the difference in the frequencies (or first derivatives) of $C_A$ and $C_{true}$, i.e.,
%\begin{equation}
%S_A(t) = C_A'(t) - C_{true}'(t).
%\end{equation}
A positive clock skew means that $C_A$ runs faster than $C_{true}$.
%, while a negative clock skew implies that $C_{A}$ runs slower than $C_{true}$.
The unit of clock skew is microseconds per second ($\mu$s/s) or parts per million (ppm). 
For example, if $C_A$ is faster by $2~\mu$s every $20$ ms w.r.t. $C_{true}$, then its clock skew relative to $C_{true}$ is $100$ ppm.

In-vehicle ECUs typically have constant clock skews \cite{Shin:2016:finger}. 
Suppose that $C_A$ has a constant clock skew $S_A$.
If $\Delta t$ is the time duration measured by $C_{true}$, the amount of time that has passed according to $C_A$ is 
%\begin{equation}
$\Delta t_A = (1+S_A)\Delta t$,
%\end{equation} 
and $\Delta t = \Delta t_{A}/(1+S_{A})$. 
Similarly, if there is a second non-true clock $B$ with a constant clock skew $S_B$ that reports a time duration of $\Delta t_B$, we have $\Delta t_B = (1+S_B)\Delta t$.
Then the clock skew of $C_B$ relative to $C_A$, denoted as $S_{BA}$, is given by
\begin{equation}
S_{BA} = \frac{\Delta t_B- \Delta t_A}{\Delta t_A} = \frac{S_B - S_A}{1+S_A} \label{eq:relative_skew}
\end{equation}
and the relationship between $S_{BA}$ and $S_{AB}$
%, that is, the clock skew of $C_A$ relative to $C_B$, 
is given by
\begin{equation}
S_{AB} = \frac{-S_{BA}}{1+S_{BA}}. \label{eq:relative_skew_conversion}
\end{equation}

In the absence of a true clock, the \textit{relative clock offset} and \textit{relative clock skew} can be defined  with respect to a reference clock. 
%For clocks A and B, the relative skew of clock A with respect to clock B is defined by $S_{AB} = C_{A}^{\prime}(t)-C_{B}^{\prime}(t)$.
Two clocks are said to be \textit{synchronized} at time $t$ if both the relative clock offset and relative clock skew are zero. 

%\begin{figure}[t!]
%	\centering
%	\includegraphics[width=1\columnwidth]{./figures/can_frame_structure.pdf}
%	\caption{Each CAN message (or frame) consists of Start of Frame (SOF) field, Arbitration field (including a 11-bit message ID), Control field, Data field (8-64 bits), CRC field, ACK field, and End of Frame (EOF) field. }
%	\label{fig:CAN_DataFrame}
%	\vspace{-0.3cm}
%\end{figure}

\subsection{Timing Model}
\label{sec:timing}
\begin{figure}[t!]
	\centering
	\includegraphics[width=0.8\columnwidth]{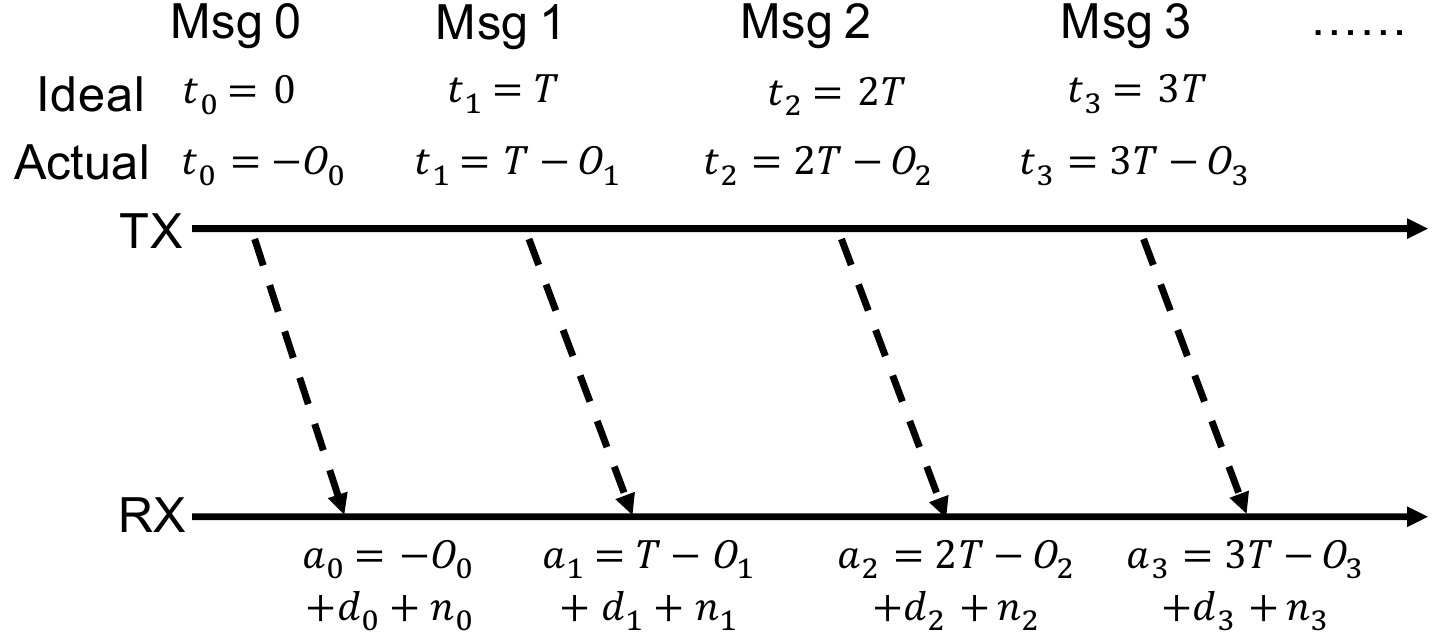}
	\caption{Timing model of message arrivals on CAN bus.}
	\label{fig:timing_model}
	\vspace{-0.3cm}
\end{figure}

We now discuss our timing model in Fig.~\ref{fig:timing_model}, in which the receiving ECU R timestamps messages that arrive periodically. 
We consider R's clock as the reference clock and refer to the relative offset and relative skew of the transmitter's clock as offset and skew, respectively.

Consider an ECU that transmits a message every $T$ seconds as per its local clock. 
If the two clocks are synchronized, the $i$-th message will be transmitted at $t_i = iT$ in R's clock. 
%\footnote{Strictly speaking, $t_i$ is the time when the transmitter puts the first bit of message $i$ into the outgoing buffer. }.
However, due to the transmitter's clock skew, there exists an \textit{accumulated offset} $O_i$ between the transmitter's clock that reports time $iT$ and R's clock that reports time $t_i$ since the transmission of message $0$, which means $O_i = iT - t_i$ according to Eq.~(\ref{eq:NTP_def_offset}).
Therefore, the actual transmission time is $t_i=iT - O_i$ in R's clock.
While the clock skew may be slowly varying due to factors like temperature, it is almost constant over short durations. 
Hence, we model the accumulated offset as a random variable $O_i = iO + \epsilon_i$, where $O$ is the clock offset induced in one period $T$ given the constant clock skew, and $\epsilon_i$ is the offset deviation due to jitters in the transmitter.
We assume that the $\epsilon_i$'s are independent and identically distributed zero-mean random variables.
After a network delay of $d_i$ (due to message transmission, propagation, and reception), the message arrives at R's incoming buffer and has a timestamp \begin{equation}
a_i = iT - iO - \epsilon_i + d_i + n_i,
\end{equation} 
where $n_i$ is the zero-mean noise introduced by R's timestamp quantization process \cite{Zander:2008:ICM:1496711.1496726}.

Let $\eta_i = - \epsilon_i + d_i + n_i$ and thus $a_i = iT - iO + \eta_i$. 
Since the data lengths of periodic CAN messages are constant over time, it is reasonable to assume constant-mean network delays, i.e., $\mathbb{E}[d_i]=d$.
Hence, we model the $\eta_i$'s as i.i.d. Gaussian random variables with  $\eta_i \sim N(d, \sigma_\eta^2)$.

The inter-arrival time between the ($i-1$)-th message and the $i$-th message %, denoted as $T_{rx,i}$, is given by 
is $T_{rx,i} = a_i - a_{i-1} = (T-O) + (\eta_i - \eta_{i-1})$.
%\begin{equation*}
%T_{rx,i} = a_i - a_{i-1} = (T-O) + (\eta_i - \eta_{i-1}).
%\end{equation*}
Hence, the inter-arrival times have a mean $\mu \triangleq \mathbb{E}[T_{rx,i}] = T-O$, and a variance $\sigma^2 \triangleq Var(T_{rx,i})= 2\sigma_{\eta}^2$.

\subsection{Adversary Model}
\label{sec:adversary}
We consider adversaries who gain access to the CAN bus of an automobile by compromising one or more ECUs. 
We adopt the following two adversary models \cite{foster15fast,Shin:2016:finger}:
% that were initially presented in \cite{Shin:2016:finger} and later used in \cite{sagong2017cloaking}: 
%for mounting attacks on the CAN network:
%As in \cite{Shin:2016:finger} and \cite{sagong2017cloaking}, we distinguish between the following two types of adversaries:
\begin{itemize}
\item \emph{Weak adversary --} A weak adversary who compromises an ECU is able to eavesdrop on all the CAN traffic and can block outgoing messages from the compromised ECU. The weak adversary, however, cannot send messages from the compromised ECU.
\item \emph{Strong adversary --} A strong adversary who compromises an ECU has complete control over the compromised ECU, including eavesdropping on all messages, blocking outgoing messages, and transmitting messages with the timing and content of the adversary's choosing.
\end{itemize}
%\textbf{*Need to explain how these levels of compromise can arise.} In what follows, we present two attacks that can be mounted by the types of adversaries described above. Both attacks aim to inject false messages in order to disrupt the safety and performance of the vehicle.

\begin{figure}[t]
	\centering
	\includegraphics[width=1\columnwidth]{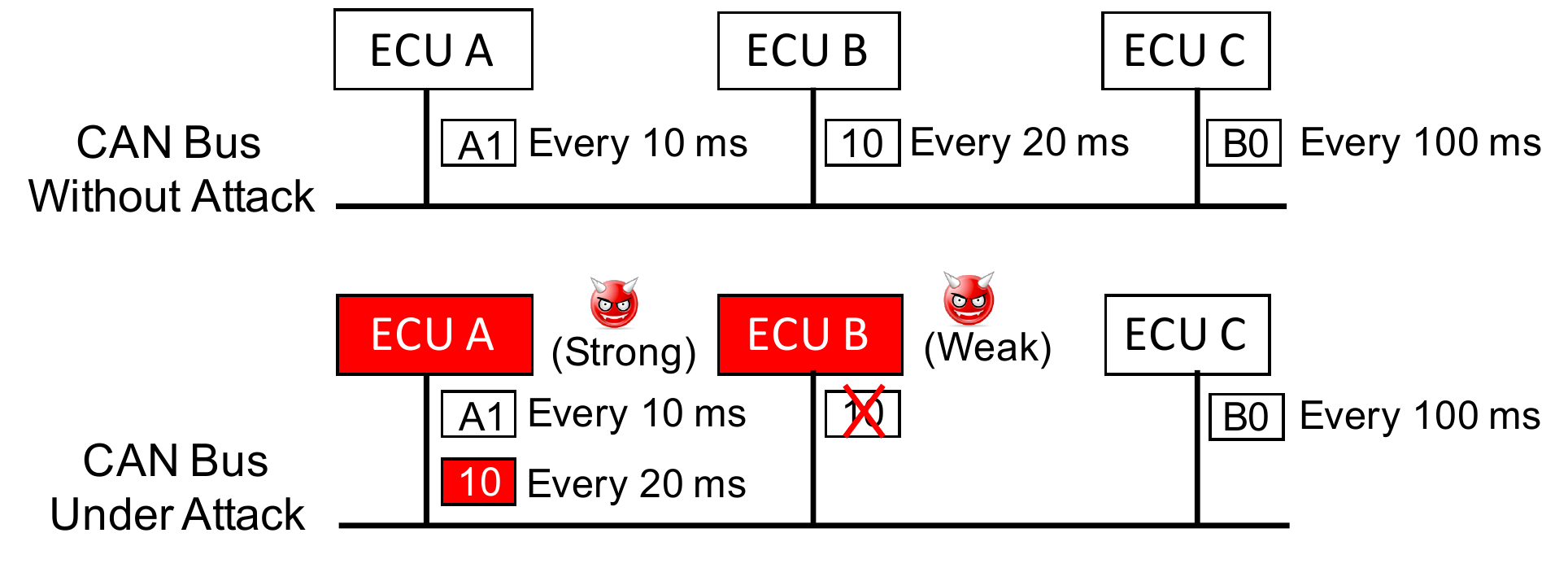}
	\caption{Illustration of a masquerade attack. Without a masquerade attack, ECU A transmits message 0xA1 every 10 ms, and ECU B transmits message 0x10 every 20 ms. During the masquerade attack, ECU B is weakly compromised and its transmission of 0x10 is blocked. %by the weak adversary. 
		Meanwhile, ECU A is strongly compromised and is used to inject the false messages 0x10 every 20 ms in addition to its original message 0xA1. 
	}
	\label{fig:masquerade_attack_model}
	\vspace{-0.4cm}
\end{figure}

%\subsection{Masquerade Attack} 
%\label{subsec:masquerade}
We consider adversaries who attempt to mount \emph{masquerade attacks}.
Fig.~\ref{fig:masquerade_attack_model} illustrates a masquerade attack that is mounted by a weak adversary and a strong adversary acting in coordination.
The strong adversary has compromised ECU A, while the weak adversary has compromised ECU B. 
The goal of the attack is to inject false messages from ECU A, so as to degrade the safety, performance, and/or functionality of the vehicle. 
This attack enables an adversary who compromises a low-priority\footnote{On the CAN bus, messages with smaller ID levels (i.e., higher priorities) will be transmitted earlier in the event of collisions through a process called arbitration. A larger ID indicates a lower priority. See \cite{Bosch:1991} for more details.} %, outward-facing 
ECU to effectively impersonate a higher-priority ECU, thus maximizing the impact of the attack.

We observe that, if ECU B were compromised by a strong adversary, the attack would be trivial. 
On the other hand, when ECU B is compromised by a weak adversary, the adversary cannot directly inject messages from ECU B itself. Instead, the weak adversary blocks the targeted messages %all outgoing messages 
from ECU B. 
The strong adversary then uses the compromised ECU A to inject false messages that are claimed to be from ECU B.

This attack exploits two vulnerabilities of CAN that have been identified in the related literature~\cite{Checkoway:2011:comprehensive,Shin:2016:finger}. First,  all ECUs have access to the same broadcast medium, allowing easily-compromised, low-priority ECUs (ECU A in Fig.~\ref{fig:masquerade_attack_model}) %(ECU A in the example above) 
to listen to and impersonate higher-priority ECUs. Second, the lack of integrity checks means that spoofed messages from ECU A are not detected as long as the normal formatting and error-correction checks of CAN messages are passed. %any ECU can broadcast messages claiming to be from 

%\subsection{Cloaking Attack}
%\label{subsec:cloaking}
%The masquerade attack enables coordinating weak and strong adversaries to inject false messages claiming to be from a victim ECU. While the CAN protocol lacks integrity checks that could be used to detect such attacks, intrusion detection systems (IDSs) have been proposed that leverage side-channel information such as message timing. 

%The cloaking attack attempts to thwart side-channel based IDSs by masking the physical invariants of the compromised ECU. In the case of message timing, the adversary can passively eavesdrop on the victim ECU (e.g., ECU A above) in order to learn the clock skew of the victim. After learning the clock skew of the victim, the adversary can deliberately alter the transmission times of the spoofed messages in order to mimic the victim's clock skew. If performed successfully, this attack will prevent a clock skew based IDS from detecting the attack. The specific details and success rate of the attack will be determined by the IDS that is used. 

%the attack leverages the open broadcast nature and lack of integrity checks in the CAN protocol, as follows.
\section{Clock Skew-Based IDS}
\label{sec:ids}
Clock skew-based IDSs leverage \rev{the} clock skew to uniquely fingerprint each ECU and detect masquerade attacks. 
\rev{Since CAN messages do not have transmit timestamps,  approaches that require transmit timestamps for clock skew estimation such as \cite{Jana:2008:on,Kohno:2005:remote,zander2008improved} are not applicable.}
%\rev{but} the lack of transmit timestamps in CAN messages renders \rev{many} clock skew-based IDSs %developed in other domains 
%\rev{that require transmit timestamps} inapplicable~\cite{Jana:2008:on,Kohno:2005:remote,Zander:2008:ICM:1496711.1496726}.
%The challenge of using clock skews for fingerprinting on CAN buses is the lack of transmit timestamps in CAN messages, which renders clock skew-based IDSs developed in other domains inapplicable~\cite{Jana:2008:on,Kohno:2005:remote,Zander:2008:ICM:1496711.1496726}. 
\rev{Similar to \cite{novak2013preventing}}, clock skew-based IDSs on CAN buses instead exploit \rev{traffic periodicity}
%the periodic nature of \rev{the} %CAN 
%traffic~
\cite{Shin:2016:finger}. 
Since almost all messages are transmitted periodically, the receiving IDS can monitor the inter-arrival times of a target message and estimate the clock skew of the transmitting ECU accordingly. We note that this approach is only viable for periodic message traffic. 
%The approach to clock skew estimation taken by the clock skew-based IDSs is described as follows.
In the rest of this section, we will review the SOTA IDS and propose an NTP-based IDS.

\subsection{Review of SOTA IDS}
The SOTA IDS in \cite{Shin:2016:finger} consists of a clock skew estimator and a CUSUM (Cumulative Sum \cite{basseville1993detection})-based detector. 
The estimator tracks the clock skew from message inter-arrival times and feeds identification errors to the CUSUM for detection.
We now describe the two components in more detail.

%\textbf{Clock skew estimator.}
\subsubsection{Clock Skew Estimator}
Incoming periodic messages are processed in batches of size $N$ to mitigate undesired impacts of quantization and other sources of noise in receive timestamps. 
Let $a_{k,i}$ be the arrival time of the $i$-th message in the $k$-th batch. 
The average offset of the $k$-th batch is given by 
\begin{equation}
\label{eq:shin_avg_offset}
O_{avg}[k]=\frac{1}{N-1}\sum_{i=2}^{N}[a_{k,i}-(a_{k,1}+(i-1)\mu[k-1])],
\end{equation}
where $\mu[k-1]$ is the mean inter-arrival time of the previous ($(k-1)$-th) batch.

The absolute value of $O_{avg}[k]$ is added to the previous accumulated offset to compute the updated value, 
\begin{equation}
\label{eq:shin_acc_offset}
O_{acc}[k]=O_{acc}[k-1]+|O_{avg}[k]|,
\end{equation}
which is modeled as $O_{acc}[k]=S[k]t[k]+e[k]$, where $S[k]$, $t[k]$, and $e[k]$ denote the clock skew estimate in batch $k$, the elapsed time until the last message of the $k$-th batch, and the (unnormalized) identification error in batch $k$, respectively. 

The estimated clock skew $S[k]$ is the output of the Recursive Least Squares (RLS) algorithm. 
Ideally, the identification error would converge to zero if clock skew is correctly estimated. 
Hence, a change in the identification error indicates a change in the clock skew.
Besides, the rate of convergence is governed by a parameter $\lambda<1$ (e.g., $\lambda=0.9995$) that exponentially weighs past samples.
More details are available in \cite{Shin:2016:finger}.

%\textbf{CUSUM-based detector.} 
\subsubsection{CUSUM-Based Detector} 
The detector tracks the mean $\mu_{\text{CUSUM}}$ and the standard deviation $\sigma_{\text{CUSUM}}$ of identification errors that are used as reference (denoted as $\{e_{ref}[k]\}$). 
In batch $k$, %the identification error 
$e[k]$ is first normalized as $e_{n}[k] = (e[k]-\mu_{\text{CUSUM}}[k-1])/\sigma_{\text{CUSUM}}[k-1]$. 
To mitigate the undesired impact of outliers, $e[k]$ will be considered as a reference error sample for updating $\mu_{\text{CUSUM}}$ and $\sigma_{\text{CUSUM}}$ only if $e_n[k]$ is less than the preset update threshold $\gamma$ (e.g., $\gamma=4$), as noted in \cite{Shin:2016:finger}.

The detector then uses %the normalized identification error 
$e_{n}[k]$ to update the upper control limit $L^{+}$ and the lower control limit $L^{-}$ in batch $k$ as follows
\begin{align}
\label{eq:cusum-1}
L^+[k] &= \max[0,L^+[k-1]+e_{n}[k]-\kappa], \\
\label{eq:cusum-2}
L^-[k] &= \max[0,L^-[k-1]-e_{n}[k]-\kappa],
\end{align}
where $\kappa$ is a sensitivity parameter that reflects the number of standard deviations to be detected. 
%The detector declares an attack if either the control limit exceeds the preset detection threshold $\Gamma$ (e.g., $\Gamma=5$  \cite{Shin:2016:finger}).
The detector declares an attack if either the control limit, \rev{$L^+$ or $L^-$}, exceeds the preset detection threshold $\Gamma$, \rev{which implies a sudden positive or negative shift in value, respectively.
	As the general rule of thumb for CUSUM, $\Gamma$ is usually set to $4$ or $5$ \cite{montgomery2007introduction}, and the SOTA IDS  chooses $\Gamma=5$.}

\subsection{Proposed NTP-based IDS}
We now present an adapted IDS that computes clock offset and clock skew as per the NTP specifications, which is referred to as the NTP-based IDS.
The motivation for our NTP-based IDS is two-fold. 
First, we note that the metric in Eq.~(\ref{eq:shin_avg_offset}) is not consistent with the NTP definition in Eq.~(\ref{eq:NTP_def_offset}), since it does not calculate the time difference between the transmitter's clock and the reference clock. 
In addition, it is assumed that $O_i$ is a random variable and $\mathbb{E}[O_i - O_{i-1}]=0$.
It implies that $\mathbb{E}[O_i]=\mathbb{E}[O_j]$ for $i\neq j$, which does not hold in general since offsets accumulate over time (if $i\gg j$, $\mathbb{E}[O_i] \gg \mathbb{E}[O_j]$).
%clocks have different skews. 
Our second motivation is the widespread use and acceptance of NTP as a timing mechanism for real-time systems, which raises the question of whether NTP definitions of clocks can be used for intrusion detection as well.  
While both the SOTA IDS \cite{Shin:2016:finger} and the proposed NTP-based IDS estimate the clock skew via the RLS and detect an attack via the CUSUM, they update average and accumulated offsets differently, as explained below.

Let $T$ be the message period and  $\hat O_i$ be the clock offset of the $i$-th period observed by the receiver.
According to the NTP clock definitions  (Section~\ref{sec:clock_related_concepts}) and the timing model (Section~\ref{sec:timing}), $\hat O_i$ is equal to 
\begin{equation}
\hat O_i = T - (a_i-a_{i-1})=O - \Delta \eta_i,
\end{equation}
where $\Delta \eta_i = \eta_i - \eta_{i-1}$.
In batch $k$, the average offset is %computed as
\begin{equation}
\label{eq:NTP_avg_offset}
O_{avg}[k] = \frac{1}{N}\sum_{i=1}^{N}{\hat{O}_{k,i}} = T - \frac{a_{k,N}-a_{k,0}}{N},
\end{equation}
where $a_{k,0}=a_{k-1,N}$ is the receive timestamp of the last message in the previous (($k-1$)-th) batch. The accumulated offset of the $k$-th batch is updated %by adding $N\times O_{avg}[k]$ to the previous accumulated offset.
as follows
\begin{equation}
\label{eq:NTP_acc_offset}
O_{acc}[k] = O_{acc}[k-1] + N O_{avg}[k].
\end{equation}  

Eq.~(\ref{eq:shin_avg_offset}) and (\ref{eq:NTP_avg_offset}) highlight the differences in how the average offset is updated by the SOTA and NTP-based IDSs, respectively.
Similarly, Eq.~(\ref{eq:shin_acc_offset}) and (\ref{eq:NTP_acc_offset}) show how the SOTA and NTP-based IDSs update the accumulated offset, respectively.
\hl{Compared to the SOTA IDS, the NTP-based IDS provides more consistent clock skew estimates for the same message across different batch sizes and data traces. See Appendix~\ref{appendix:estimation_consistency} for a detailed discussion.}
As we will show in Section~\ref{sec:evaluation}, the NTP-based IDS is more effective in detecting masquerade attacks than the SOTA IDS.

\section{Proposed Cloaking Attack}
\label{sec:cloaking_attack}
In this section, we propose a new masquerade attack called the \textit{cloaking attack}, in which the adversary adjusts the inter-transmission times of the spoofed messages in order to manipulate the estimated clock skew and bypass an IDS.

Consider a message transmitted by the targeted ECU B every $T$ seconds in its own clock, which corresponds to every $\hat{T}=T/(1+S_B)$ seconds in the receiver R's clock, where $S_B$ is B's clock skew.
For the ease of discussion, we ignore offset deviations and the noise in arrival timestamps due to network delay and quantization.
Then B's clock skew as estimated by R is given by $\hat{S} = (T-\hat{T})/\hat{T} = S_B$.

In a masquerade attack, the weak adversary prevents ECU B from transmitting the targeted message, and the strong adversary controlling ECU A transmits the spoofed message every $T$ seconds as per A's local clock $C_A$. 
Hence, ECU R receives messages every $\hat{T}^{\prime} = T/(1+S_{A})$ seconds, as measured by $C_{R}$, where $S_{A}$ is A's clock skew. 
The clock skew measured by ECU R %for the messages injected by the attacker 
will then be $\hat{S}^{\prime} = (T-\hat{T}')/\hat{T}'= S_{A}$. 
Hence, if $S_{A} \neq S_{B}$, then the IDS will detect a change in the estimated clock skew after the adversary launches the attack.

The insight underlying our attack is that, while clock skew is a physical invariant, clock skew estimation in an IDS is based entirely on message inter-arrival times, which can be easily manipulated by the transmitter (i.e., the strong adversary controlling ECU A) adjusting the message inter-transmission times. Effectively, the adversary \emph{cloaks} the skew of its hardware clock, % by using the software clock, 
thus motivating the term \emph{cloaking attack}. Under the cloaking attack, instead of transmitting every $T$ seconds, the compromised ECU A transmits every $\tilde{T}=T + \Delta T_0$ seconds, in order to match the clock skew observed at R. 

We now discuss the choice of $\Delta T_0$. Under the cloaking attack, the inter-arrival time observed by R is 
\begin{equation*}
\hat{T}'' = \frac{\tilde{T}}{1 + S_A} = \frac{T+\Delta T_0}{1 + S_A}
\end{equation*}
and the transmitter's clock skew estimated by R is
\begin{equation}
\hat{S}'' = \frac{T - \hat{T}''}{\hat{T}''} = \frac{S_A \cdot T - \Delta T_0}{T+\Delta T_0}.
\end{equation}
Hence, to bypass the IDS, the adversary needs to choose  $\Delta T_0$ such that $\hat{S}''=\hat{S}$, or equivalently $\hat{T}'' = \hat{T}$, which means
\begin{equation}
\Delta T_0 = \frac{(S_A - S_B)}{1+S_B} \cdot T = S_{AB} \cdot T = \frac{-S_{BA}}{1+S_{BA}}\cdot T, \label{eq:delta_T}
\end{equation}
where $S_{AB}$ is A's clock skew relative to B's clock, and the last two equalities are due to Eq.~(\ref{eq:relative_skew}) and Eq.~(\ref{eq:relative_skew_conversion}), respectively.

Therefore, the message inter-transmission time $\tilde{T}$ would be
\begin{equation*}
	\tilde{T} = T + \Delta T_0 = T - \frac{S_{BA}}{1+S_{BA}} T = \frac{T}{1+S_{BA}},
\end{equation*}
which is the period of the message from B (weak adversary) measured by the local clock of A (strong adversary). 

To summarize, the cloaking attack is performed as follows. 
After the adversary compromises two ECUs as strong and weak adversaries, the strong adversary estimates the period of the targeted message $\tilde{T}$ using its local clock. 
During the cloaking attack, the strong adversary transmits  spoofed messages every $\tilde{T}$ seconds. 
While the preceding analysis ignores the noise in the system, our results in Section \ref{sec:evaluation} show that the cloaking attack is effective in a realistic environment.
%\footnote{\hl{It is worth noting that the SOTA IDS in \cite{choi2016identifying} contains a correlation detector as an optional feature, and the cloaking attack is also able to thwart the correlation detector, as discussed in Appendix~\ref{sec:cloaking_on_correlation_analysis}.}}.

In practice, however, the adversary may not be able to achieve the exact value of $\Delta T_0$ due to hardware limitations and possible measurement inaccuracy.
Let the total amount of the actual inter-transmission delay added by the adversary be $\Delta T + \Delta T_0$, where $\Delta T$ is the amount of deviation from $\Delta T_0$.
When $\Delta T$ is closer to zero, the attack will be successful with a higher probability.
Hence, the attack success probability $P_{s}$ is a function of $\Delta T$ (an attack parameter), parameters of the detector (e.g., $\lambda$, $\gamma$, and $\Gamma$), and the hardware platform.
In order to %accurately 
predict and characterize the impact of the cloaking attack on a CAN bus and IDS without having to solely rely on extensive experiments, we aim to  formally model $P_s$ for both the SOTA and NTP-based IDSs, as presented below.

\section{Formal Analysis}
\label{sec:formal-analysis}
\subsection{Formal Analysis of SOTA IDS}
\label{sec:state-of-art}
In this section, we develop a formal model for the probability of a successful cloaking attack $P_s$ as a function of  parameters including the distribution of message inter-arrival times, the message period, the added inter-transmission delay, and the detection parameters of the IDS. 
We first present our modeling assumptions and observations. We then formulate our formal model and derive $P_s$ %the attack success probability 
for the SOTA IDS.

\subsubsection{Assumptions for SOTA IDS}
\label{subsec:soa-assumptions}
For the SOTA IDS, the detection parameters including batch size $N$ and CUSUM parameters $\Gamma$ (the detection threshold) and $\kappa$ (the sensitivity parameter) are known to the IDS.
Since the IDS records all message arrival timestamps, it knows the message period $T$ and can measure the mean $\mu$ and standard deviation $\sigma$ of the message inter-arrival times.

Our analysis takes as input a ``snapshot'' of the IDS right before the attack that begins in the $m$-th batch. 
This means that the following parameters maintained by the IDS are readily available: the mean $\mu_{\text{CUSUM}}$ and standard deviation $\sigma_{\text{CUSUM}}$ of the reference identification errors in the CUSUM, % before the attack, 
the average inter-arrival time $\mu[m-1]$, the accumulated offset $O_{acc}[m-1]$, the estimated skew $S[m-1]$, and the elapsed time $t[m-1]$.

\subsubsection{Observations}
\label{subsec:soa-obs}
\begin{figure*}[t!]
	\centering
	\begin{subfigure}[h]{0.32\textwidth}
		\includegraphics[width=\textwidth]{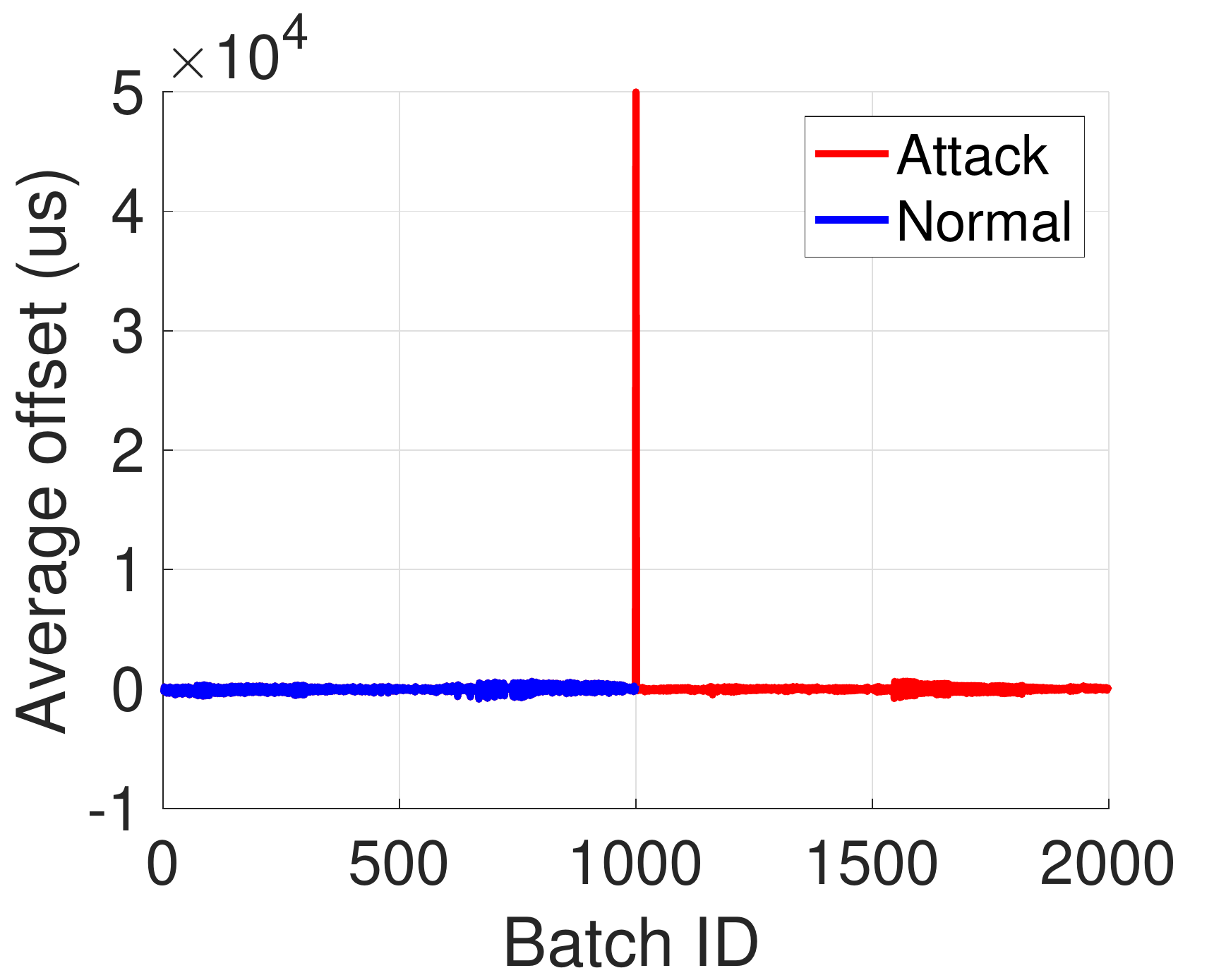}
		\caption{}
		\label{fig:SoA-offset}
	\end{subfigure}
	\begin{subfigure}[h]{0.32\textwidth}
		\includegraphics[width=\textwidth]{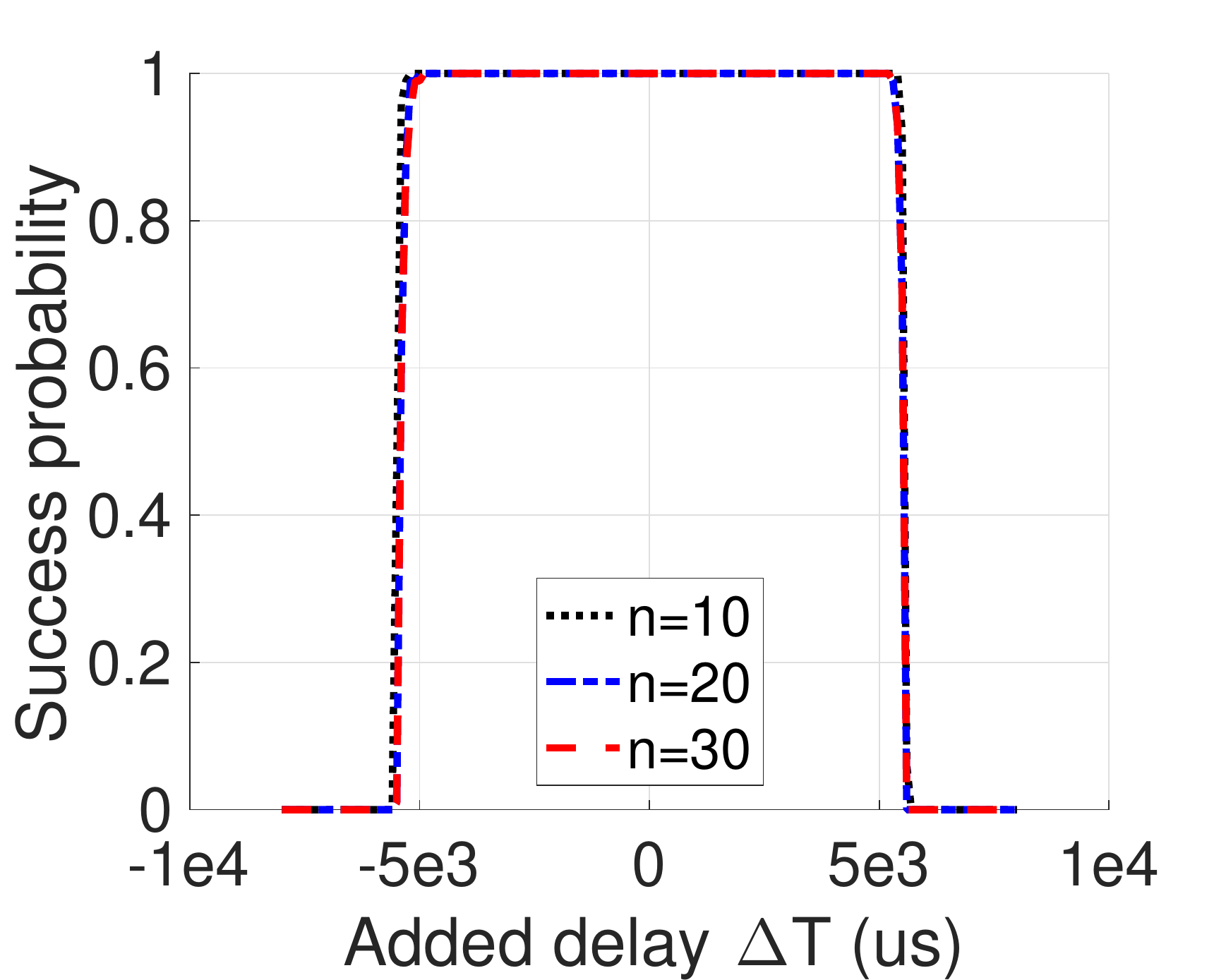}
		\caption{}
		\label{fig:SoA-success}
	\end{subfigure}
	\begin{subfigure}[h]{0.32\textwidth}
		\includegraphics[width=\textwidth]{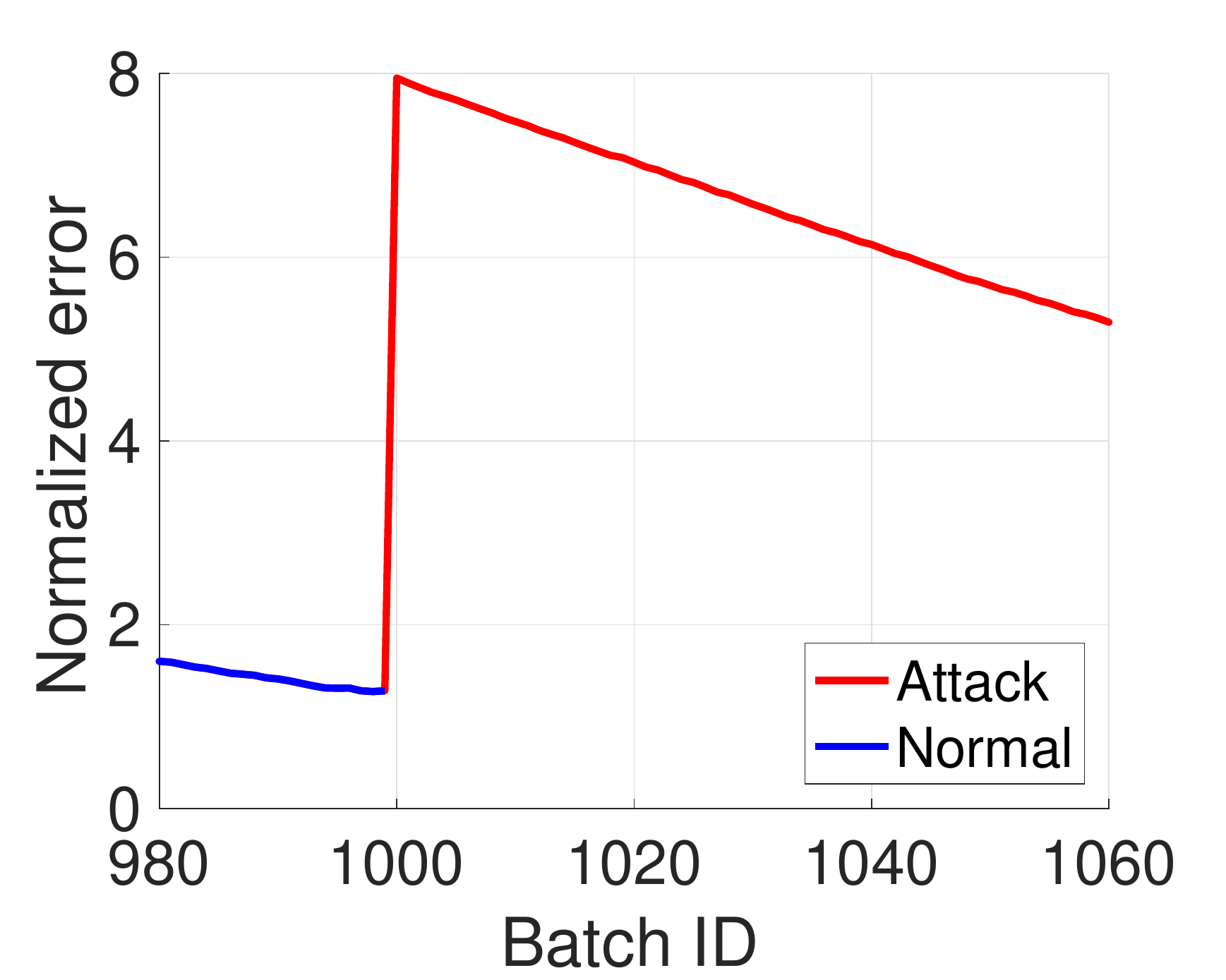}
		\caption{}
		\label{fig:SoA-error}
	\end{subfigure}
	\caption{Impact of the cloaking attack on the SOTA IDS.
		(a) Average offset as a function of batch ID. Only the first attack batch has a large average offset. 
		%The masquerade attack starts from the $1000$-th batch. 
		%(b) Observations used in derivation of error distribution for state-of-the-art IDS. 
		(b) The attack success rates are roughly the same for $n=10$, $20$, and $30$ attack batches. 
		%, suggesting that attacks are typically detected in the first ten batches. 
		(c) The normalized identification error suddenly increases when the attack begins, and it then starts decreasing at an almost constant rate. 
		Note that the figures are generated using the data for the $20$ ms message 0x185 collected from the UW EcoCAR testbed. 
		We set $N=20$, $\gamma=4$, $\Gamma=5$, and $\kappa=8$.
		The attack data is obtained by adding $5$ ms to the inter-arrival times of the cloaking data collected from the UW EcoCAR testbed, and the attack starts from batch $1000$. }
\end{figure*}

Our modeling and analysis of the SOTA IDS are based on the following observations. 
As shown in Fig.~\ref{fig:SoA-offset}, the first batch after the attack begins is the only batch that has a large average offset, and all subsequent batches have small offsets. 
This is because the average offset of the current batch is computed from the mean inter-arrival time of the previous batch (Eq.~(\ref{eq:shin_avg_offset})). 
The first attack batch has a very different mean inter-arrival time from the last normal batch due to $\Delta T$, whereas adjacent batches before and after the attack have close mean inter-arrival time.

As a result, for an attack that begins in the $m$-th batch\footnote{We assume that the first attack message appears as the 1st message of the $m$-th batch.}, the identification error  will be larger due to the sudden change in the mean inter-arrival time and will decrease over time due to clock skew update. 
In fact, we observe that the attack is usually either detected during the first tens of batches following the attack, or is not detected at all (Fig. \ref{fig:SoA-success}).

If we take a closer look at the first tens of batches after the attack begins, we observe a linear decrease in the normalized identification error (Fig.~\ref{fig:SoA-error}). 
These observations motivate the following model of the normalized identification error $e_n[k]$ at batch $k\geq m$ 
\begin{equation}
e_n[k] \approx e_n[m] - \tau(k-m),
\label{eq:SoA_norm_error_model}
\end{equation}
where $\tau>0$ is a constant slope representing the rate of decrease of the normalized identification error.

\subsubsection{Attack Success Probability}
\label{subsec:initial-error}
Based on the observations of Section \ref{subsec:soa-obs}, we divide our formal analysis into three stages: 1) modeling the distribution of the normalized identification error in the first attack batch $e_n[m]$, 2) estimating the rate of decrease $\tau$ of the normalized identification error, and 3) computing the attack success probability from estimated distributions of $\{e_n[k] : k \geq m\}$. Each stage is described as follows.

\textbf{Distribution of the normalized identification error in the first attack batch.}
We now examine the identification error $e[m]$ at the first attack batch $m$, which is $$e[m] = O_{acc}[m] - S[m-1]t[m].$$ 
The clock skew value $S[m-1]$ is known, but the parameters $O_{acc}[m]$ and $t[m]$ are to be modeled. From the definitions of accumulated offset and elapsed time, we have 
\begin{align}
\label{eq:sota-e}
e[m] &= O_{acc}[m-1] + |O_{avg}[m]| \nonumber\\
&~~- S[m-1](t[m-1] + T_{m,0} + a_{m,N}-a_{m,1}),
\end{align}
where $T_{m,0}$ is the inter-arrival time between the last message of the previous (($m-1$)-th) batch and the first message of the current ($m$-th) batch.
Next, we will compute the mean and standard deviation of $e[m]$.

Based on our timing model (Section \ref{sec:timing}), the average offset under an attack with a delay of $\Delta T$ (i.e., the equivalent total amount of added delay is $\Delta T + \Delta T_0$) is 
\begin{align}
O_{avg}[m] &= \frac{1}{N-1}\sum_{i=2}^N [ (i(T+\Delta T-O)+\eta_{m,i}) \nonumber\\
&~~~~- ((T+\Delta T-O)+\eta_{m,1} + (i-1)\mu[m-1])] \nonumber \\
&= \frac{1}{N-1}\sum_{i=2}^N [(i-1)(\mu+\Delta T - \mu[m-1]) \nonumber \\
&~~~~+ (\eta_{m,i}-\eta_{m,1})] \nonumber,
\end{align}
where $\mu = T-O$ is the mean inter-arrival time before an attack\footnote{Strictly speaking, the resulting offset due to the added delay of $\Delta T$ is $O'=(T+\Delta T)/T \cdot O$. However, $\Delta T$ is usually much smaller than $T$, and thus we can approximate $O'$ as $O$. }. 
Although the statistics of $\eta$ after the attack may be different from those before the attack due to different characteristics of transmitting ECUs, such information is not available at batch $(m-1)$. Therefore, we assume the same statistics of $\eta$ before and after the attack, namely, $\eta_{m,i}\sim N(d,\sigma^2_\eta)$ for $1\leq i\leq N$, which yields
\begin{equation}
O_{avg}[m] \sim N \left(\frac{N}{2}(\mu + \Delta T - \mu[m-1]), \frac{N}{N-1}\sigma_\eta^2 \right).
\label{eq:O_avg_dist_sota}
\end{equation}
Since $\sigma_{\eta}^2 = \sigma^2/2$ (Section~\ref{sec:timing}),  the variance of $O_{avg}[m]$ is also equal to $\frac{N}{2(N-1)}\sigma^2$, where $\sigma$ is the standard deviation of inter-arrival times. 
For $\Delta T$ sufficiently large, the $\Delta T$ term will dominate $(\eta_{m,i}-\eta_{m,1})$, and hence we have
\begin{align}
|O_{avg}[m]| &\approx \frac{1}{N-1}\sum_{i=2}^{N}[(i-1) \cdot |\mu + \Delta T - \mu[m-1]| \nonumber \\
&~~~~+ (\eta_{m,i}-\eta_{m,1})].
\label{eq:O_avg_approx}
\end{align}

Next, we can substitute the $|O_{avg}[m]|$ term in Eq.~(\ref{eq:sota-e}) with Eq.~(\ref{eq:O_avg_approx}) and compute the mean and standard deviation of $e[m]$, as described in the following lemma.
\begin{lemma}
\label{lemma:sota-initial-error}
Under the assumption (\ref{eq:O_avg_approx}), the identification error $e[m]$ of the first attack batch is Gaussian with mean 
\begin{align}
\label{eq:sota-mu-e}
\mu_{e} &= O_{acc}[m-1] + \frac{N}{2}(|\mu + \Delta T-\mu[m-1]|) \nonumber\\
 &- S[m-1](t[m-1] + T_{m,0} + (N-1)(\mu+\Delta T))
\end{align}
and variance
\begin{displaymath}
\sigma_{e}^{2} = \frac{1}{2}\left(\frac{N - 2S[m-1]}{N-1} + 2S[m-1]^{2} - 2S[m-1]\right)\sigma^{2}.
\end{displaymath}
\end{lemma}
A proof can be found in Appendix~\ref{appendix:lemma1}.
 
The distribution of the normalized identification error in the first attack batch is Gaussian and satisfies 
 \begin{displaymath}
e_n[m] = \frac{e[m] - \mu_{\text{CUSUM}}}{ \sigma_{\text{CUSUM}}} \sim N \left(\frac{\mu_e - \mu_{\text{CUSUM}}}{\sigma_{\text{CUSUM}}}, \frac{\sigma_e^2}{\sigma^2_{\text{CUSUM}}} \right).
\end{displaymath}

After obtaining the distribution of the normalized identification error in the first attack batch, our next task is to model the rate of decrease $\tau$ of the normalized identification error $e_n[k]$ in Eq.~(\ref{eq:SoA_norm_error_model}), which will give us an approximation of $e_n[k]$ for $k\geq m+1$.

%The distribution of the initial identification error is shown in Figure \ref{fig:sota-analysis-initial-distribution}. For comparison, when $\Delta T$ is $200 \mu s$, the predicted mean of initial identification error for the message 0x185 was $11092 \mu s$, compared to the actual value of $10948 \mu s$, while the predicated and actual standard deviations were $221 \mu s$ and $145 \mu s$, respectively. The system parameters were the same as in Section \ref{subsec:soa-obs}. 
%\begin{figure}[!ht]
%\centering
%$\begin{array}{cc}
%\includegraphics[width=2in]{figures/sota-error-distribution-200us-ecocar-0x185.pdf} %&
%\includegraphics[width=1.5in]{figures/sota-error-distribution-400us.pdf} \\
%(a) & (b)
%\end{array}$
%\caption{Comparison of distribution of identification error following attack $e[k]$ for our formal model and experimental data. \sy{(SY: I think figures for observations are enough for motivating modeling. A figure that show validation results is not needed here.  )}}
%\label{fig:sota-analysis-initial-distribution}
%\end{figure}
 
\textbf{Rate of decrease of the normalized identification error.}
According to Eq.~(\ref{eq:sota-e}), the identification error $e[k+1]$ after an attack begins (i.e., $k\geq m$) is given by
\begin{align*}
e[k+1] &= O_{acc}[k+1] - S[k]t[k+1],\\
&= O_{acc}[k] + |O_{avg}[k+1]| \\
&~~~~ - S[k](t[k] + T_{k+1,0} + (a_{k+1,N}-a_{k+1,1})),
\end{align*}
where $T_{k+1,0}\approx \mu + \Delta T$ is the inter-arrival time between the last message in the $k$-th batch and the first message of the ($k+1$)-th batch during the attack.

Since skew updating is slow in the first tens of batches due to the slow convergence of the RLS algorithm, we may assume that $S[k]=S$ is a constant.
Then we have
\begin{align*}
e[k+1] &= O_{acc}[k] + |O_{avg}[k+1]| - St[k] \\
&~~~~- S((\mu+\Delta T) + (a_{k+1,N}-a_{k+1,1}))\\
&= e[k] + |O_{avg}[k+1]| \\
&~~~~- S( N(\mu + \Delta T) + (\eta_{k+1,N}-\eta_{k+1,1})).
\end{align*}
According to Eq.~(\ref{eq:O_avg_dist_sota}), the average offset $O_{avg}[k+1]$ is Gaussian with mean $\frac{N}{2}(\mu + \Delta T- \mu[k])$ and variance $\frac{N}{2(N-1)}\sigma^2$. 
Although the value of $\mu[k]$ for $k\geq m$ is not available at batch $(m-1)$, we have $\mathbb{E}(\mu[k]) = \mu + \Delta T$, which means $O_{avg}[k+1]$ can be approximated as zero.

Therefore, we can derive a linear approximation to $e[k]$ by taking the expectation of $(e[k+1]-e[k])$.
Since $|O_{avg}[k+1]|$ is the absolute value of a Gaussian random variable with mean zero and variance $\frac{N}{2(N-1)}\sigma^2$, we have
\begin{displaymath}
\mathbb{E}(|O_{avg}[k+1]|) = \sqrt{\frac{N}{2(N-1)}\sigma^2} \cdot \sqrt{\frac{2}{\pi}} = \sigma \sqrt{\frac{N}{\pi(N-1)}}.
\end{displaymath}

Since the normalized identification error is computed as $e_n[k]=(e[k]-\mu_{\text{CUSUM}})/\sigma_{\text{CUSUM}}$, the rate of decrease $\tau$ of $e_n[k]$ can be approximated as
\begin{align*}
\tau &\approx \left| \mathbb{E}(e_n[k+1]-e_n[k]) \right| = \left|\mathbb{E}\left(\frac{e[k+1]-e[k]}{\sigma_{\text{CUSUM}}}\right)\right| \\
&= \left|\frac{1}{\sigma_{\text{CUSUM}}} \left( \sigma \sqrt{\frac{N}{\pi(N-1)}} - S( N(\mu + \Delta T)) \right)\right|. 
\end{align*}
Note that the fixed $\sigma_{\text{CUSUM}}$ is used, since $e_n[k]$ is usually larger than $\gamma$ and thus $\sigma_{\text{CUSUM}}$ will not be updated.

Now that we have distributions of normalized identification errors $\{e_n[k]:k\geq m\}$, we can compute the distribution of the maximum value of control limits $L^{+}$ and $L^{-}$, and derive the attack success probability.

\textbf{Computation of the attack success probability.}
In order to derive the attack success probability, let us take a closer look at how the control limits are updated. 
Without loss of generality, we consider positive $\Delta T$ and assume that the upper control limit $L^+$ is zero before the attack. 
From Eq.~(\ref{eq:cusum-1}), we can see that if $e_n[m] \geq \kappa+\Gamma$, the attack will be detected immediately in the first batch; if $e_n[m]\leq \kappa$, it will not be detected at all. 
If $e_n[m]$ lies in $(\kappa,\kappa+\Gamma)$ and $L^+[k]=\sum_{k\geq m}(e_n[l]-\kappa)$ is greater than $\Gamma$ for some $k$, the attack will still be detected after several batches. 
Hence, we can first compute the maximum value of $L^+$, which depends on $e_n[m]$ and $\tau$, and then relate the attack success probability $Pr(L^+_{\max} \leq \Gamma)$ to the distribution of $e_n[m]$, as shown in the following theorem. 

%Let $e_n[l]$ be the normalized identification error at batch $l$ when the attack begins at batch $k$.  Let $\tau$ denote the decreasing rate derived above. 
%Without loss of generality, we consider a large positive $\Delta T$ and a decreasing rate of $\tau$. 
%Then we have 
%\begin{equation*}
%e_n[l+1] \approx e_n[l] - \tau, \quad e_n[l] \approx e_n[k] - (l-k)\tau. 
%\end{equation*}

%The following theorem gives the probability of successful attack for the adversary.
\begin{theorem}
\label{theorem:sota-success}
The attack success probability satisfies 
\begin{multline}
\label{eq:sota-success}
P_s = Pr\left( \frac{\tau - \sqrt{\tau^2 + 8 \tau \Gamma}}{2} - \kappa \leq e_n[m] \right.\\
\left. \leq \frac{-\tau + \sqrt{\tau^2 + 8 \tau \Gamma}}{2} + \kappa \right). 
\end{multline}
\end{theorem}
The proof can be found in Appendix~\ref{appendix:theory}.

By Theorem \ref{theorem:sota-success}, we can see that the attack success probability can be computed by evaluating the cumulative density function of a Gaussian random variable.

%\subsection{Validation}
%\label{subsec:sota-validation}

\subsection{Formal Analysis of NTP-Based IDS}
\label{sec:ntp}

%\textcolor{blue}{1. Consistency in terms: successful attack probability, identification error, reference identification error}

We then formally analyze the probability of a successful cloaking attack for the NTP-based IDS, given the system parameters immediately before the attack.
%\textcolor{red}{The goal is to provide a theoretic estimate for detection probability for the IDS immediately before the attack happens.  }

\subsubsection{Assumptions for NTP-Based IDS}
For the NTP-based IDS, the batch size $N$ and CUSUM parameters including $\gamma$ (the update threshold), $\Gamma$ (the detection threshold), $\kappa$ (the sensitivity parameter) and $\lambda$ (the parameter in the RLS), are known to the IDS.
Since the IDS records the receive timestamps of the target message, it knows the period $T$ and can also measure the mean $\mu$ and standard deviation $\sigma$ of inter-arrival times.

As mentioned in Section~\ref{sec:ids}, the NTP-based IDS tracks the accumulated offset $O_{acc}[k]$ and elapsed time $t[k]$ in each batch $k$, and maintains the reference identification errors. Hence, it is reasonable to assume that the values of $\{O_{acc}[k]:k<m\}$, $\{t[k]:k<m\}$, and $\{e_{ref}[k]:k<m\}$ are known to the NTP-based IDS prior to the attack.

\begin{figure*}[t!]
	\centering
	\begin{subfigure}[h]{0.32\textwidth} %{0.5\columnwidth}
		\includegraphics[width=\textwidth]{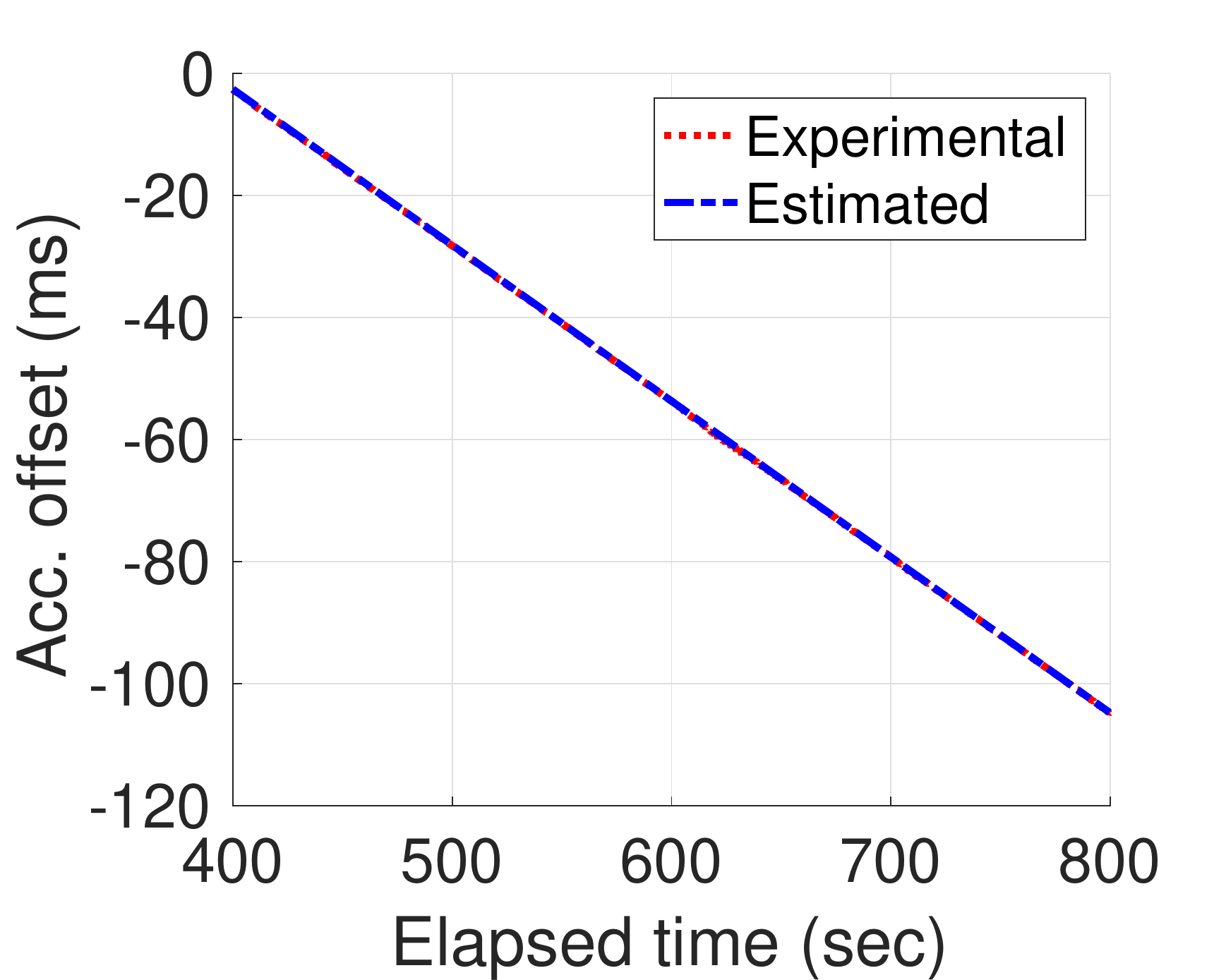}
		\caption{}
		\label{fig:NTP-offset}
	\end{subfigure}
	\begin{subfigure}[h]{0.32\textwidth}
		\includegraphics[width=\textwidth]{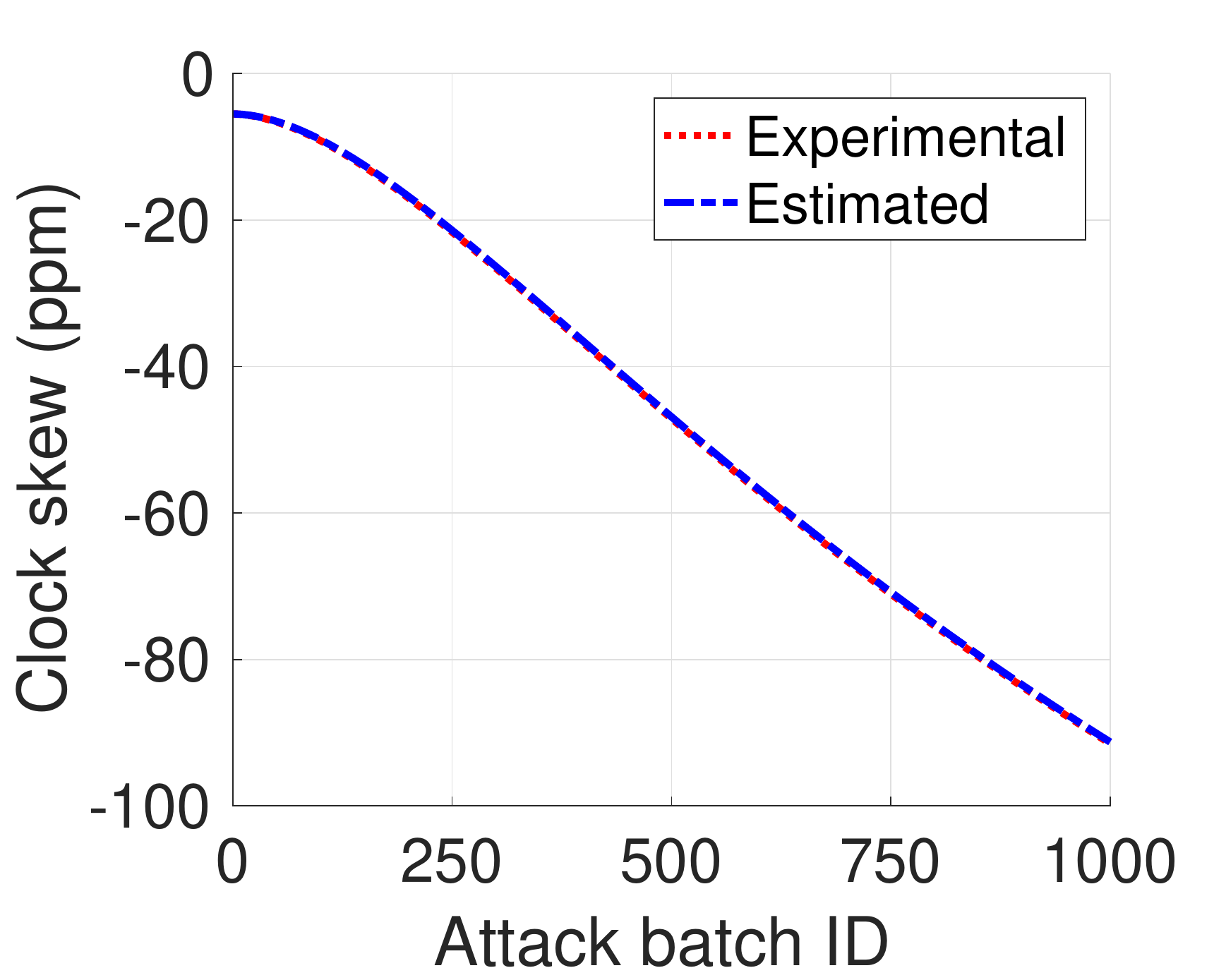}
		\caption{}
		\label{fig:NTP-skew}
	\end{subfigure}
	\begin{subfigure}[h]{0.32\textwidth}
		\includegraphics[width=\textwidth]{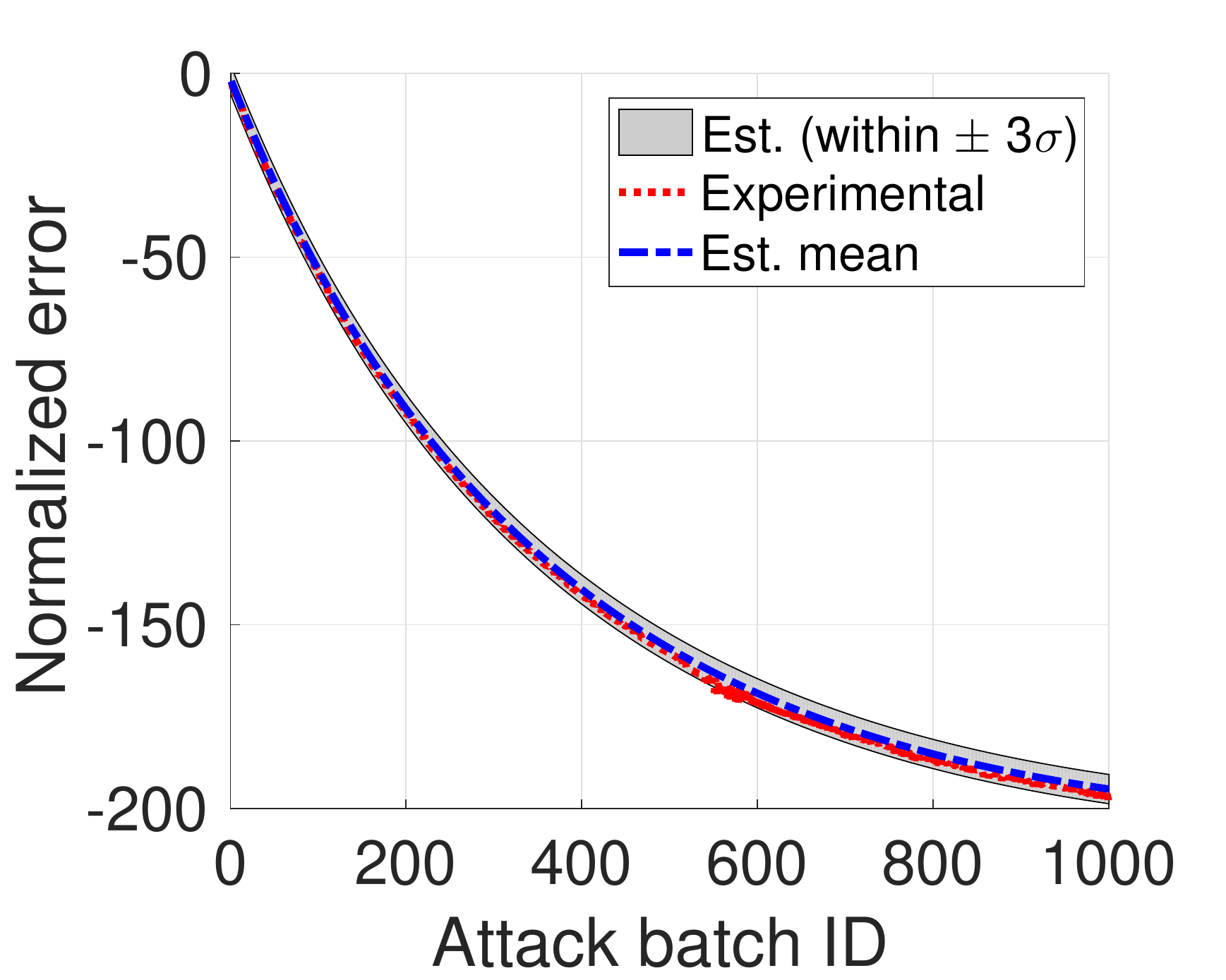}
		\caption{}
		\label{fig:NTP-error}
	\end{subfigure}
	\caption{Experimental versus estimated (a) accumulated offset and elapsed time, (b) clock skew, and (c) normalized identification error.
		The estimated values match closely with the experimental values.  
		Note that figures are generated using data for the $20$ms message 0x185 collected from our testbed with $N=20$, $\gamma=4$, $\Gamma=5$, $\kappa=8$, and $\Delta T = 5~\mu s$. 
	}
	\label{fig:NTP-example}
	%	\vspace{-0.2cm}
\end{figure*}

\subsubsection{Observations}\label{sec:ntp-observations}
%\textcolor{blue}{I am not sure about this subsection right now. The first para just summarizes Section 5.2. The second paragraph does not give observations per se, but rather summarizes the results of 6.3.}
%As introduced in Section~\ref{sec:ids}, the NTP-based IDS performs the following tasks for each incoming batch: 
%1) computing identification error $e[k]$ based on the newly obtained elapsed time $t[k]$, accumulated offset $O_{acc}[k]$ and  previous skew $S[k-1]$, 
%2) updating skew $S[k]$ via RLS based on $t[k]$ and $O_{acc}[k]$, 
%3) computing normalized error $e_n[k]$ using CUSUM statistics $\mu_{CUSUM}[k-1]$ and $\sigma_{CUSUM}[k-1]$, and updating control limits as in Eq.~(\ref{eq:cusum-1}) and (\ref{eq:cusum-2}) accordingly, and 
%4) updating CUSUM statistics if $e_n[k]$ is less than or equal to the update threshold $\gamma$. 

Our modeling and analysis of the NTP-based IDS are based on the following observations. 
First, if the attack with an added delay of $\Delta T$ starts in the $k$-th batch, the resulting $O_{acc}[k]$, $t[k]$, and $e[k]$ can be estimated from $\mu$, $T$, $S[k-1]$, and $\Delta T$.
Second, although the IDS keeps track of the slowly changing clock skew via the RLS based on newly obtained $t[k]$ and $O_{acc}[k]$, the output of the RLS converges to that of a non-RLS estimator that minimizes the weighted mean squared error. 
Third, with the estimated value of $e[k]$, the IDS can further estimate the CUSUM statistics following its updating rule, as well as the mean value and distribution of normalized errors. 

\subsubsection{Attack Success Probability}
Based on the observations in Section~\ref{sec:ntp-observations}, we divide our formal analysis into four stages: 1) estimating the accumulated offset $O_{acc}[k]$ and the elapsed time $t[k]$ after the attack begins at batch $m$, 2) approximating the clock skew $S[k]$ estimated by the RLS, 3) modeling the distributions of normalized identification errors $\{e_n[k]:k\geq m\}$, and 4) computing the probability of control limits exceeding $\Gamma$ to obtain the attack success probability. 

\textbf{Accumulated offset and elapsed time.} 
For the NTP-based IDS, the accumulated offset before the attack is 
\begin{align}
O_{acc}[k] &= \sum_{i=1}^k N O_{avg}[i] = \sum_{i=1}^k N \left(T - \frac{a_{i,N}-a_{i-1,N}}{N}\right)\nonumber \\
%&=  kNT - (a_{k,N}-a_{0,N}) \nonumber \\
%&= kNT - (kN(T-O) + \eta_{k,N} - \eta_{0,N}) \nonumber \\
&= kNO - (\eta_{k,N}-\eta_{0,N}),
\end{align}
where $a_{0,N}$ is the arrival timestamp of the last message in the initialization batch, and $O=T-\mu$ is the average offset in each period $T$.
%, whose estimated value is given by 
%\begin{equation}
%\hat{O} = T - \frac{a_{k,N} - a_{0,N}}{kN} = O + \frac{\eta_{k,N} - \eta_{0,N}}{kN}.
%\end{equation}
The elapsed time is
\begin{align}
t[k] = a_{k,N} - a_{0,N} = kN(T-O) + \eta_{k,N} - \eta_{0,N}.
\end{align}
%and the relationship between $O_{acc}[k]$ and $t[k]$ is 
%\begin{equation}
%O_{acc}[k] = kNT - t[k].
%\end{equation}

We assume that the attack starts from the first message of batch $m$, and the inter-arrival time between the last normal message and the first attack message is roughly equal to $\mu+\Delta T$.
Then for $k\geq m$, we have
\begin{align}
t[k] &= kN(T-O) + (k-m+1)N\Delta T + \eta_{k,N} - \eta_{0,N} \nonumber \\
&= t[m-1] + (k-m+1)N (T-O+\Delta T) \nonumber \\
&~~~~+ \eta_{k,N} - \eta_{m-1,N}. \label{eq:elapsed_time_under_attack}
\end{align}
Since $O_{acc}[k] = kNT - t[k]$, we also have
\begin{align}
O_{acc}[k] &= O_{acc}[m-1] - (k-m+1)N(-O+\Delta T) \nonumber \\
&~~~~- (\eta_{k,N} - \eta_{m-1,N}). \label{eq:acc_offset_under_attack}
\end{align}
Note that in the above equations, the amount of network delay and noise as captured by $\eta_{m-1,N}$ is given at batch $(m-1)$, and thus $\eta_{k,N}$ is the only random variable. 

With more attack batches arriving, the estimated clock skew will gradually change over time. Hence, it is important to model the process of clock skew updating, which is our next step of modeling.

\textbf{Approximation of the estimated clock skew.} 
%The main challenge of modeling the clock skew updating process is that the RLS is an online algorithm that updates the clock skew estimate with non-linear equations, which \rev{can be} difficult to keep track of. %, if not possible.
%Nevertheless, 
While the RLS is an online algorithm that \rev{recursively} updates the clock skew estimate with non-linear equations, it has been shown in \cite{goodwin1977dynamic} that the clock skew estimated via the RLS would converge to the value $S$ that minimizes the following quadratic function, 
\begin{equation}
J_k(S) = \sum_{i=1}^k \lambda^{k-i} (O_{acc}[i] - S \cdot t[i])^2,
\end{equation}
where $\lambda<1$ is the parameter in the RLS, and the optimal value is given by
\begin{equation}
\hat{S}[k] = \arg \min_S J_k(S) = \frac{\sum_{i=1}^k \lambda^{k-i} O_{acc}[i] \cdot t[i] }{ \sum_{i=1}^k \lambda^{k-i} t^2[i] }.\label{eq:estimate_skew_non_RLS}
\end{equation}
Let the mean of $t[k]$ in Eq.~(\ref{eq:elapsed_time_under_attack}) and $O_{acc}[k]$ in Eq.~(\ref{eq:acc_offset_under_attack}) be $\hat{t}[k]$ and $\hat{O}_{acc}[k]$, respectively. 
Given $\hat{t}[k]$ and $\hat{O}_{acc}[k]$, we can estimate the output of RLS as $\hat{S}[k]$ based on Eq.~(\ref{eq:estimate_skew_non_RLS}). 

As shown in Fig.~\ref{fig:NTP-offset} and Fig.~\ref{fig:NTP-skew}, the estimated values of accumulated offset, elapsed time, and clock skew are closely matched with the experimental values.

\textbf{Distribution of the normalized identification errors.} 
With the estimated clock skew values $\{\hat{S}[k]\}$, the identification error $e[k]$ is given as
\begin{align}
e[k] &= O_{acc}[k] - \hat{S}[k-1]t[k] \nonumber\\
&= (\hat{O}_{acc}[k] - \eta_{k,N}) - \hat{S}[k-1](\hat{t}[k] + \eta_{k,N}) \nonumber \\
&= \hat{e}[k] - (1+\hat{S}[k-1])\eta_{k,N}, \nonumber
\end{align}
where $\hat{e}[k]=\hat{O}_{acc}[k] - \hat{S}[k-1] \hat{t}[k]$.
Since $\eta_{k,N}$ is Gaussian, the identification error $e[k]$ is also Gaussian with mean
$\hat{e}[k]$ and variance $(1+\hat{S}[k-1])^2 \sigma_{\eta}^2$. 

In order to estimate the distribution of $e_n[k]$, we need to model the updating process of CUSUM statistics, i.e., $\hat{\mu}_{\text{CUSUM}}$ and $\hat{\sigma}_{\text{CUSUM}}$. 
Hence, given $\{\hat{e}[k]\}$, we can compute  $\hat{e}_n[k]= (\hat{e}[k]-\hat{\mu}_{\text{CUSUM}}[k-1])/\hat{\sigma}_{\text{CUSUM}}[k-1]$. 
If $|\hat{e}_n[k]| \leq \gamma$, we add $\hat{e}[k]$ to $\{e_{ref}[k]\}$ and re-compute $\hat{\mu}_{\text{CUSUM}}[k]$ and $\hat{\sigma}_{\text{CUSUM}}[k]$ from $\{e_{ref}[k]\}$.
Then we increment $k$ by $1$ and repeat the above steps.

Since $e_n[k]= (e[k]-\hat{\mu}_{\text{CUSUM}}[k-1])/\hat{\sigma}_{\text{CUSUM}}[k-1]$, it implies
\begin{equation*}
e_n[k] \sim N\left(\frac{\hat{e}[k]-\hat{\mu}_{\text{CUSUM}}[k-1]}{\hat{\sigma}_{\text{CUSUM}}[k-1]}, \frac{(1+\hat{S}[k-1])^2 \sigma_{\eta}^2}{\hat{\sigma}_{\text{CUSUM}}[k-1]^2}\right).
\end{equation*}
As shown in Fig.~\ref{fig:NTP-error}, the estimated mean values of $e_n[k]$ match closely with the experimental values.
Based on the distributions of $\{e_n[k]:k\geq m\}$ derived above, we can now compute the attack success probability.

\textbf{CUSUM analysis.} 
Let the probability density function of $e_{n}[k]$ be $f_{k}$, and the number of attack batches used for detection be $n$. We assume that $\kappa \geq \Gamma$, which is consistent with the NTP-based IDS and our simulations. A detection takes place in the $k$-th attack batch if $L^{+}[k] > \Gamma$ or $L^{-}[k] > \Gamma$. 
Let $\alpha = \min{\{k : \max{\{L^{+}[k], L^{-}[k]\}} > \Gamma\}}$, which is the attack batch ID when control limits first exceed the detection threshold. In other words, if $\alpha>n$, it means that the attack is not detected within $n$ batches.
Hence, the attack success probability is equal to $Pr(\alpha > n)$, and the following lemma shows how to compute  $$g_{n,k}(z^{+},z^{-}) \triangleq Pr(\alpha > n | L^{+}[k] = z^{+}, L^{-}[k] = z^{-}).$$ %\textbf{We assume throughout that $\kappa \geq \Gamma$, as is the case for the Cho-Shin CIDS and our simulations. While the basic approach is still sound, the derivation will have to modified if this assumption does not hold.}

\begin{lemma}
	\label{lemma:NTP-CUSUM}
	The probability of a successful cloaking attack for the CUSUM-based detector satisfies 
	\begin{align*}
	g_{n,k}(z^{+},z^{-}) &= \int_{z^{-}-\kappa-\Gamma}^{z^{-}-\kappa}{g_{n,k+1}(0,z^{-}-r-\kappa)f_{k}(r) \ dr} \\
	&~~~~+ g_{n,k+1}(0,0)Pr(e_n[k] \in [z^{-} - \kappa, \kappa - z^{+}]) \\
	&~~~~+ \int_{\kappa - z^{+}}^{\kappa-z^{+}+\Gamma}{g_{n,k+1}(z^{+} + r - \kappa, 0)f_{k}(r) \ dr}.
	\end{align*}
\end{lemma}
From Lemma \ref{lemma:NTP-CUSUM}, we can  take a discrete approximation of $g_{n,k}(z^+,z^-)$ as 
\begin{align*}
g_{n,k}\left(\frac{i\Gamma}{m}, \frac{j\Gamma}{m}\right) &\approx \frac{\Gamma}{m}\sum_{l=0}^{m}{g_{n,k+1}\left(0, \frac{l\Gamma}{m}\right)f_{k}\left(\frac{(j-l)\Gamma}{m} - \kappa\right)} \\
& + g_{n,k+1}(0,0)Pr(e_n[k] \in [z^{-}-\kappa, \kappa - z^{+}]) \\
& + \frac{\Gamma}{m}\sum_{l=0}^{m}{g_{n,k+1}\left(\frac{l\Gamma}{m},0\right)f_{k}\left(\frac{(l-i)\Gamma}{m} + \kappa\right)}.
\end{align*}
A proof can be found in Appendix~\ref{appendix:lemma2}.

Therefore, the value of $g_{n,k}$, that is, the probability of a successful cloaking attack within $n$ attack batches predicted at the $k$-th attack batch ($k\leq n$), can be computed as a linear function of the values of $g_{n,k+1}$. The attack success probability is equal to $g_{n,0}(0,0)$.

\section{Evaluation}
\label{sec:evaluation}
In this section, we evaluate the proposed cloaking attack on two CAN bus testbeds and demonstrate that the cloaking attack is able to bypass both the SOTA and NTP-based IDSs.
We then validate our formal analysis through extensive experiments.

\subsection{Testbeds}
We build two CAN bus testbeds: a CAN bus prototype and a CAN testbed on a real vehicle (the UW EcoCAR, a 2016 Chevrolet Camaro \cite{ecocar}).
Compared with the prototype that consists of three ECUs, the UW EcoCAR hosts 8 stock ECUs and two experimental ECUs.
A total of 2500+ messages with 89 different IDs are being exchanged every second. 
%The EcoCar testbed provides a real CAN environment to evaluate and demonstrate the proposed cloaking attack. 

\subsubsection{CAN Bus Prototype}
As shown in Fig.~\ref{fig:arduino_testbed}, each ECU on the CAN bus prototype consists of an Arduino UNO board %\cite{arduino} 
and a Sparkfun CAN bus shield %\cite{sparkfun} 
that uses a Microchip MCP2515 CAN controller with a MCP2551 CAN transceiver.
The bus speed is set to $500$ Kbps as in typical CAN buses.

\subsubsection{UW EcoCAR testbed}
\begin{figure}[t!]
	\centering
	\begin{subfigure}[h]{0.4\columnwidth} % {0.48\columnwidth}
		\includegraphics[width=\columnwidth]{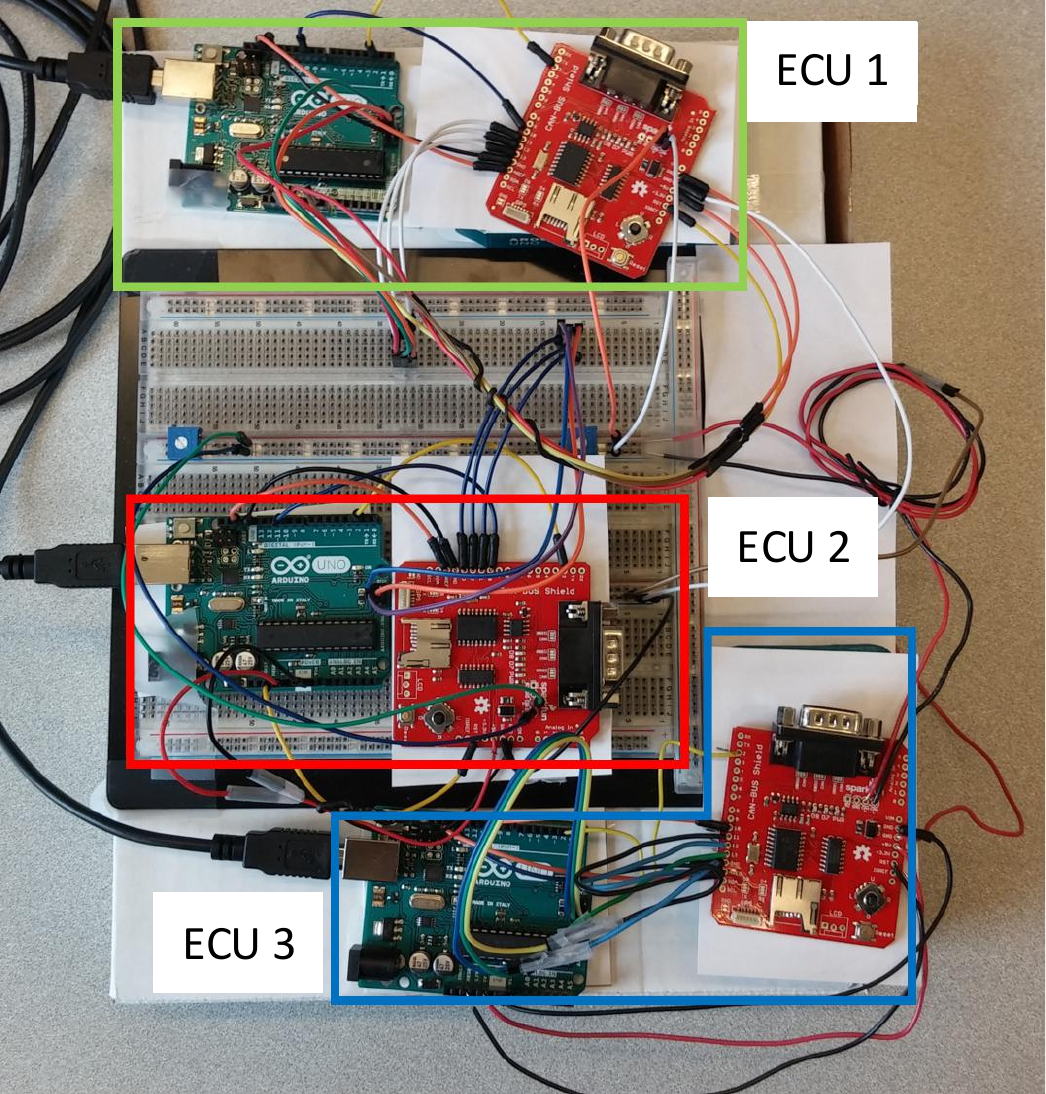}
		\caption{CAN bus prototype}
		\label{fig:arduino_testbed}
	\end{subfigure}
	\begin{subfigure}[h]{0.52\columnwidth} % {0.48\columnwidth}
		\includegraphics[width=\columnwidth]{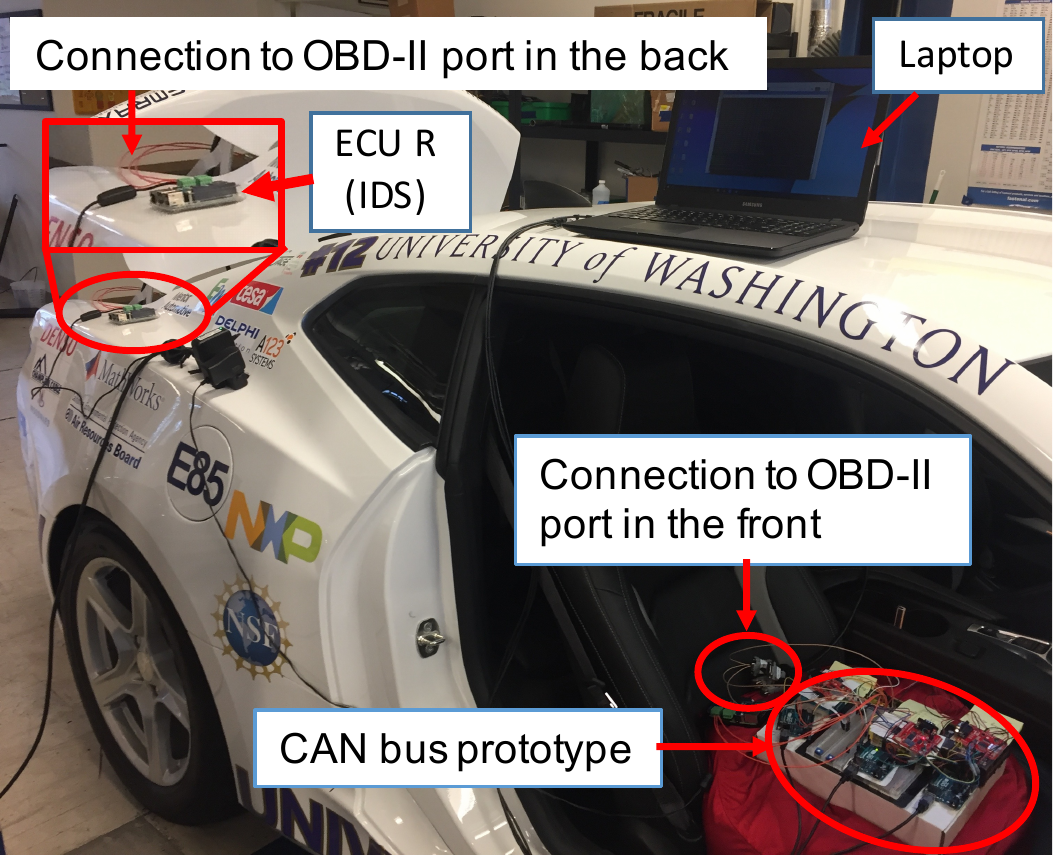}
		\caption{UW EcoCAR testbed}
		\label{fig:real_vehicle_testbed}
	\end{subfigure}
	\caption{Setup of CAN bus testbeds. 
	(a) The CAN bus prototype consists of three testbed ECUs, each of which consists of an Arduino board and a Sparkfun CAN bus shield. 
	(b) The CAN bus prototype and Raspberry Pi-based ECUs are connected to the CAN bus of the UW EcoCAR via the OBD-II ports to build the UW EcoCAR testbed.  }
\end{figure}

The CAN bus prototype is connected to the CAN bus of the UW EcoCAR via the On-Board Diagnostics (OBD-II) port to build the UW EcoCar testbed (Fig. \ref{fig:real_vehicle_testbed}). 
During our experiments, the UW EcoCAR was in the park mode in an isolated and controlled environment for safety purposes, but all ECUs were functional and actively exchange CAN messages.
We noticed that ECUs in the park mode had very close clock skews as in the drive mode.

%\textcolor{red}{In the drive mode, the clock skew of an ECU is still constant over a short period of observation time as shown in Fig. \ref{fig:cho_shin_impact_of_batch_size} in which the data is collected while Toyota Camry is in the drive mode \cite{Ruth:2012:Accuracy}. Hence, without loss of generality, we can use the park mode for the data collection.}

Due to the large CAN traffic and limited computing capability, Arduino-based ECUs are not able to log all CAN messages on the bus or transmit high frequency messages. 
Therefore, we build additional ECUs that consist of a Raspberry Pi 3 %\cite{raspberrypi} 
and a PiCAN 2 board % \cite{pican2}, 
and used SocketCAN \cite{socketCAN} 
to enable the interaction between the added ECUs and the UW EcoCAR. 

%A Raspberry Pi-based ECU is used as the receiving ECU (the IDS).
%A stock ECU is considered as the targeted ECU (the weak attacker) which transmits message 0x184 every $100$ ms ($10$ Hz), and the Arduino-based ECU $3$ acts as the strong attacker that injects spoofed messages.

\subsection{Evaluation of Cloaking Attack}
\label{sec:evaluation_cloaking_attack}
We first demonstrate and evaluate the cloaking attack on both the CAN bus prototype and the UW EcoCAR testbed.

\subsubsection{Setup} 
On the CAN bus prototype, ECU 1 acts as the IDS that logs all messages, ECU 2 is the targeted ECU that transmits message 0x11 every $100$ ms ($10$ Hz), and ECU 3 is the strong adversary that impersonates ECU 2.
On the UW EcoCAR testbed, a stock ECU that transmits message 0x184 every $100$ ms is treated as the targeted ECU and the same ECU 3 acts as the strong adversary that injects spoofed messages.

When launching the cloaking attack, the impersonating ECU 3 transmits every $100040~\mu$s ($\Delta T_0=40~\mu$s) to spoof message 0x11 on the CAN bus prototype and every $99971~\mu$s ($\Delta T_0=-29~\mu$s\footnote{While Arduino's time resolution is $4~\mu$s , we set $\Delta T_0$ to $-28~\mu$s and changed it to $-32~\mu$s every five messages so that $\Delta T_0 \approx -29~\mu$s on average.}) to spoof message 0x184 on the UW EcoCAR testbed.
During our experiments, we collected 8.5 hours of attack data from the CAN bus prototype and the UW EcoCAR testbed separately.

We set batch size $N=20$ for both the SOTA and the NTP-based IDSs.
For the SOTA IDS, the update threshold $\gamma$ is $3$ and the detection threshold $\Gamma$ is $5$, which is consistent with \cite{Shin:2016:finger}.
For the NTP-based IDS, we use $\gamma=4$ and $\Gamma=5$. 
For the data collected from the CAN bus prototype, the sensitivity parameter $\kappa$ is set to $5$ for both IDSs, while it is set to $8$ for the UW EcoCAR data to avoid false alarms. 

To simulate the cloaking attack, the IDS is fed with $1000$ batches of normal data, followed by $n$ batches of attack data in each experiment\footnote{We assume perfect timing for the cloaking attack, that is, the first attack message is received at the next expected time instant of the targeted message.  \hl{The impact of mistiming on the cloaking attack is studied in Appendix~\ref{appendix:mistiming}.}}.
An attack is successful if it is undetected by the IDS and fails otherwise. 
A total of $100$ independent experiments are performed to compute the attack success probability $P_s$.

% Successful Rates
\begin{figure}[t!]
	\centering
	\begin{subfigure}[h]{0.42\columnwidth} % {0.48\columnwidth}
		\captionsetup{justification=centering}
		\includegraphics[width=\columnwidth]{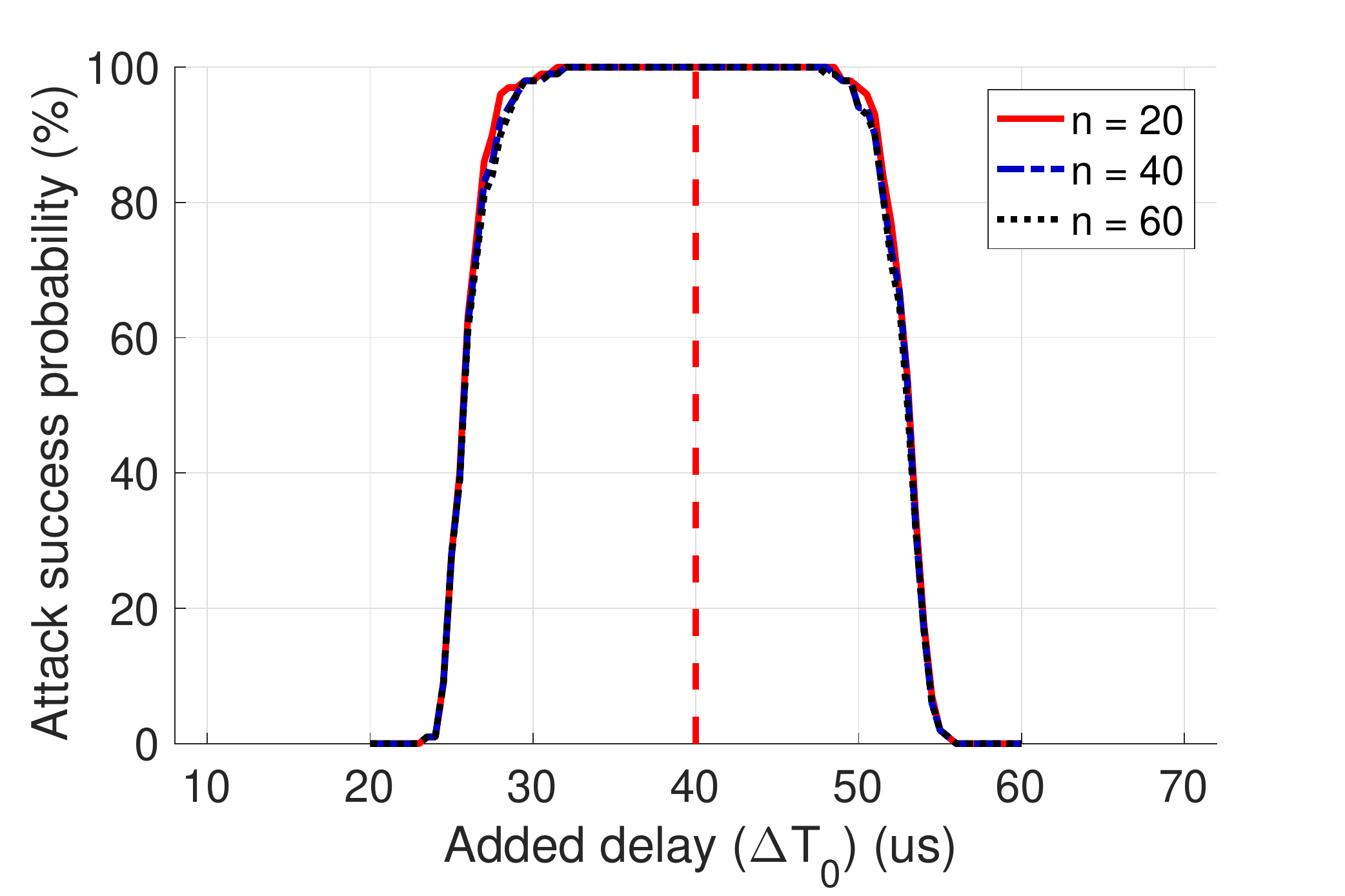}
		\caption{CAN prototype, SOTA}
		\label{fig:arduino_clock_skew_attack_success_rate_Cho}
	\end{subfigure}
	\begin{subfigure}[h]{0.42\columnwidth} % {0.48\columnwidth}
		\captionsetup{justification=centering}
		\includegraphics[width=\columnwidth]{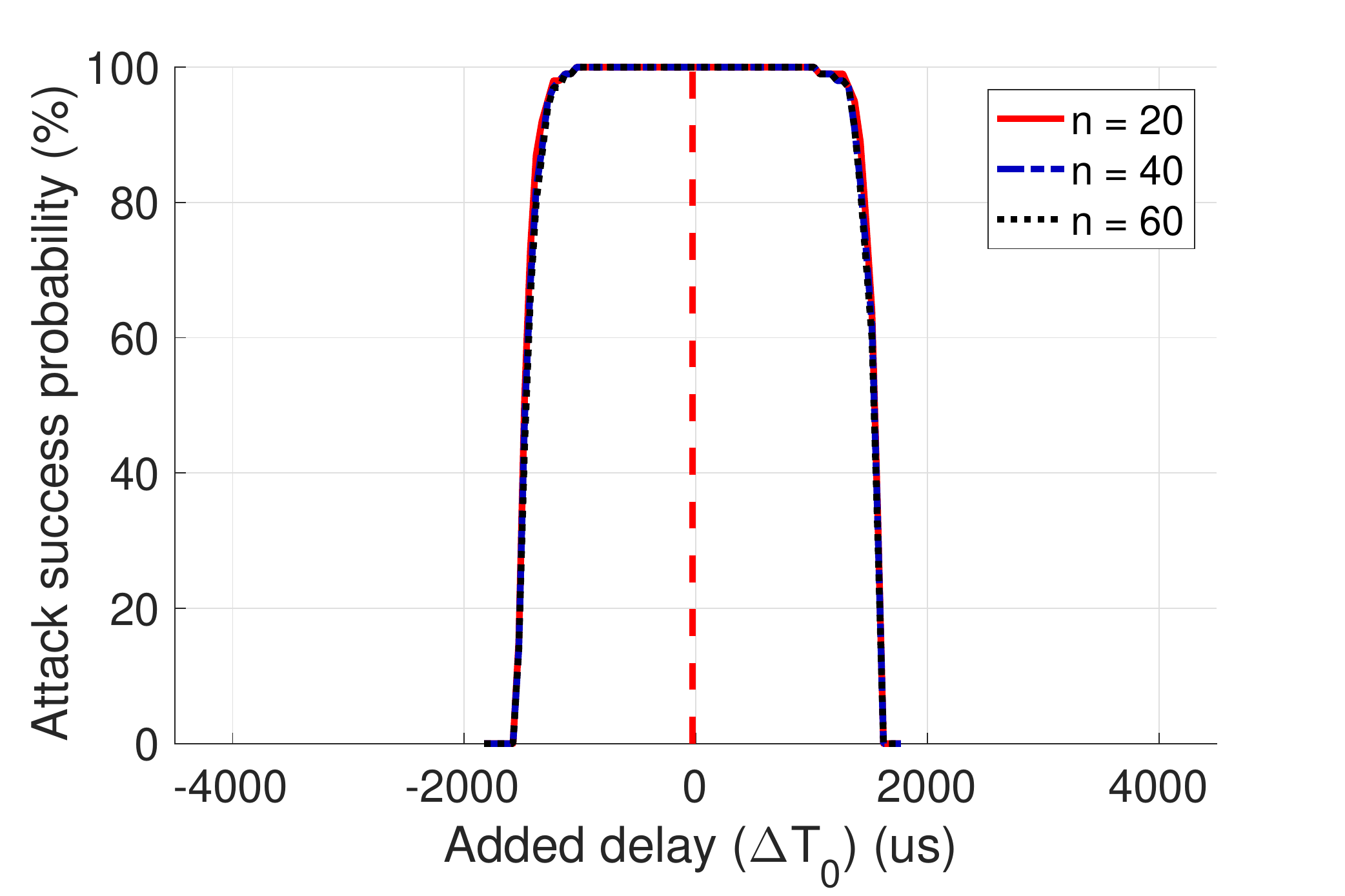}
		\caption{EcoCAR testbed, SOTA}
		\label{fig:ecocar_clock_skew_attack_success_rate_Cho}
	\end{subfigure}
	\\
	\begin{subfigure}[h]{0.42\columnwidth} % {0.48\columnwidth}
		\captionsetup{justification=centering}
		\includegraphics[width=\columnwidth]{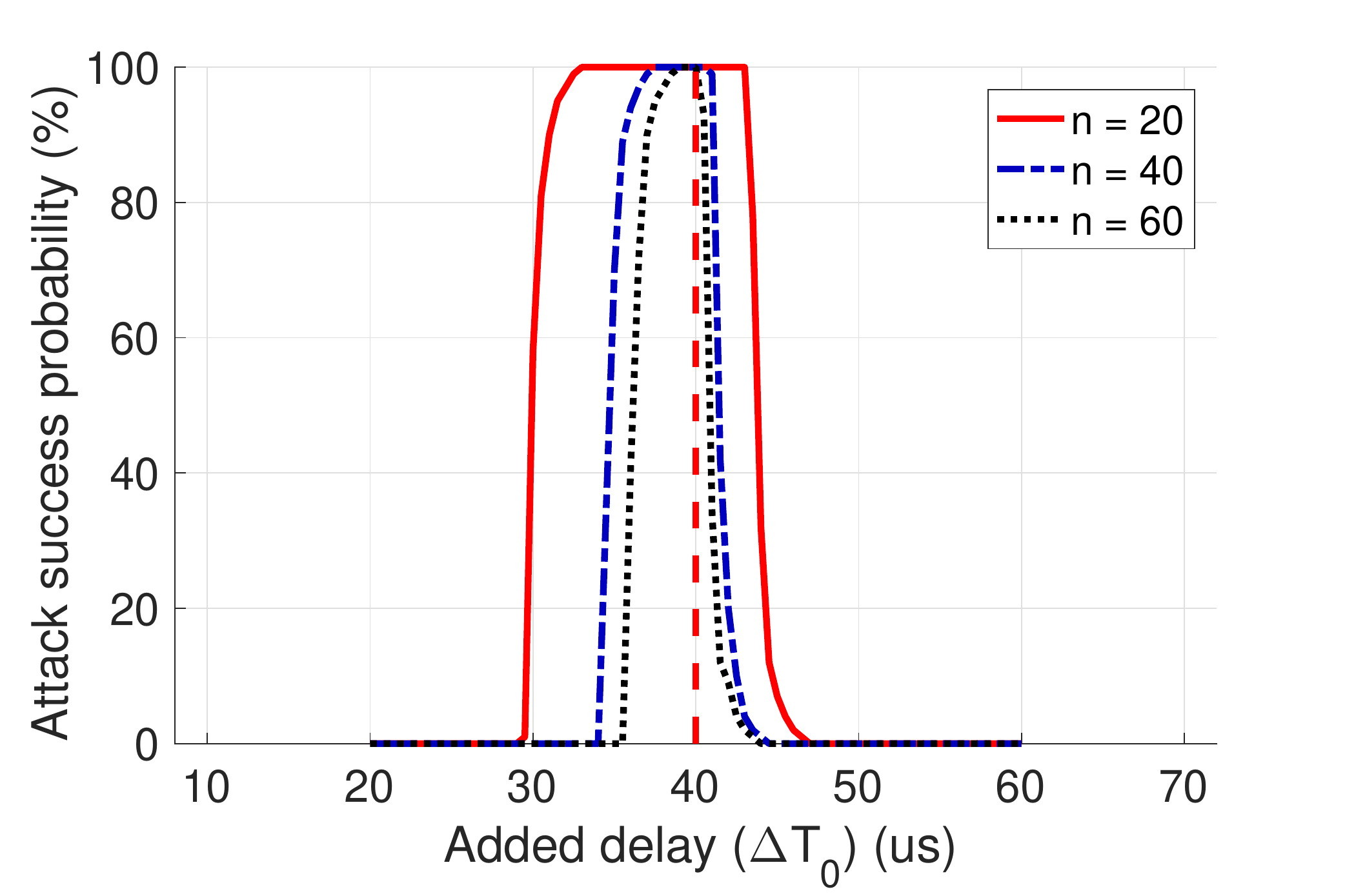}
		\caption{CAN prototype, NTP-based}
		\label{fig:arduino_clock_skew_attack_success_rate_ntp}
	\end{subfigure}
	\begin{subfigure}[h]{0.42\columnwidth} % {0.48\columnwidth}
		\captionsetup{justification=centering}
		\includegraphics[width=\columnwidth]{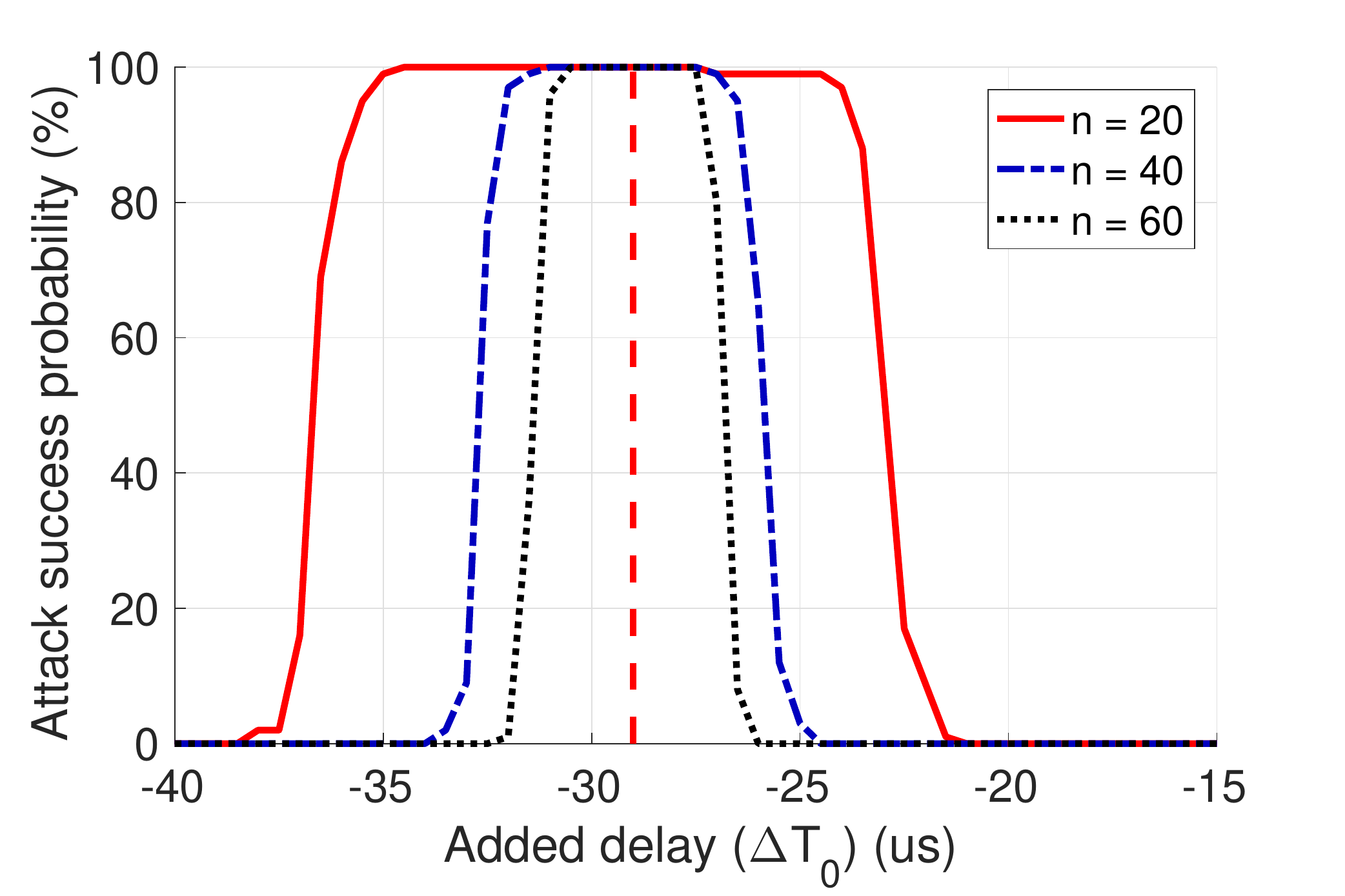}
		\caption{EcoCAR testbed, NTP-based}
		\label{fig:ecocar_clock_skew_attack_success_rate_ntp}
	\end{subfigure}
	\caption{Attack success probability on the SOTA IDS and the NTP-based IDS on the CAN bus prototype and the UW EcoCAR testbed with message period $100$ ms. 
	For the $\Delta T_0$ values achieved in our hardware experiments (red dashed line), the cloaking attack was successful in all test cases. 
	%The width of each curve is equal to the $\epsilon\text{-MSI}$ for the given detector.
	}
	\label{fig:clock_skew_attack_success_rate}
\end{figure}

\subsubsection{Results}
For the $\Delta T_0$ values achieved in our evaluation, $P_s$ is $1$ against both the SOTA and NTP-based IDSs (Fig. \ref{fig:clock_skew_attack_success_rate}, dashed line). 
In order to gain additional insight into the performance of each IDS under cloaking attack, we generate additional datasets by adding different values of $\Delta T_0$ to the message inter-arrival times and then analyze both IDSs using the new datasets.

In order to quantify the effectiveness of an IDS against the masquerade (cloaking) attack, we define a metric called $\epsilon$-\textit{Maximum Slackness Index} (MSI), which measures the interval of $\Delta T_0$ that an adversary can introduce while remaining undetected with a probability of $(1-\epsilon)$. 
We first let $P_s(\Delta T_0)$ be the attack success probability when the added delay is $\Delta T_0$. 
We define the upper and lower limits of $\Delta T_0$ for a successful attack as 
$(\Delta T_0)_{\max} (\epsilon) = \max\{\Delta T_0: P_s(\Delta T_0) > 1-\epsilon\}$ and $(\Delta T_0)_{\min} (\epsilon) = \min\{\Delta T_0: P_s(\Delta T_0) > 1-\epsilon\}$, respectively.
%\begin{align}
%(\Delta T_0)_{\max} (\epsilon) &= \max\{\Delta T_0: P_s(\Delta T_0) > 1-\epsilon\} \nonumber \\
%(\Delta T_0)_{\min} (\epsilon) &= \min\{\Delta T_0: P_s(\Delta T_0) > 1-\epsilon\}. \nonumber 
%\end{align}
We then define $\epsilon\text{-MSI}=(\Delta T_0)_{\max}(\epsilon) - (\Delta T_0)_{\min} (\epsilon)$. 
Intuitively, a smaller value of $\epsilon$-MSI signifies a more effective detector and less freedom for the adversary, since the adversary's clock skew must closely match with that of the targeted ECU in order to remain undetected.

On the CAN bus prototype, with $n=20$ and $\epsilon=0.05$, %$0.05-MSI(state-of-the-art \, IDS)$
the $\epsilon\text{-MSI}$ value for the SOTA IDS is $22.5~\mu$s  (Fig.~\ref{fig:arduino_clock_skew_attack_success_rate_Cho}), but only $11.5~\mu$s for the NTP-based IDS (Fig.~\ref{fig:arduino_clock_skew_attack_success_rate_ntp}).
Hence, it is much easier for the cloaking attack to bypass the SOTA IDS than the NTP-based IDS. We also found that increasing $n$ has little impact on $\epsilon\text{-MSI}$ for the SOTA IDS, which is $20.5~\mu$s for $n=40$ or $60$, but significantly impacts $\epsilon\text{-MSI}$ of the NTP-based IDS, which varies from $11.5~\mu$s to $2.5~\mu$s as $n$ is increased from $20$ to $60$. This result suggests that the performance of the NTP-based IDS improves over the attack duration.
Another interesting observation is that the $P_s$ curves are skewed instead of symmetric. 
This is because when the Arduino-based ECU starts operating, its clock skew slowly decreases due to the temperature change in hardware. As a result, the IDS tends to overestimate the clock skew, and is more sensitive to a larger positive delay (that would further decrease the clock skew). 

$\epsilon\text{-MSI}$ for the SOTA IDS increases significantly for a real vehicle, as shown in Fig.~\ref{fig:ecocar_clock_skew_attack_success_rate_Cho}, due to the significantly heavier CAN traffic compared to the prototype, which reduces the effectiveness of the detection. 
As an example, a cloaking attack with $\Delta T_0$ between $-1029~\mu$s and $1021~\mu$s can bypass the SOTA IDS with $100$\% probability regardless of $n$.
For the NTP-based IDS with $\epsilon=0.01$, $\epsilon\text{-MSI}$ is $10.5~\mu$s for $n=20$ and $3~\mu$s for $n=60$. Hence, in the real vehicle, as in the CAN prototype, the NTP-based IDS is more effective in detecting masquerade attacks than the SOTA IDS.
The proposed cloaking attack, however, is still able to thwart both detection schemes when $\Delta T_0$ is chosen to be within the interval $[(\Delta T_0)_{min}(\epsilon), (\Delta T_0)_{max}(\epsilon)]$.

\subsection{Evaluation of Formal Analysis}\label{sec:evaluation_ecocar}
We now validate the proposed formal models using the data collected from the UW EcoCAR testbed.

\textbf{Data Collection.} 
Since it is both labor and time intensive to collect data for all periodic messages, we select a subset of $6$ representative messages with different periods, message ID levels, and transmitting ECUs, which are listed in Table~\ref{table:representative_messages}.

When collecting the cloaking attack data for a targeted message of period $T$, the strong adversary (a Raspberry Pi-based ECU A that is connected to the OBD-II port) transmits messages every $T$ seconds, using a non-conflicting message ID to avoid any undesirable impact on the vehicle. 
The IDS at ECU R records the timestamps of all received messages.
The targeted and spoofed messages will be filtered and later used as the normal and attack data, respectively.

In order to determine a suitable amount of added delay for the cloaking attack, we first set $\Delta T_0$ to be the difference between the average inter-arrival time of the targeted message observed by ECU A and the nominal period $T$ (Section~\ref{sec:cloaking_attack}). 
We then experimentally tune $\Delta T_0$ so that the observed clock skews at the IDS residing in ECU R become the same.
Hence, the collected data corresponds to the cloaking attack with $\Delta T=0$, where $\Delta T$ is the difference between the actual delay and the ideal delay $\Delta T_0$ (Section~\ref{sec:cloaking_attack}).

\begin{table}[t!]
	\centering
	\caption{Selected subset of representative messages from Electronic Brake Control Module (EBCM), Electronic Power Steering (EPS), Electronic Parking Brake (EPB), and Body Control Module (BCM).}
	\label{table:representative_messages}
	\footnotesize
	\begin{tabular}{|c|c|c|c|}
		\hline
		Message ID & Period (ms) & Transmitter & Data Size (hours) \\ \hline
		0x0D1      & 10          & EBCM     & 0.53 hours        \\ \hline
		0x185      & 20          & EBCM     & 1.01 hours      \\ \hline
		0x1FC      & 50          & EBCM     & 2.01 hours      \\ \hline
		0x184      & 100         & EPS      & 4.44 hours      \\ \hline
		0x22A      & 100         & EPB      & 3.84 hours      \\ \hline
		0x3C9      & 100         & BCM      & 4.48 hours       \\ \hline
	\end{tabular}
	\normalsize
	%\vspace{-0.3cm}
\end{table}

\textbf{Post-processing.} Due to the limited capability of the receiver to capture all messages in the vehicle, some messages are missed sporadically. 
To maintain message periodicity, missing messages are inserted during post-processing.

\textbf{Setup.}  
To obtain the predicted attack success probability curve, we feed $1000$ batches of normal data to the IDS and computed the attack success probability $P_s$ for different $\Delta T$ using equations in Sections \ref{sec:state-of-art} and \ref{sec:ntp}. 
For consistency, we set $N=20$, $\gamma=4$, $\Gamma=5$, and $\kappa=8$ for the IDS in all experiments, and there are no false alarms. 

To obtain the experimental attack success probability curve, the same normal data is fed to the IDS, followed by $n$ batches of attack data in each experiment. 
%The same setting used for the SOTA IDS was used for the NTP-based IDS.
A total of $100$ independent experiments are performed, and the ratio of experiments where the attack is successful (undetected) is computed as $P_s$ for $\Delta T=0$.

Since collecting data for each $\Delta T$ value would be prohibitively time-consuming, we generate attack data for other $\Delta T$ values by adding a fixed offset equal to $\Delta T$ to the inter-arrival times of the collected data, as in Section~\ref{sec:evaluation_cloaking_attack}.
By repeating the previous process, we obtain the experimental attack success probability curve.

\textbf{Metric for quantifying the prediction error.} 
In order to quantify the prediction error of the proposed  models, we define a metric called \textit{Area Deviation Error (ADE)} as
%In order to amplify the error of modeling the experimental success probability curve using a high-fidelity model, we define a metric called \textit{area percentage difference} $\rho$ as 
\begin{equation}
\label{eq:ratio_area}
\text{ADE} = \frac{\int_{-\infty}^{\infty}|P_{s,pred.}(\Delta T)-P_{s,exp.}(\Delta T)|d\Delta T}{\int_{-\infty}^{\infty}P_{s,exp.}(\Delta T)d\Delta T} \times 100\%,
\end{equation}
where $P_{s,pred.}(\Delta T)$ and $P_{s,exp.}(\Delta T)$ are the predicted and experimental attack success probabilities, % based on formal analysis and experiments, 
respectively. 
In other words, ADE is the ratio of the absolute difference of the areas under the predicted and experimental attack success probability curves to the area under the experimental curve. Hence, a smaller ADE value implies a smaller deviation from the experimental curve (the ground truth) and  better prediction accuracy.
Note that the ADE can be larger than $100\%$, when the area under the experimental curve is small.

\begin{figure}[t!]
	\centering
	\begin{subfigure}[h]{0.45\columnwidth}
		\includegraphics[width=\columnwidth]{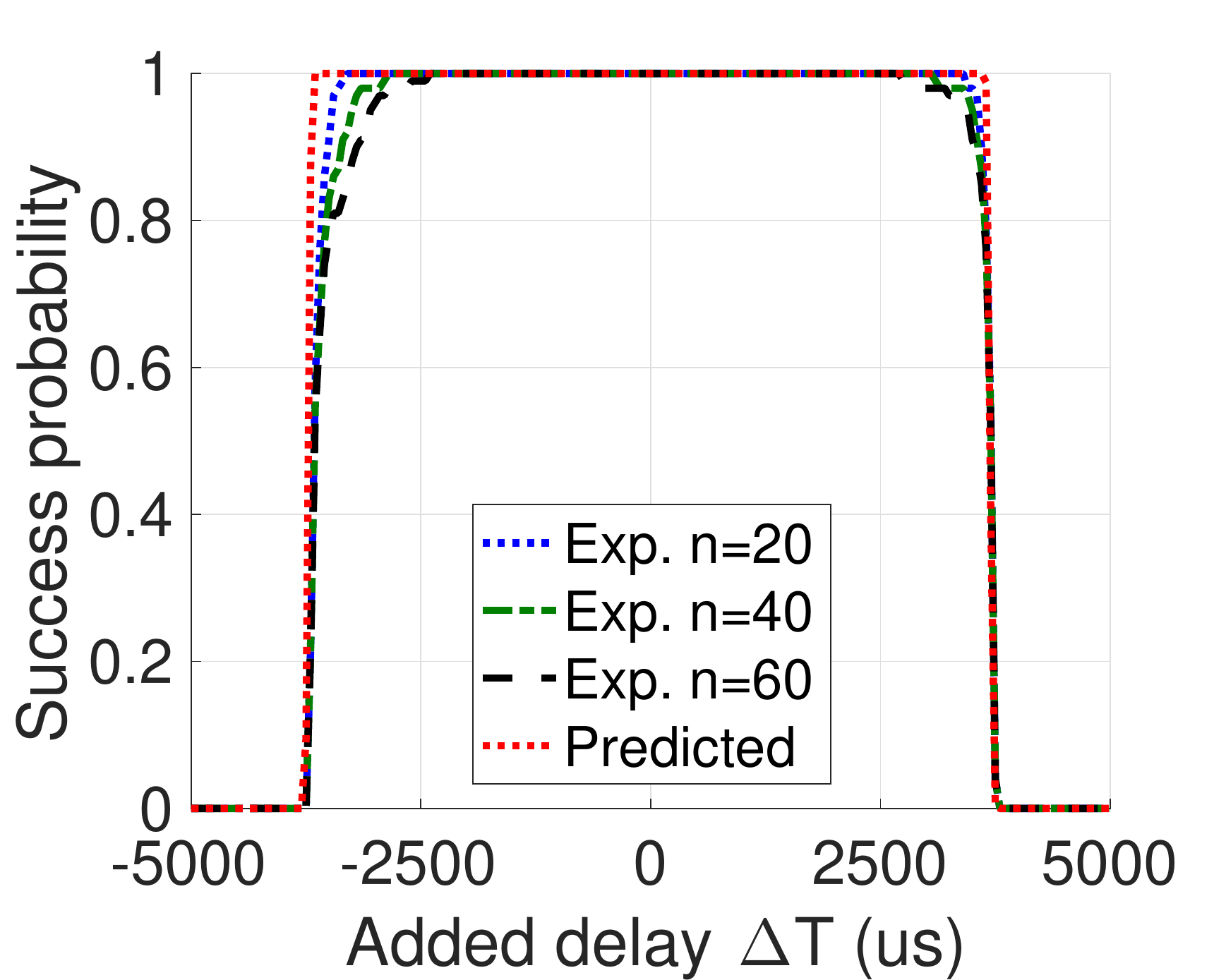}
		\caption{0x0D1}
		\label{fig:comparison_SoA_0x0D1}
	\end{subfigure}
	\begin{subfigure}[h]{0.45\columnwidth}
		\includegraphics[width=\columnwidth]{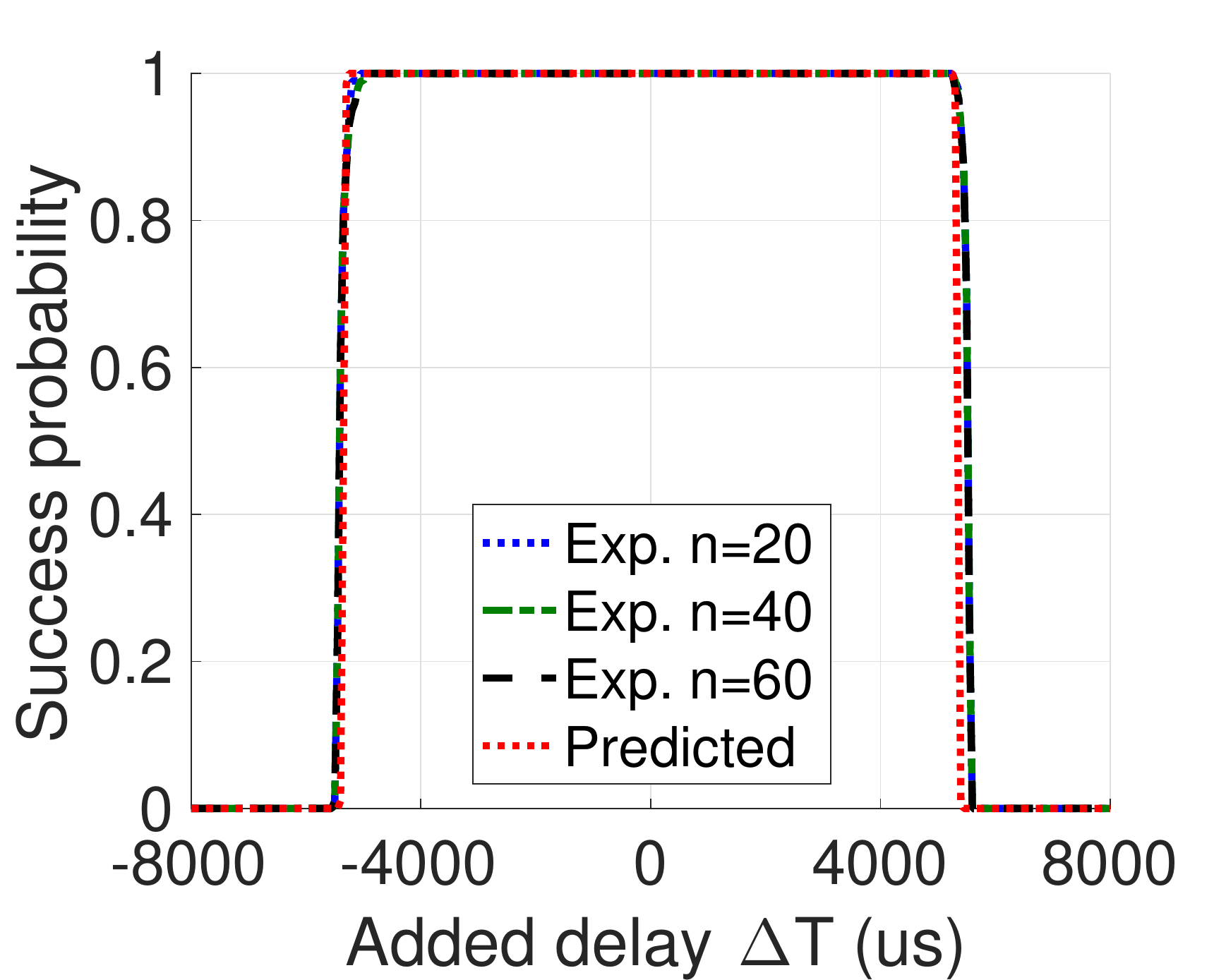}
		\caption{0x185}
		\label{fig:comparison_SoA_0x185}
	\end{subfigure}
	\begin{subfigure}[h]{0.45\columnwidth}
		\includegraphics[width=\columnwidth]{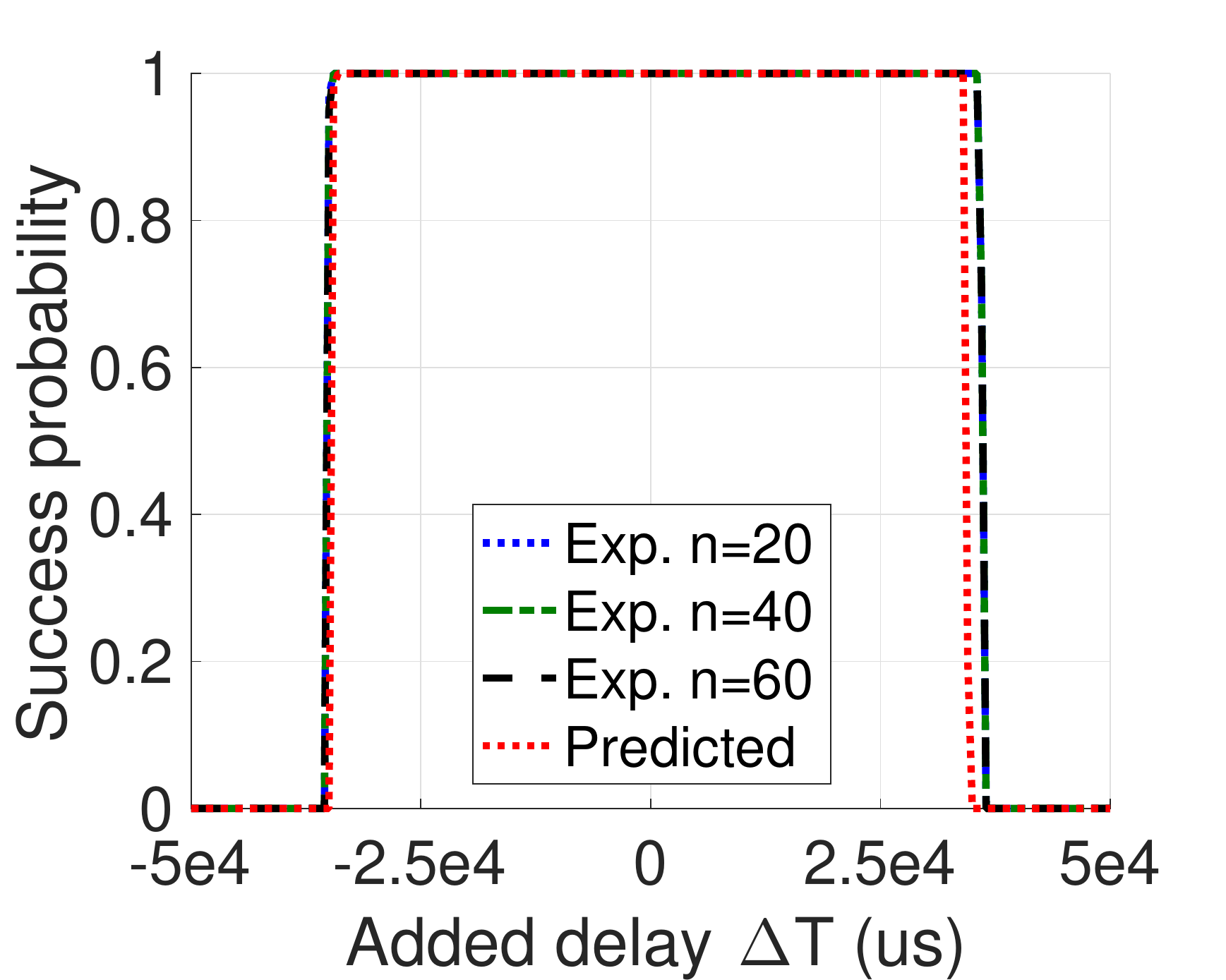}
		\caption{0x1FC}
		\label{fig:comparison_SoA_0x1FC}
	\end{subfigure}
	\begin{subfigure}[h]{0.45\columnwidth}
		\includegraphics[width=\columnwidth]{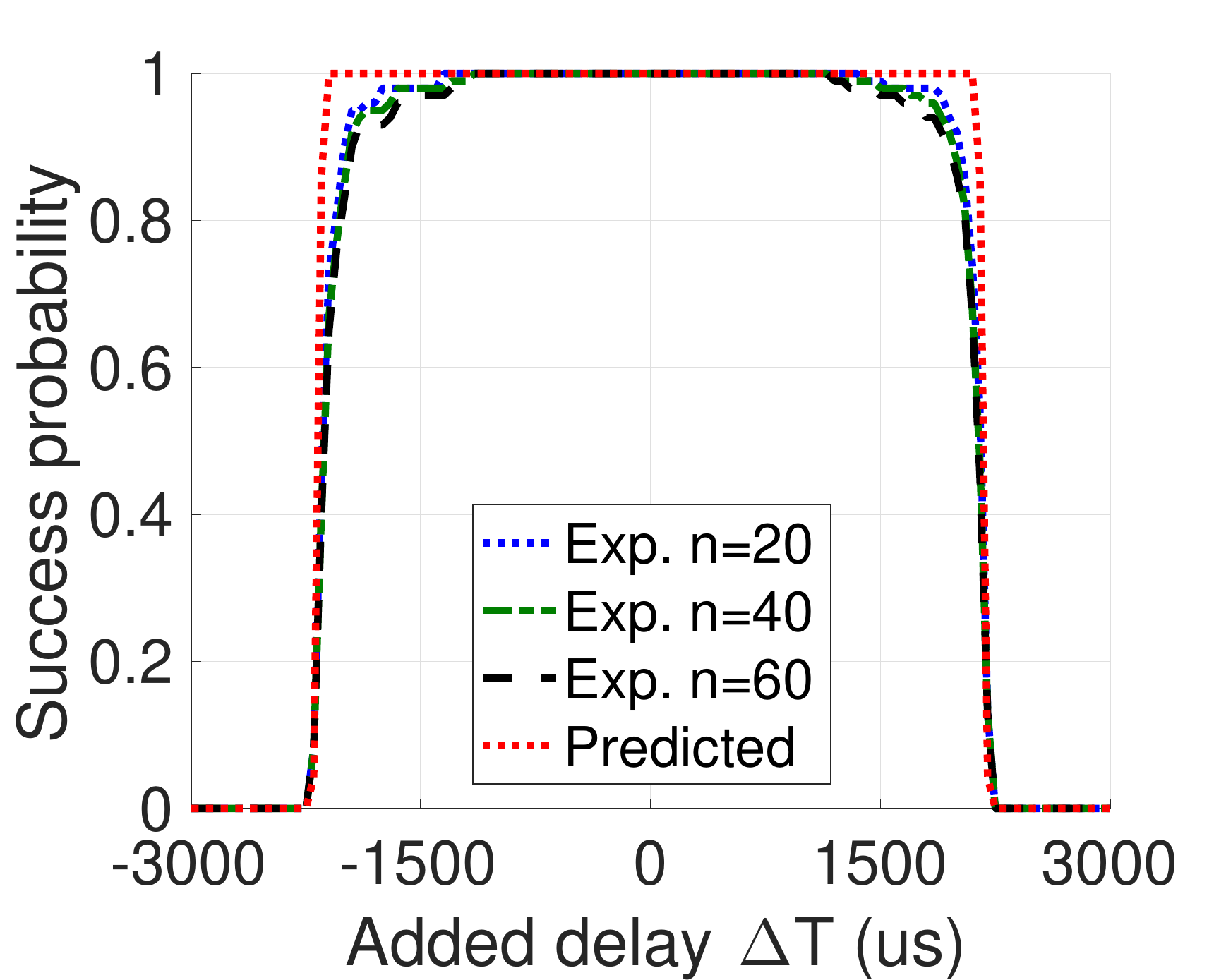}
		\caption{0x184}
		\label{fig:comparison_SoA_0x184}
	\end{subfigure}
	\begin{subfigure}[h]{0.45\columnwidth}
		\includegraphics[width=\columnwidth]{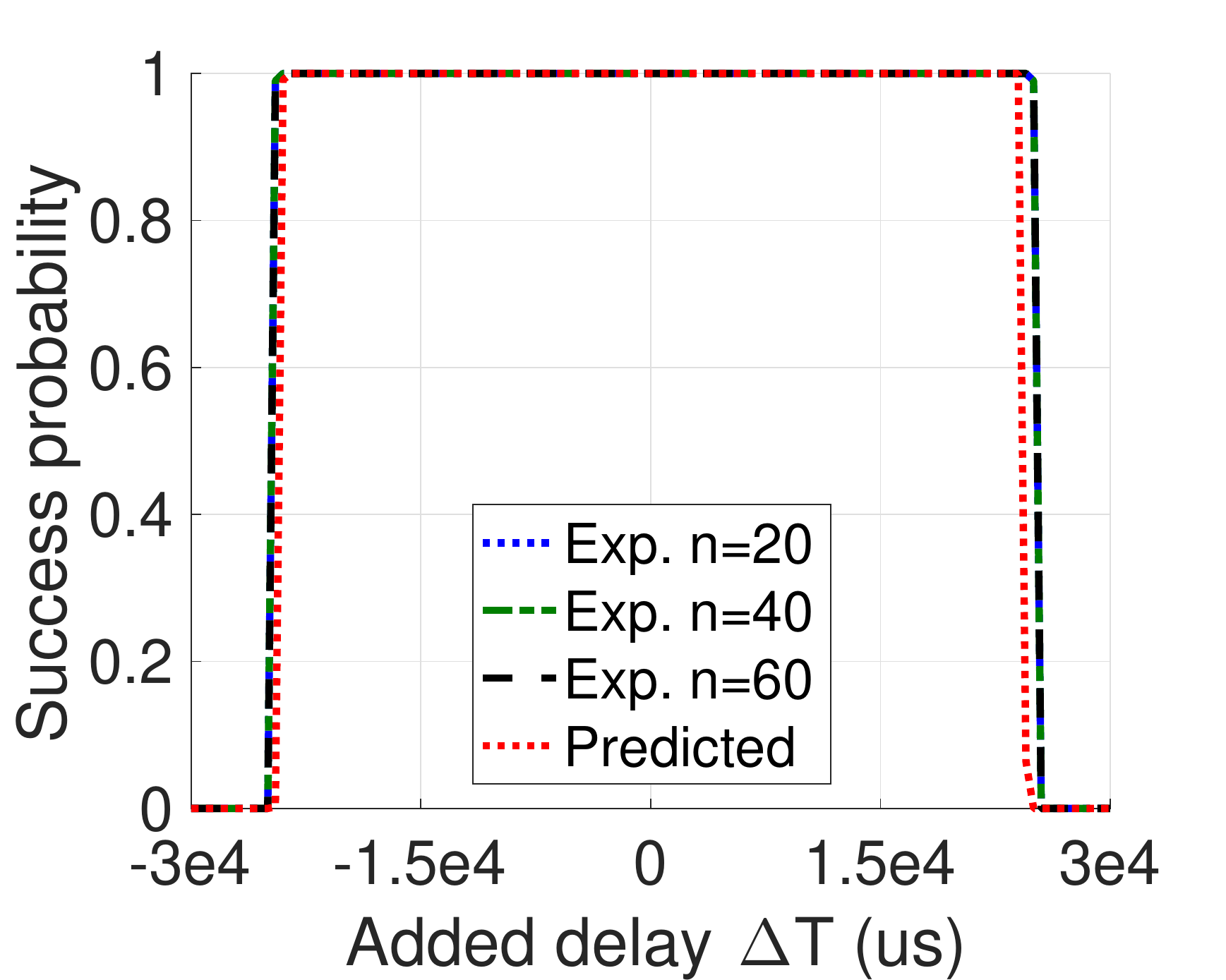}
		\caption{0x22A}
		\label{fig:comparison_SoA_0x22A}
	\end{subfigure}
	\begin{subfigure}[h]{0.45\columnwidth}
		\includegraphics[width=\columnwidth]{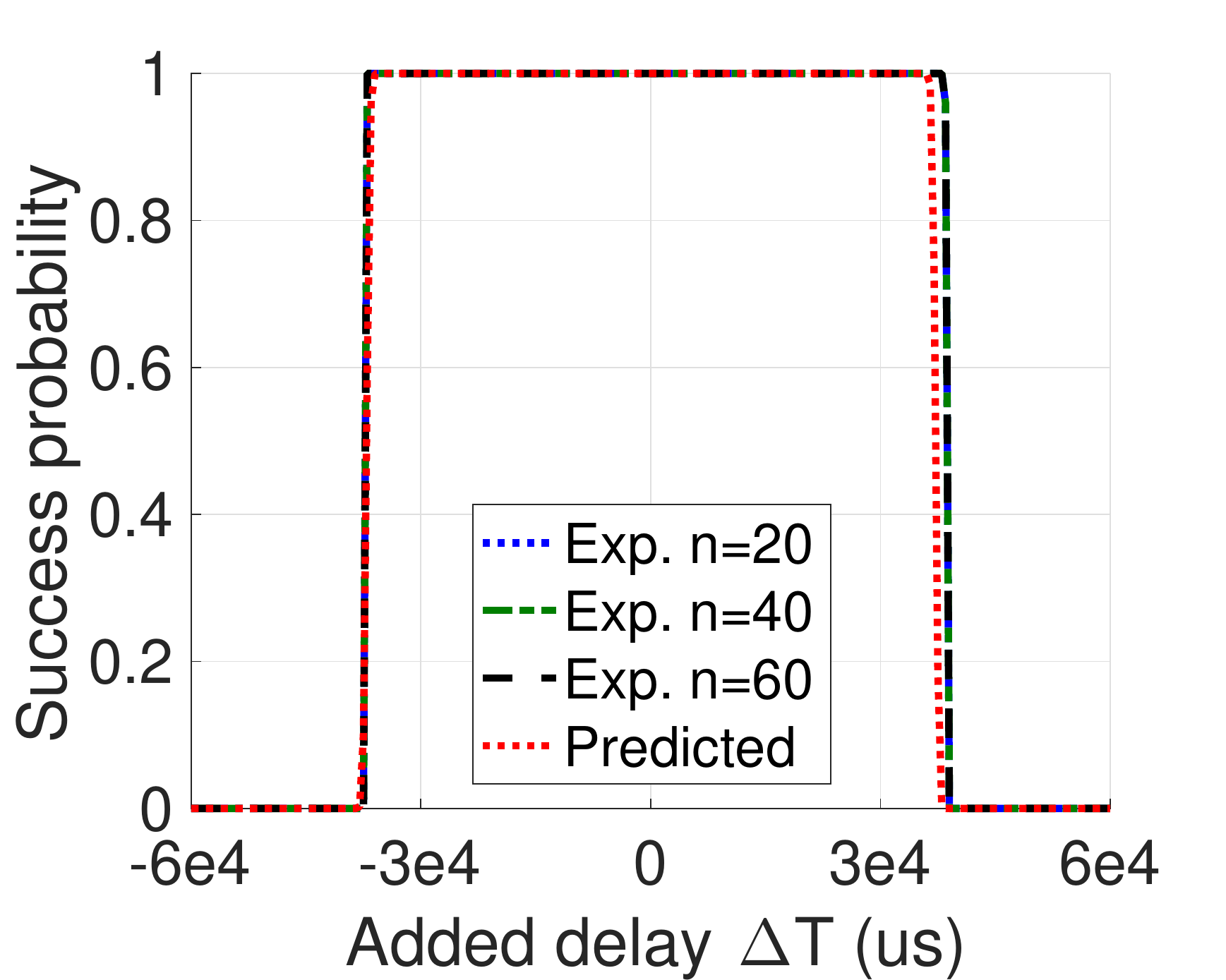}
		\caption{0x3C9}
		\label{fig:comparison_SoA_0x3C9}
	\end{subfigure}
	\caption{Experimental versus predicted attack success probability curves for the SOTA IDS with different numbers of attack batches $n$ for different messages. 
	The predicted curves are closely matched with the experimental curves for all messages. 
	Note that the proposed model for the SOTA IDS is insensitive to $n$, thus providing the same prediction for different $n$ values.
	}
	\label{fig:success_rate_comparison_SoA}
\end{figure}

\begin{figure}[t!]
	\centering
	\begin{subfigure}[h]{0.45\columnwidth}
		\includegraphics[width=\columnwidth]{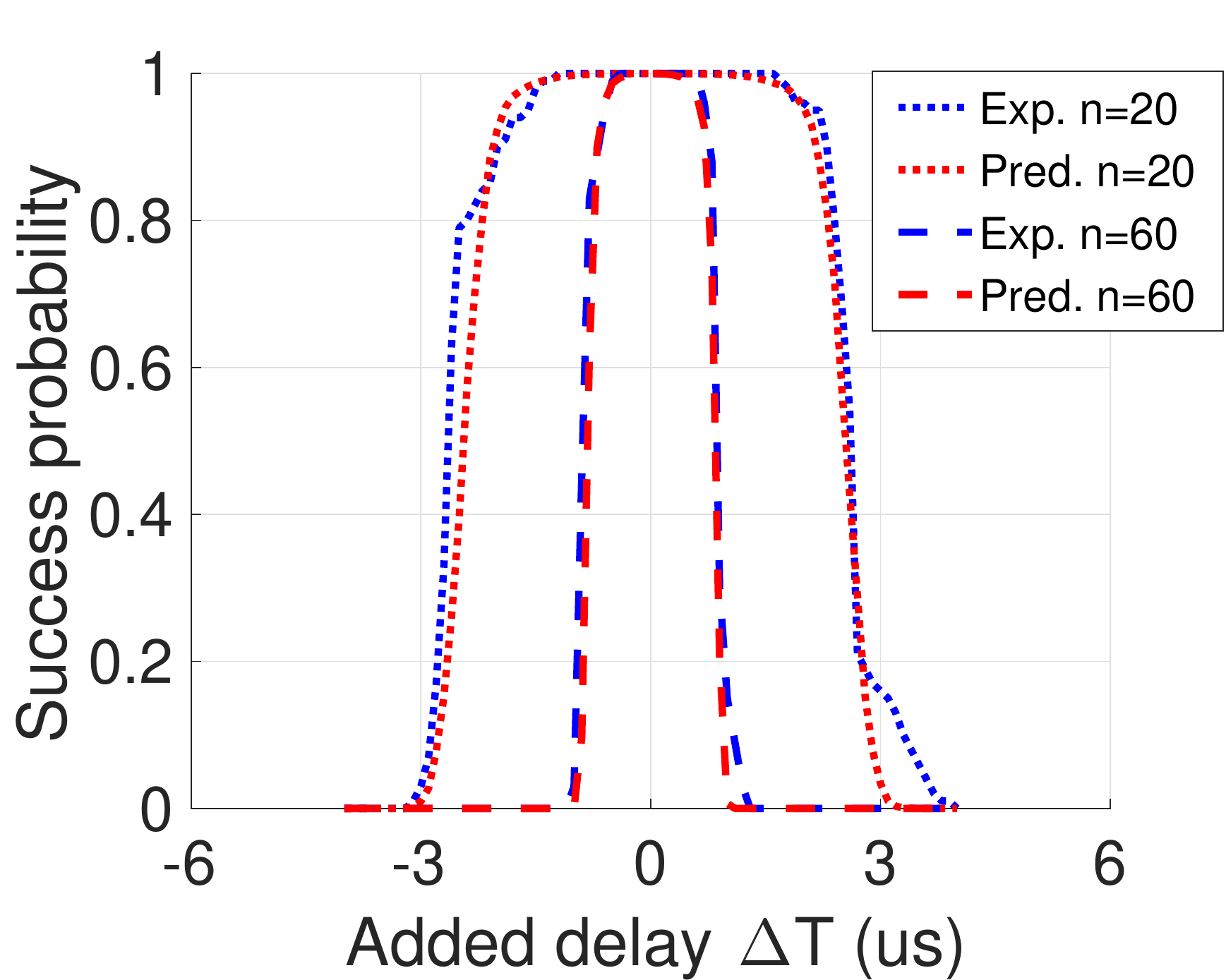}
		\caption{0x0D1}
		\label{fig:comparison_ntp_0x0D1}
	\end{subfigure}
	\begin{subfigure}[h]{0.45\columnwidth}
		\includegraphics[width=\columnwidth]{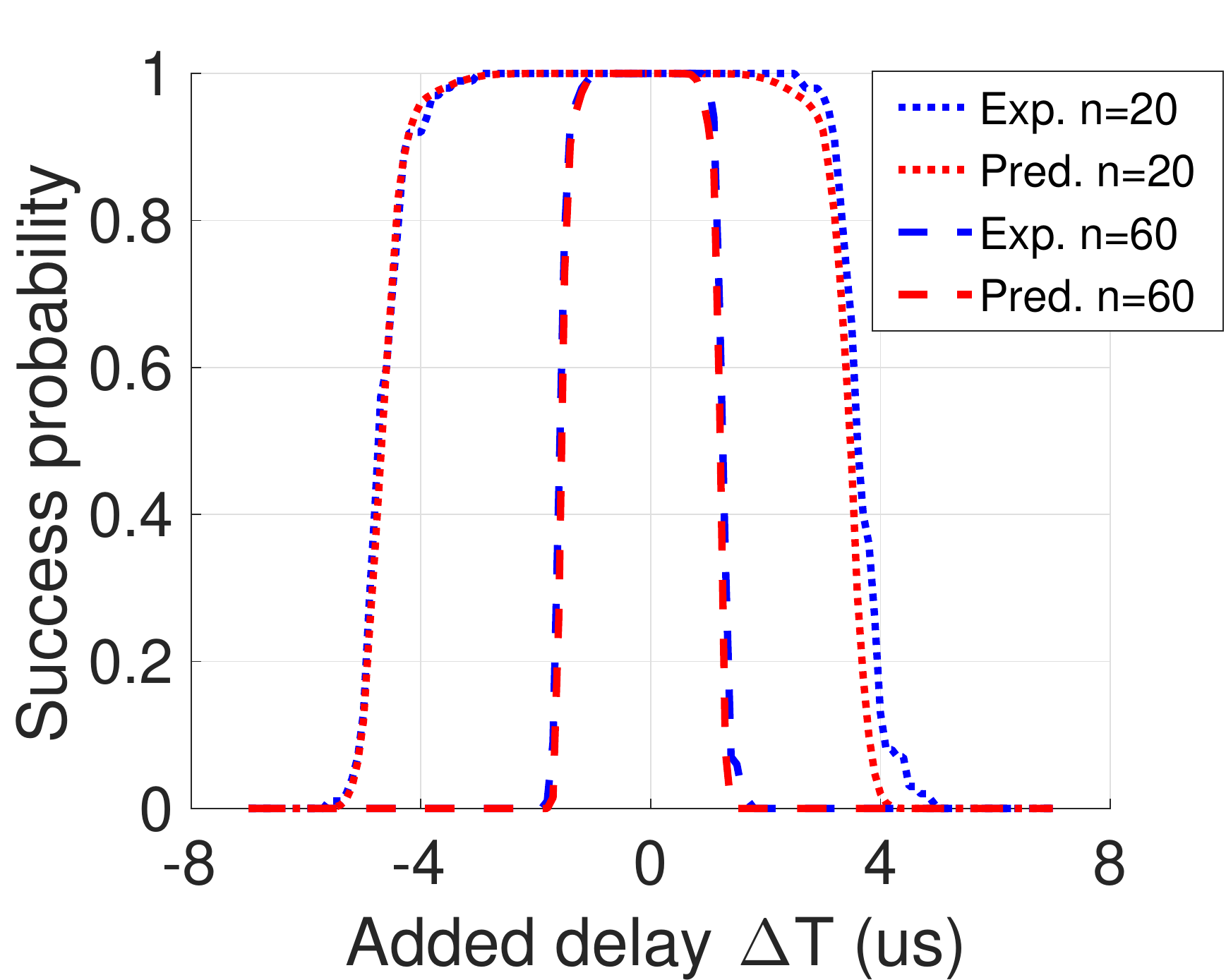}
		\caption{0x185}
		\label{fig:comparison_ntp_0x185}
	\end{subfigure}
	\begin{subfigure}[h]{0.45\columnwidth}
		\includegraphics[width=\columnwidth]{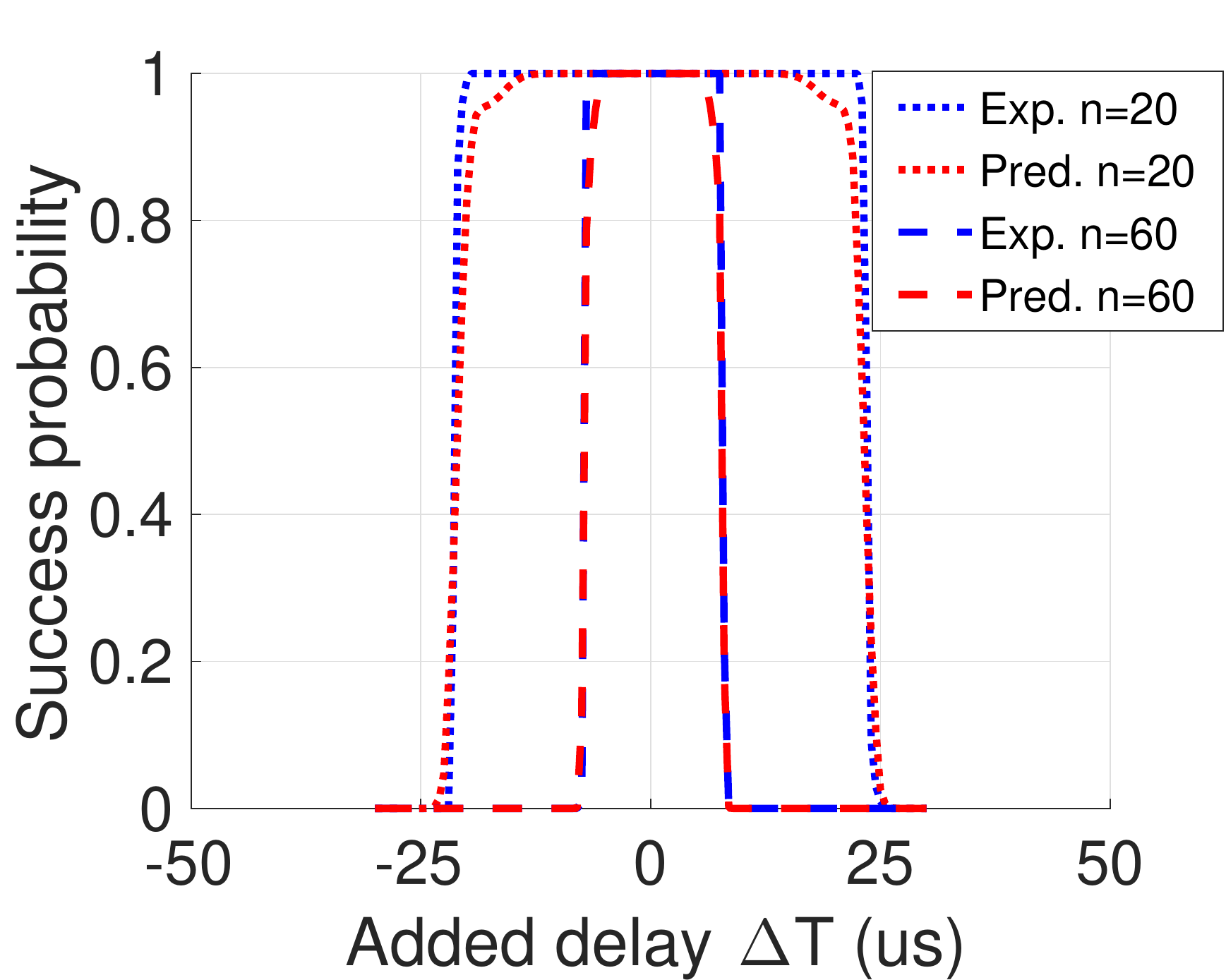}
		\caption{0x1FC}
		\label{fig:comparison_ntp_0x1FC}
	\end{subfigure}
	\begin{subfigure}[h]{0.45\columnwidth}
		\includegraphics[width=\columnwidth]{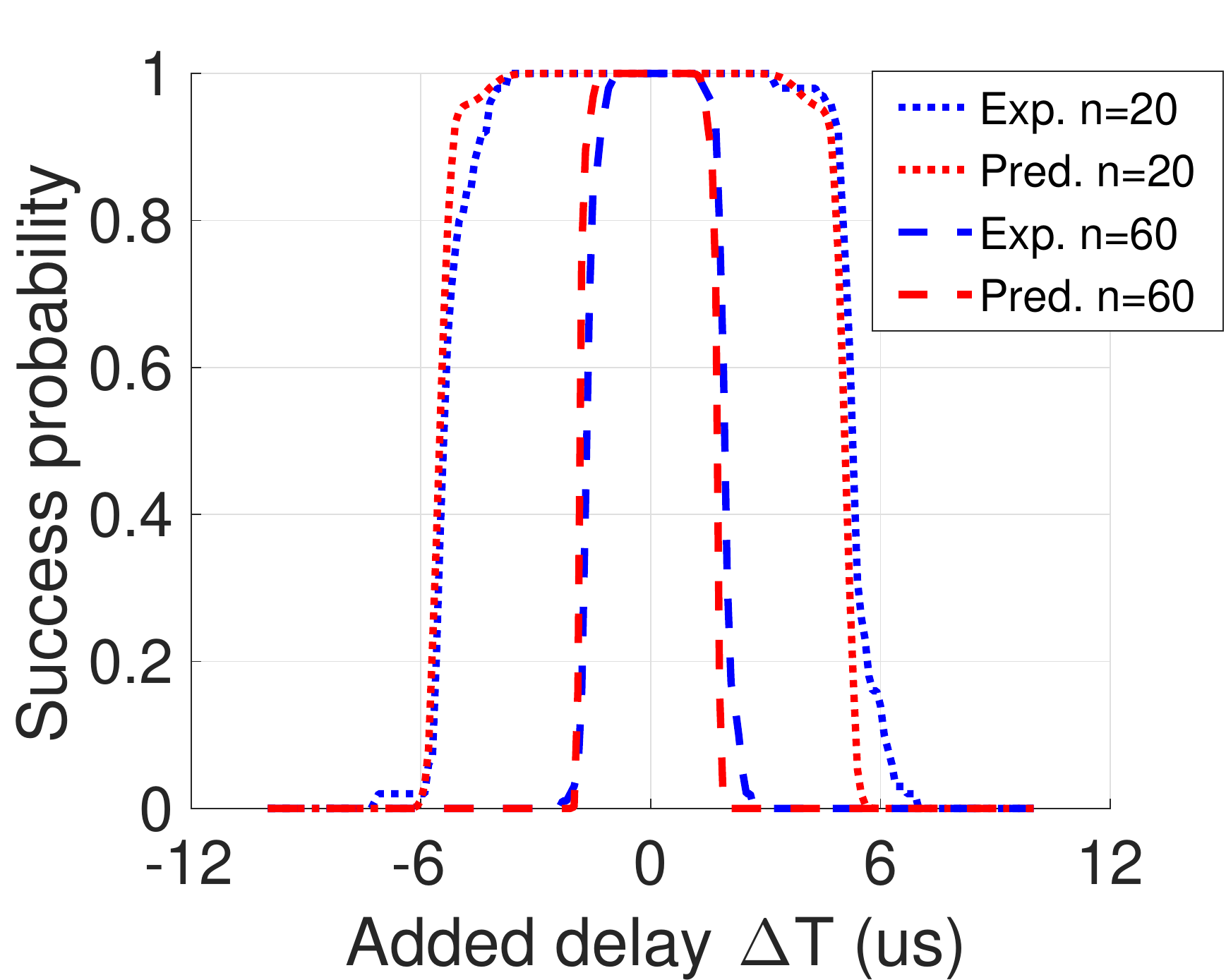}
		\caption{0x184}
		\label{fig:comparison_ntp_0x184}
	\end{subfigure}
	\begin{subfigure}[h]{0.45\columnwidth}
		\includegraphics[width=\columnwidth]{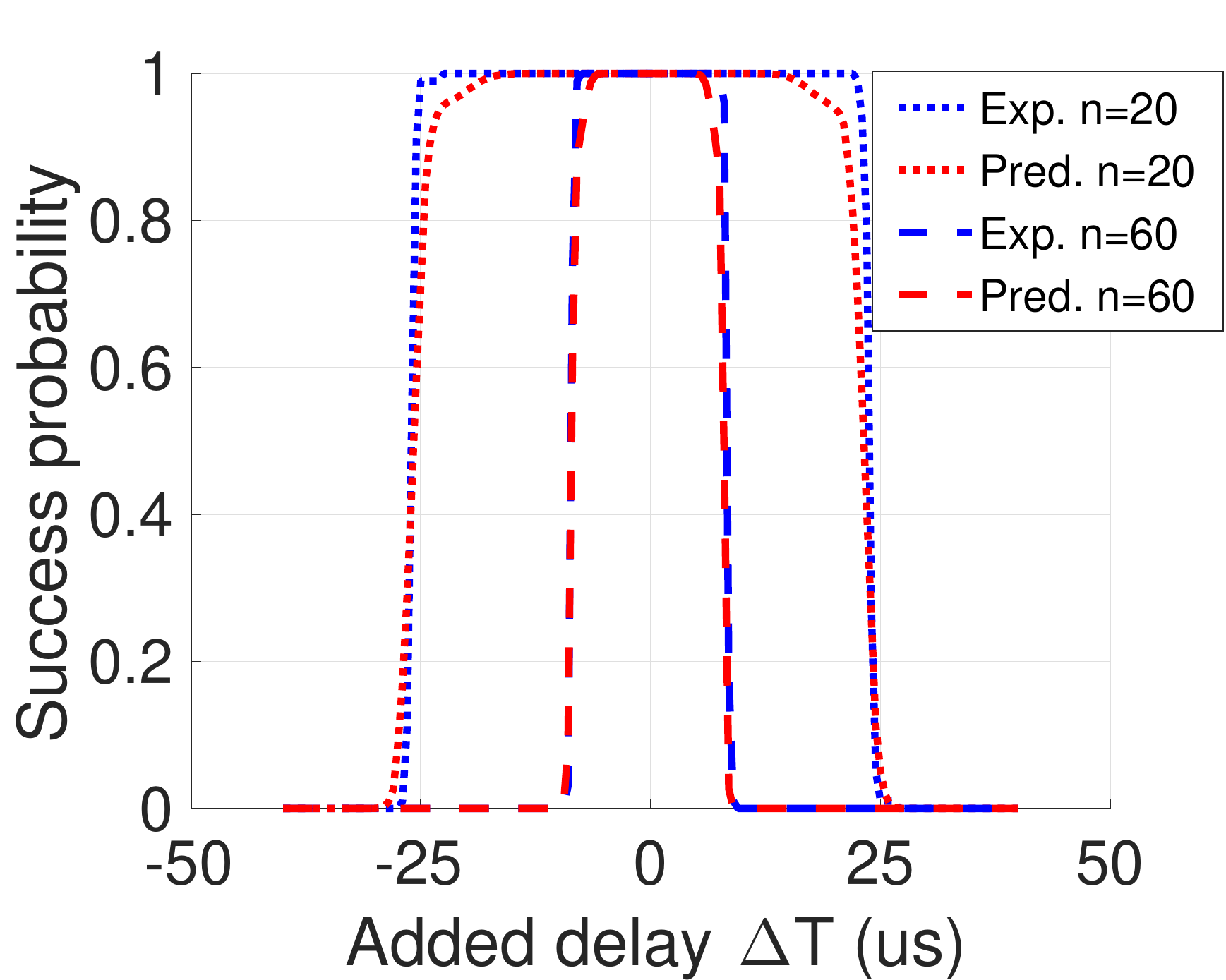}
		\caption{0x22A}
		\label{fig:comparison_ntp_0x22A}
	\end{subfigure}
	\begin{subfigure}[h]{0.45\columnwidth}
		\includegraphics[width=\columnwidth]{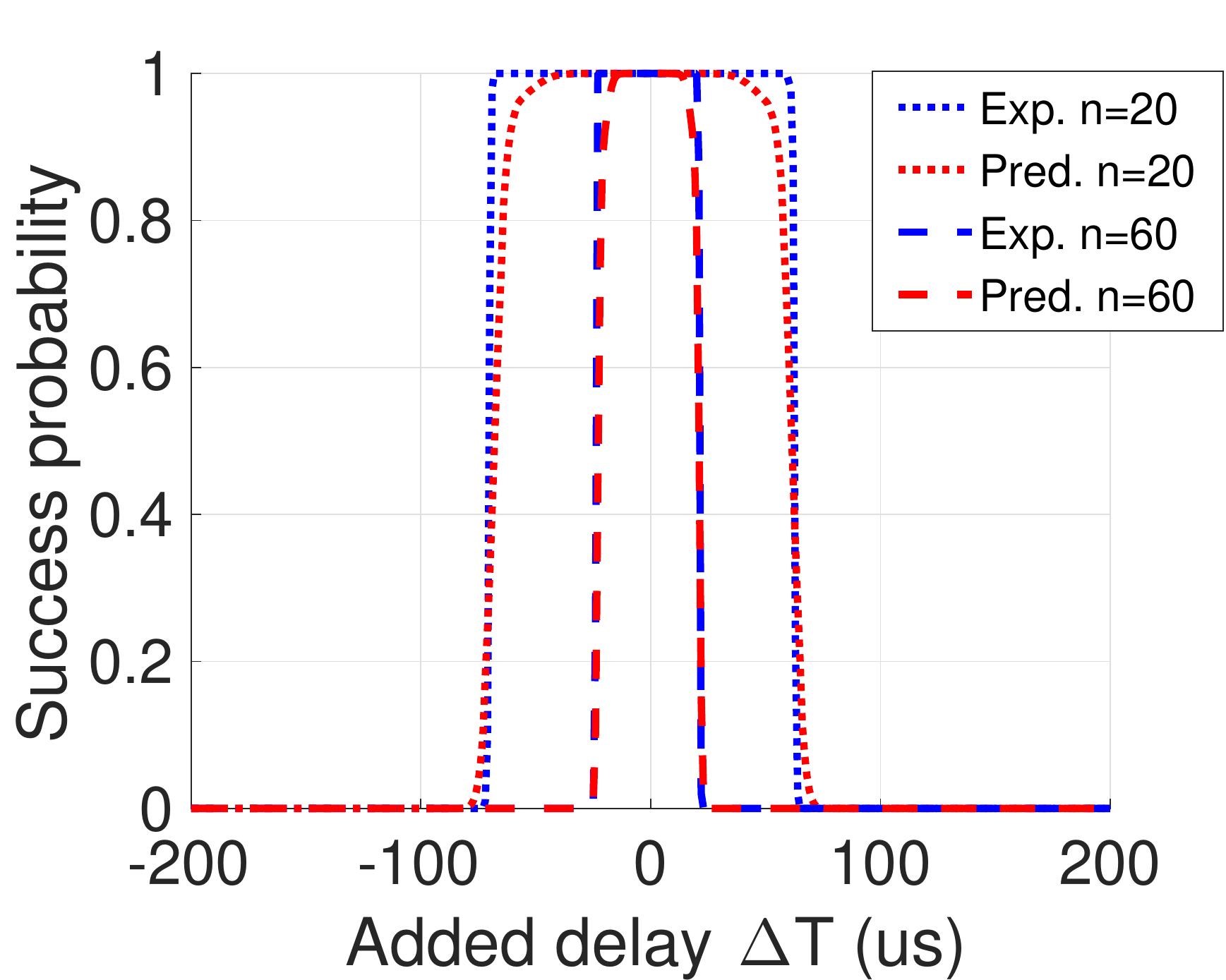}
		\caption{0x3C9}
		\label{fig:comparison_ntp_0x3C9}
	\end{subfigure}
	\caption{Experimental versus predicted attack success probability curves for the NTP-based IDS with $n=20$ and $60$ batches of attack data for different messages. 
	The predicted curves match well in general with the experimental curves for all messages and have a closer agreement for larger $n$. 
	%Note that results for $n=40$ are not provided due to space limit.
	}
	\label{fig:success_rate_comparison_NTP}
	%\vspace{-0.3cm}
\end{figure}

\textbf{Evaluation of SOTA IDS analysis.} 
As shown in Fig.~\ref{fig:success_rate_comparison_SoA}, we can see a close match between the predicted and experimental curves. %which means that the proposed model is able to capture the shape of $P_{s,exp.}(\Delta T)$ very well.
%Note that 
For a given $\Delta T$, the proposed model provides the same  attack success probability for different $n$.
This is because our analysis focuses on modeling the normalized identification error in the first attack batch and its rate of decrease using the system parameters, which are independent of $n$.
In fact, the closeness of the curves agrees with the observation that the SOTA IDS is insensitive to $n$.
%For messages like 0x0D1, we also observe small discrepancies at the corners of the curves, which are probably caused by outliers in collected data. 
%Despite of the close agreement in terms of shape, there exists discrepancies at the corners of the curve for messages like 0x0D1 and 0x184, which is mainly caused by outliers in collected data. 
%That is, the statistics of attack data used in several experiments have large deviations from the average.
%In contrast, for messages like 0x185, we can see a noticeable mismatch when $\Delta T$ is between 4ms and 5ms. This could be caused by the fact that we are taking the linear approximation of the expected value of $\tau$, while the true value of $\tau$ may deviate from its mean value.

\rev{In addition, for messages like 0x0D1 and 0x184, we observe small discrepancies at the corners of the curves, which may caused by outliers in collected data. In the meanwhile, the assumption of linearly decreasing normalized identification error may also cause the proposed model to overestimate the attack success probability. We observe that for some messages with less noise in timestamps, the normalized identification error of the SOTA IDS may not strictly decrease. In this case, the error lasts for a longer duration and causes the attack to be detected at a later time, which could explain why the experimental attack success probability is smaller than the predicted value. Improving our formal model for the SOTA IDS for messages with less noise is left as future work.}

\textbf{Evaluation of NTP-based IDS analysis.}
As shown in Fig.~\ref{fig:success_rate_comparison_NTP}, we can also see a close agreement in shape between the predicted and experimental attack success probability curves.
The fact that the attack success probability decreases as $n$ increases for the NTP-based IDS can also be captured by the proposed model.
Although there are discrepancies due to outliers, the gap between the predicted and experimental curves becomes smaller when $n$ is increased from $20$ to $60$.
%These results imply that the proposed model is able to provide better prediction for larger $n$, which is desirable as the attack is more likely to last for a longer time in reality. 

\begin{table}[t!]
	\footnotesize
	\centering
	\caption{ADE (\%) between the predicted and experimental attack success probability curves for the SOTA and NTP-based IDSs with different numbers of attack batches.}
	\begin{tabular}{|p{0.25in}|c|c|c|c|c|c|}
		\hline
		\multirow{2}{0.25in}{Msg ID} & \multicolumn{3}{c|}{SOTA IDS} & \multicolumn{3}{c|}{NTP-based IDS}\\
		\cline{2-7}
		& 20 &40 & 60 & 20 & 40 & 60\\
		\hline
		0x0D1 & 1.5 & 2.3 & 3.1 & 6.1 & 7.7 & 6.9 \\ \hline
		0x185 & 2.1 & 2.2 & 2.2 & 3.3 & 4.0 & 3.8 \\ \hline
		0x1FC & 3.2 & 3.1 & 3.1 & 3.4 & 3.6 & 2.8 \\ \hline
		0x184 & 2.9 & 3.8 & 4.5 & 5.3 & 9.3 & 11.9 \\ \hline
		0x22A & 2.9 & 2.9 & 2.9 & 3.7 & 3.9 & 3.9 \\ \hline
		0x3C9 & 2.4 & 2.4 & 2.4 & 5.6 & 5.2 & 5.1 \\ \Xhline{2\arrayrulewidth}
		%\textbf{Mean} & \sy{4.1} & \sy{4.4} & \sy{4.6} & \sy{10.1} & \sy{8.5} & \sy{8.9}  % This are old results
		\textbf{Mean} & 2.5 & 2.8 & 3.0 & 4.6 & 5.6 & 5.7 \\ \hline
	\end{tabular}
	\label{table:ratio_comparison}
	\normalsize
	%\vspace{-0.5cm}
\end{table}

\textbf{Prediction accuracy.}
As shown in Table~\ref{table:ratio_comparison}, the prediction error in terms of ADE is message-dependent for both IDSs.
For the SOTA IDS, the average ADE is within $3.0\%$ for the SOTA IDS, and it is within $5.7\%$ for the NTP-based IDS. 
We also note that there is no explicit relationship between ADE and $n$ for both IDSs.

\hl{
\textbf{Applicability to IDSs with customized settings.}
In the above experiments, we used the same IDS setting for consistency. 
In practice, IDS settings may be customized to optimize the detection performance.
In order to validate that the proposed models are not limited to a specific IDS setting, we validate our models for both SOTA and NTP-based IDSs with varying parameters of batch size $N$, as well as CUSUM parameters $\gamma$, $\Gamma$, and $\kappa$.
In this experiment, we chose message 0x185 as an example and used $N=20$, $\gamma=4$, $\Gamma=5$, and $\kappa=8$ as the baseline setting.
We then changed one of the parameters to a different value every time, while keeping others the same. 

\begin{figure}[t!]
	\centering
	\begin{subfigure}[h]{0.44\columnwidth}
		\captionsetup{justification=centering}
		\includegraphics[width=\columnwidth]{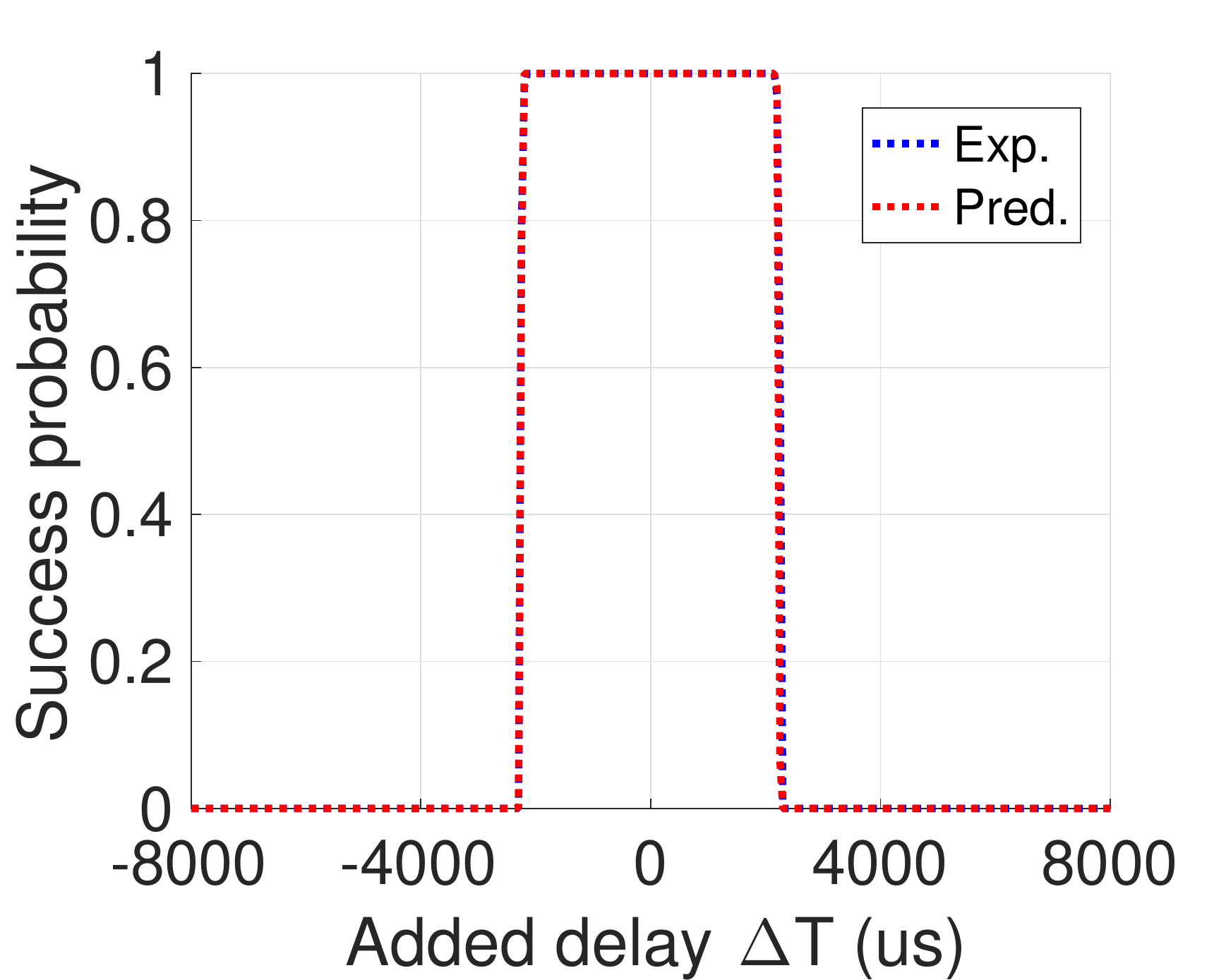}
		\caption{$\boldsymbol{N}=30$, $\gamma=4$, $\Gamma=5$, \\$\kappa=8$; $\text{ADE}=0.7$\%}
	\end{subfigure}
	\begin{subfigure}[h]{0.44\columnwidth}
		\captionsetup{justification=centering}
		\includegraphics[width=\columnwidth]{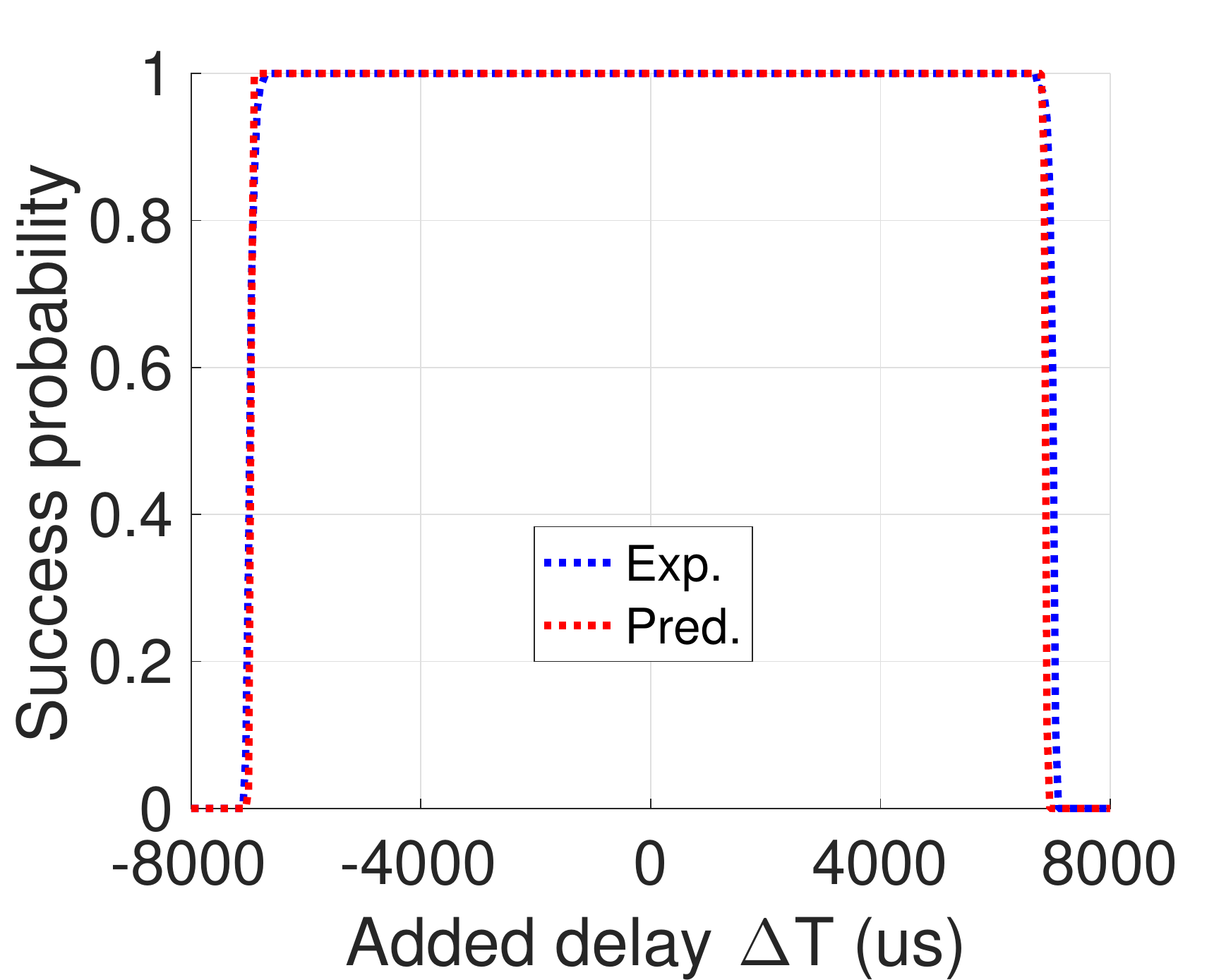}
		\caption{$N=20$, $\boldsymbol{\gamma}=3.6$, $\Gamma=5$,\\ $\kappa=8$; $\text{ADE}=1.1\%$}
	\end{subfigure}
	\begin{subfigure}[h]{0.44\columnwidth}
		\captionsetup{justification=centering}
		\includegraphics[width=\columnwidth]{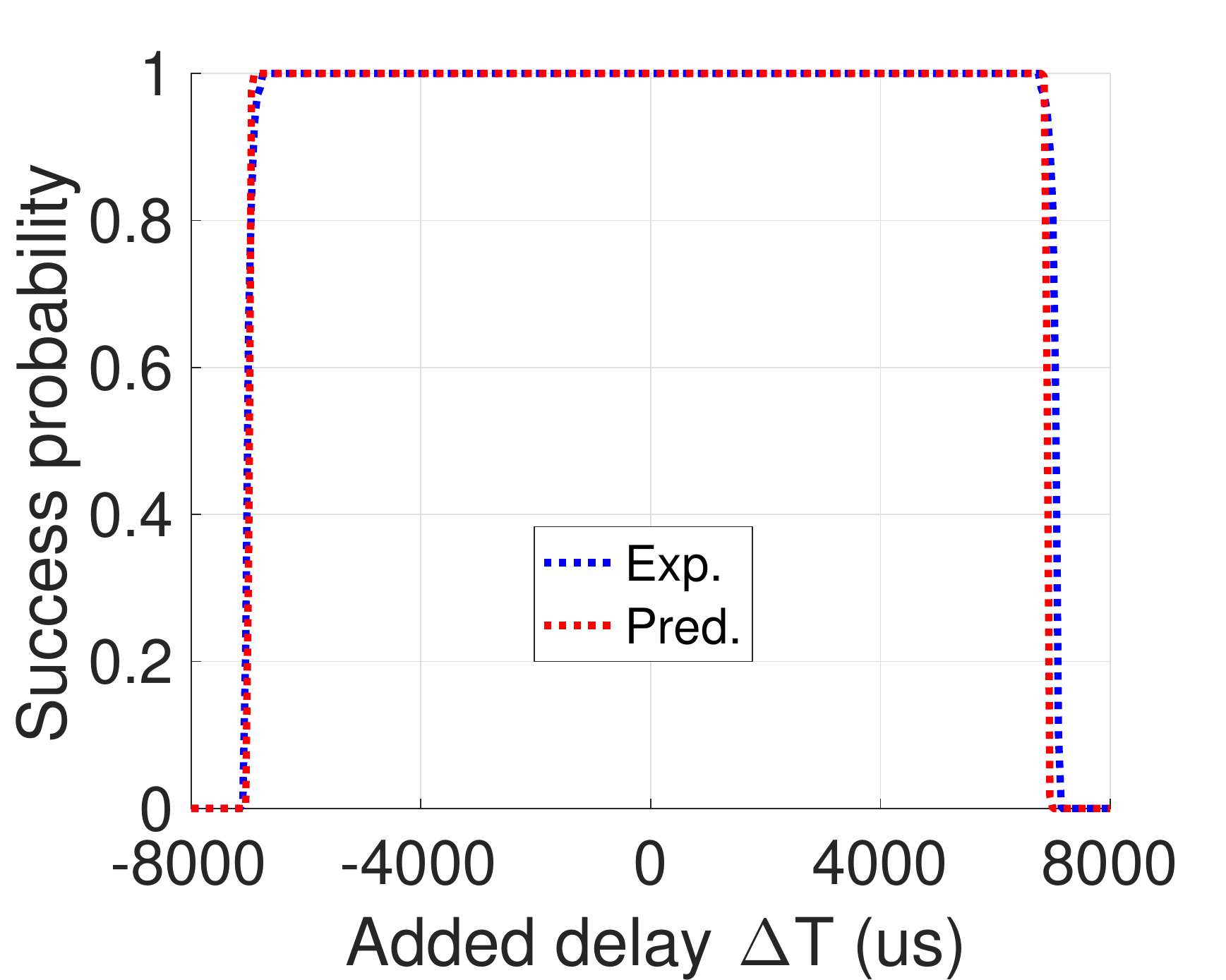}
		\caption{$N=20$, $\gamma=4$, $\boldsymbol{\Gamma}=6$,\\ $\kappa=8$; $\text{ADE}=1.3\%$}
	\end{subfigure}
	\begin{subfigure}[h]{0.44\columnwidth}
		\captionsetup{justification=centering}
		\includegraphics[width=\columnwidth]{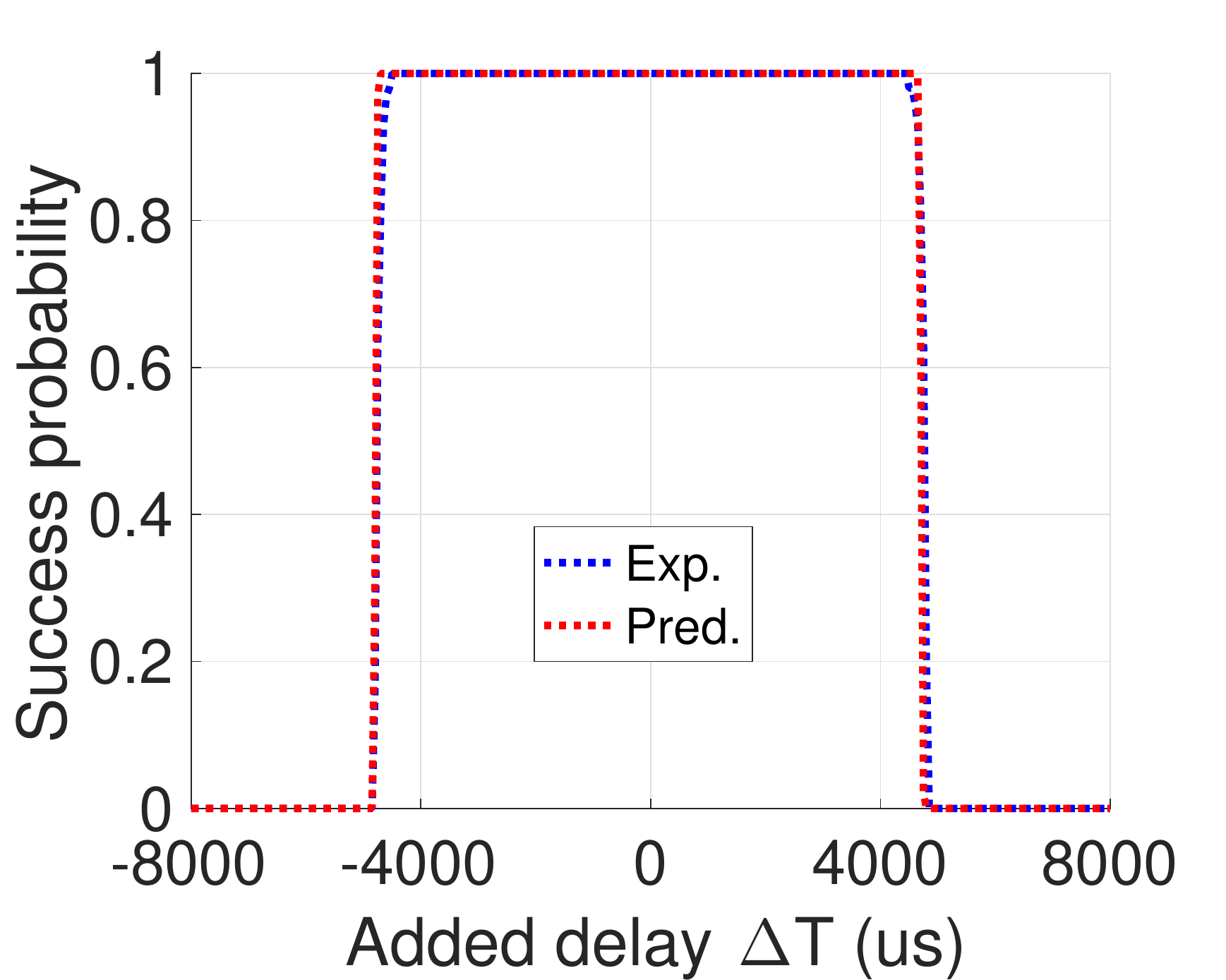}
		\caption{$N=20$, $\gamma=4$, $\Gamma=5$, \\ $\boldsymbol{\kappa}=5$; $\text{ADE}=1.0\%$}
	\end{subfigure}
	\caption{Experimental versus predicted attack success probability curves for the SOTA IDS with different settings. 
		They are closely matched in all cases.}
		%The analytical curves are closely matched with the experimental curves in all cases.}
	\label{fig:0x185_SoA_customized_IDS}
\end{figure}

\begin{figure}[t!]
	\centering
	\begin{subfigure}[h]{0.45\columnwidth}
		\captionsetup{justification=centering}
		\includegraphics[width=\columnwidth]{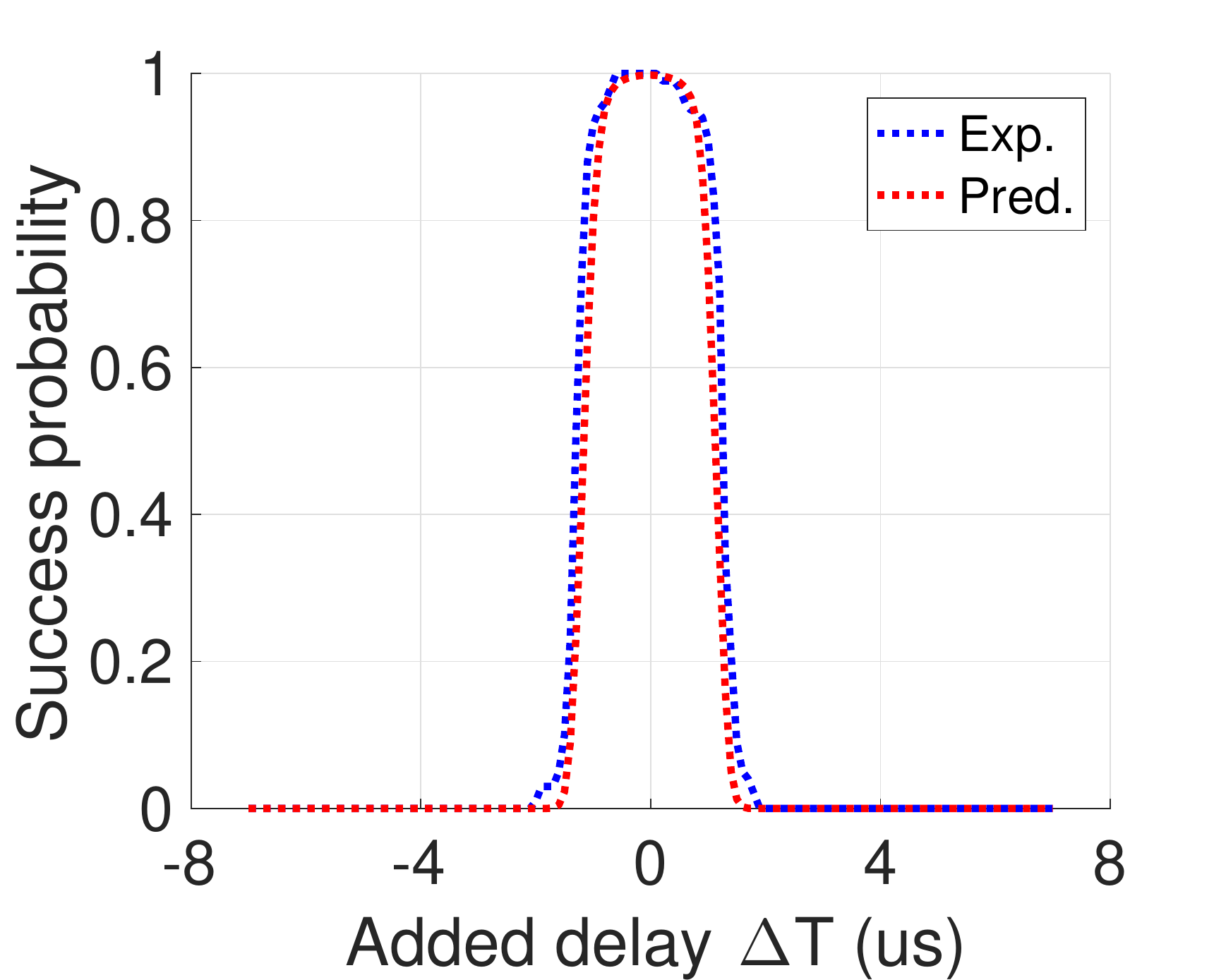}
		\caption{$\boldsymbol{N}=30$, $\gamma=4$, $\Gamma=5$,\\ $\kappa=8$; $\text{ADE}=11.6\%$}
	\end{subfigure}
	\begin{subfigure}[h]{0.45\columnwidth}
		\captionsetup{justification=centering}
		\includegraphics[width=\columnwidth]{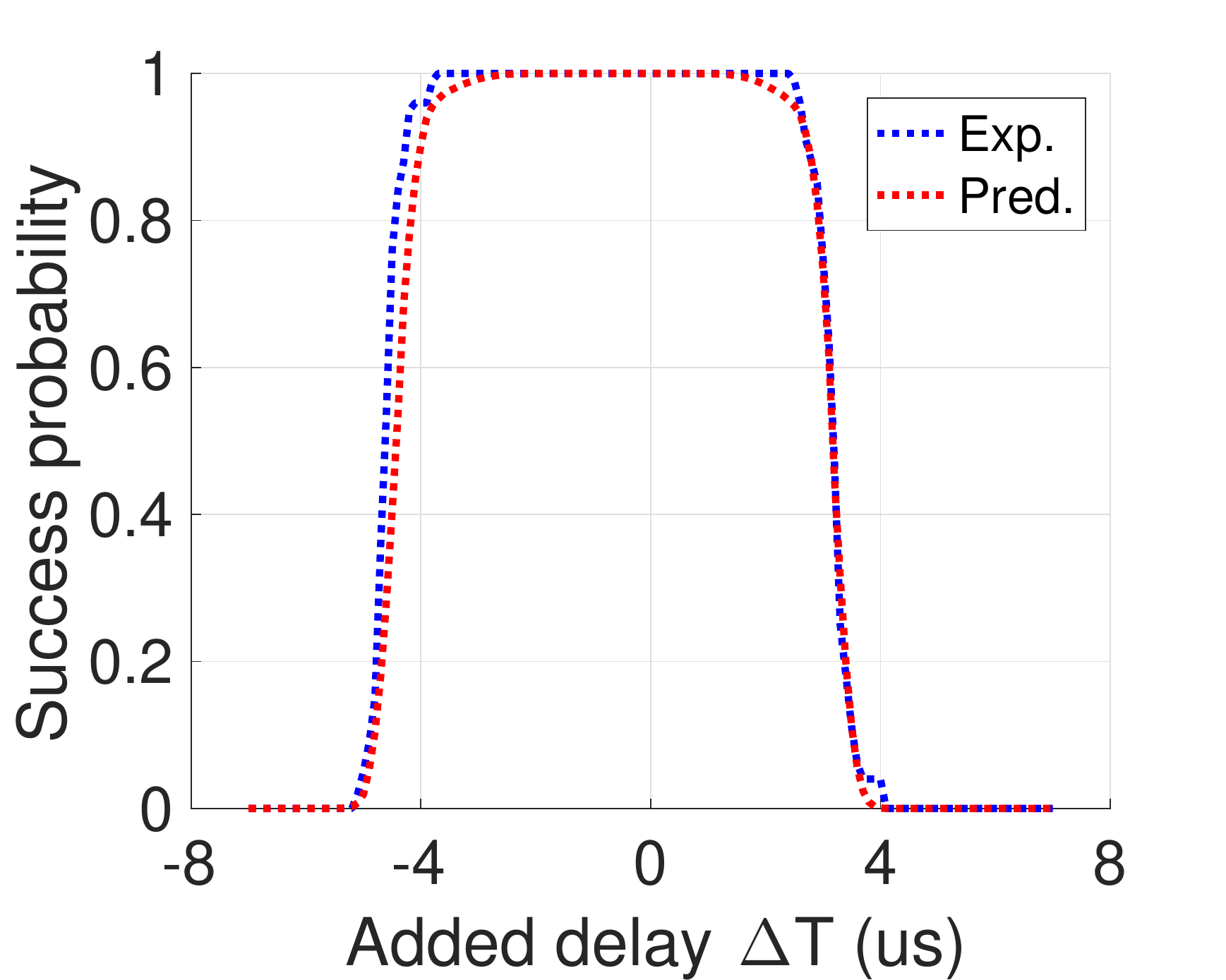}
		\caption{$N=20$, $\boldsymbol{\gamma}=3.6$, $\Gamma=5$,\\ $\kappa=8$; $\text{ADE} = 3.2\%$}
	\end{subfigure}
	\begin{subfigure}[h]{0.45\columnwidth}
		\captionsetup{justification=centering}
		\includegraphics[width=\columnwidth]{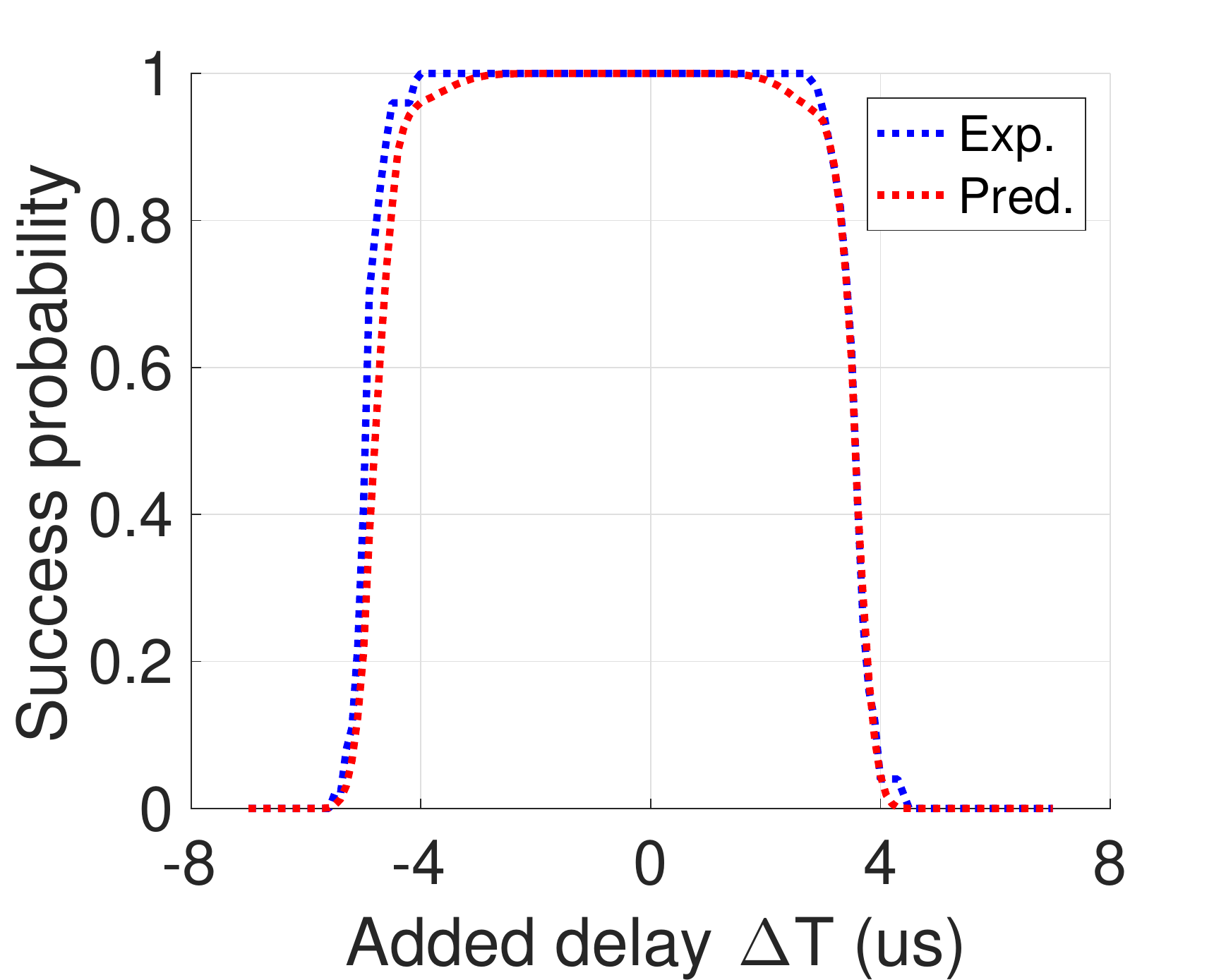}
		\caption{$N=20$, $\gamma=4$, $\boldsymbol{\Gamma}=6$,\\ $\kappa=8$; $\text{ADE}=2.9\%$}
	\end{subfigure}
	\begin{subfigure}[h]{0.45\columnwidth}
		\captionsetup{justification=centering}
		\includegraphics[width=\columnwidth]{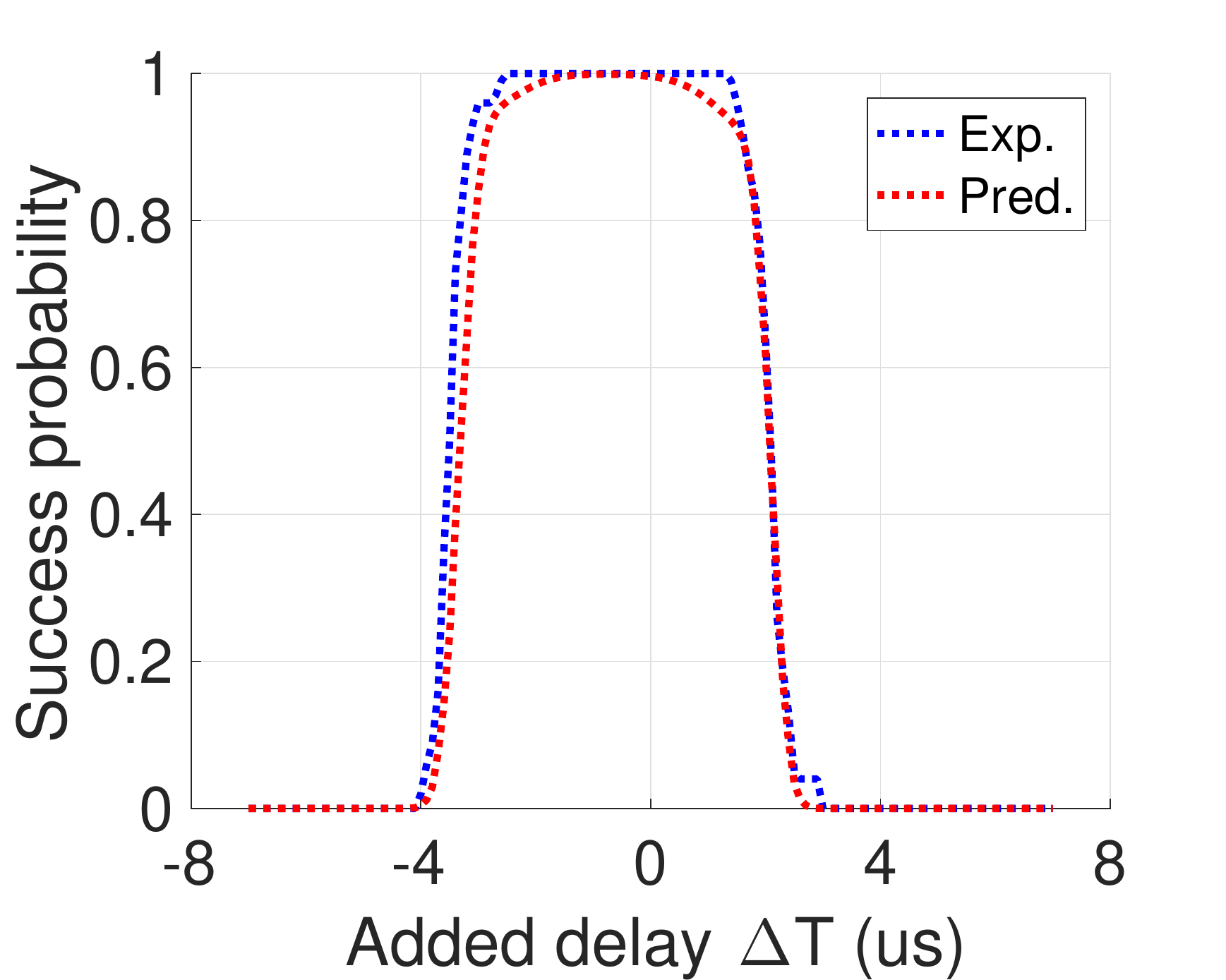}
		\caption{$N=20$, $\gamma=4$, $\Gamma=5$,\\ $\boldsymbol{\kappa}=5$; $\text{ADE}=5.4\%$}
	\end{subfigure}
	\caption{Experimental versus predicted attack success probability curves for the NTP-based IDS with different IDS settings. They are closely matched in general.}
	\label{fig:0x185_NTP_customized_IDS}
\end{figure}

As shown in Figures~\ref{fig:0x185_SoA_customized_IDS} and \ref{fig:0x185_NTP_customized_IDS}, when IDS parameters are changed, we can still see a close match between the predicted and experimental attack success probability curves for both SOTA and NTP-based IDSs. The average ADEs are $1.0\%$ and $5.8\%$, respectively.

\subsection{Applicability of Formal Analysis to Other Vehicles}
In order to validate that the proposed models are also applicable to other vehicles, we evaluate our analysis using the CAN traffic data from a Toyota Camry 2010 \cite{ruth2012accuracy}.

\textbf{Dataset.} The Toyota dataset contains $42$ distinct messages, whose periods vary from $10$ ms to $5$ sec. 
There are a total of $14$ log files with duration ranging from $70$ sec to $276$ sec. 
We concatenated the timestamps of the same message across all log files to obtain sufficient data for experiments.
As compared to the vehicle in our testbed, the CAN traffic is much less busy in the Toyota dataset, with roughly $800$ messages being exchanged per second. 

\textbf{Setup.} In this experiment, we chose three 10 ms messages of the same clock skew with IDs 0x020, 0x025, and 0x0B2. 
We first used 0x020 as the normal data, and 0x0B2 as the attack data. We then used 0x025 as the normal data, followed by the same attack data from 0x0B2.
For consistency, we again set $N=20$, $\gamma=4$, $\Gamma=5$, and $\kappa=8$ for the IDS in all experiments.

\textbf{Results.} For the SOTA IDS, the proposed model provides close predictions but tends to overestimate the attack success probability (Figure~\ref{fig:results_toyota}). This is because for a CAN bus with less traffic, the noise in the inter-arrival times becomes much smaller.
It means that the accumulated offset curve would exhibit greater linearity, implying a smaller standard deviation of identification errors. As a result, the SOTA IDS is more sensitive and allows a smaller range of $\Delta T$. 
Hence, the normalized identification error may not strictly decreasing for attacks with smaller $\Delta T$ on a less busy CAN bus, and we expect the predicted attack success probability to be larger than the experimental value.
Improving our formal model for the SOTA IDS for less busy CAN buses is left as future work. 

For the NTP-based IDS, the proposed model achieves comparable prediction accuracy for different messages as on our testbed, which demonstrates the applicability of our formal analysis to other vehicles.

\begin{figure}[t!]
	\centering
	\begin{subfigure}[h]{0.45\columnwidth}
		\captionsetup{justification=centering}
		\includegraphics[width=\columnwidth]{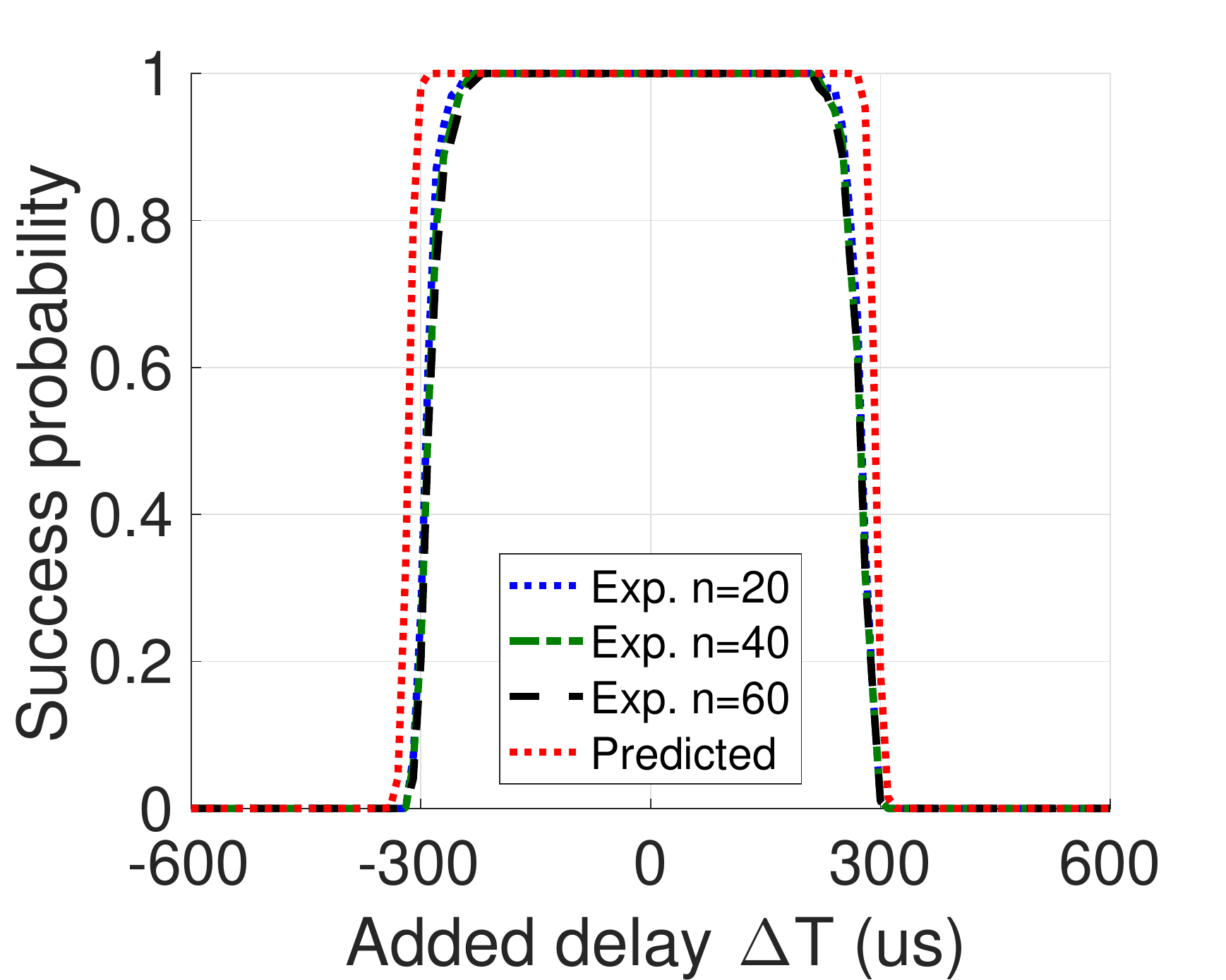}
		\caption{0x020 as normal data,\\SOTA; $\text{ADE}=9.2\%$} %n=20,rho=7.7%; n=40,rho=9.8%; n=60,rho=9.2%
		\label{fig:sota_toyota_normal_0x020_attack_0x0B2}
	\end{subfigure}
	\begin{subfigure}[h]{0.45\columnwidth}
		\captionsetup{justification=centering}
		\includegraphics[width=\columnwidth]{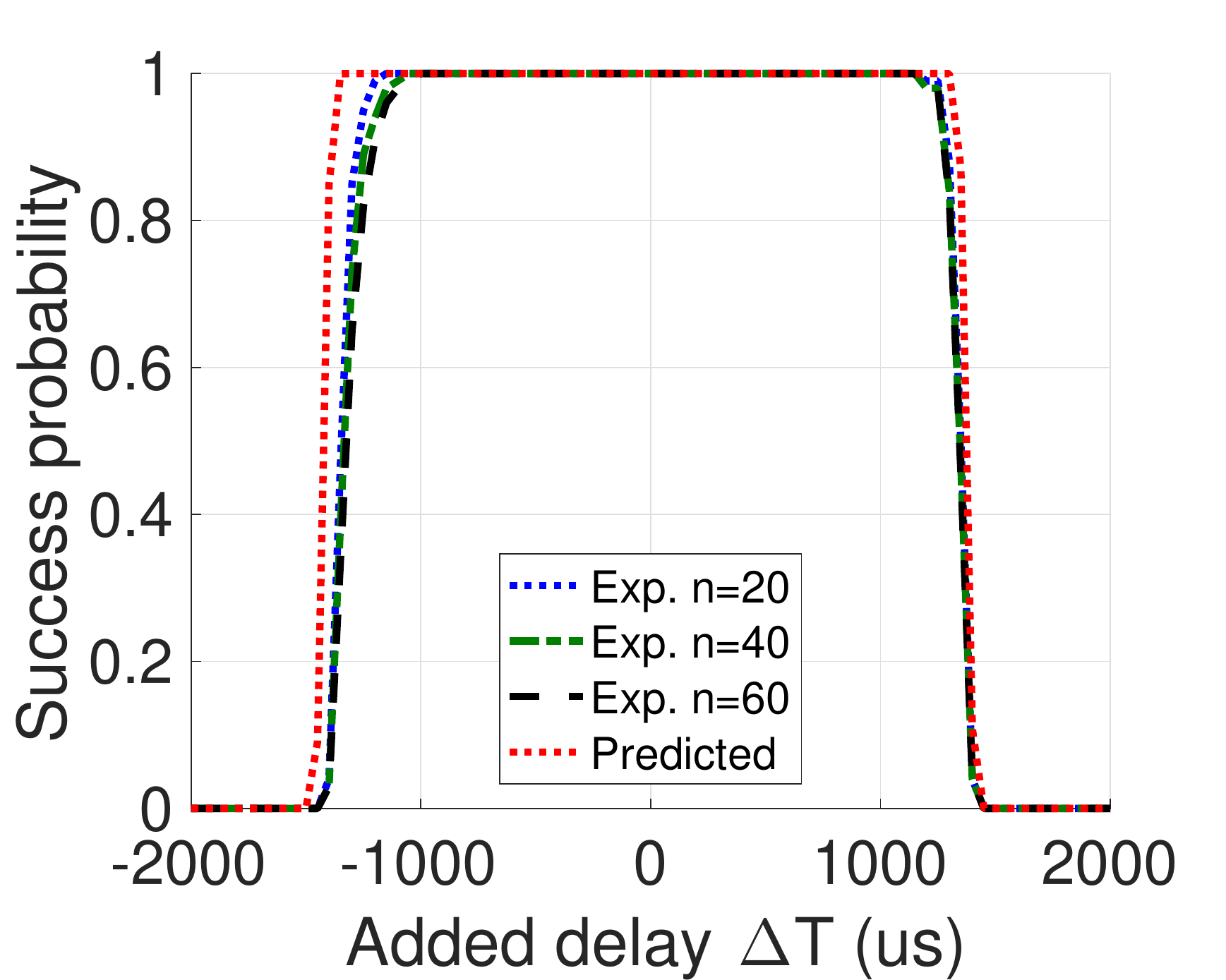}
		\caption{0x025 as normal data,\\SOTA; $\text{ADE}=5.6\%$} %n=20,rho=4.1%; n=40,rho=5.0%; n=60,rho=5.6%
		\label{fig:sota_toyota_normal_0x025_attack_0x0B2}
	\end{subfigure}
	\begin{subfigure}[h]{0.45\columnwidth}
		\captionsetup{justification=centering}
		\includegraphics[width=\columnwidth]{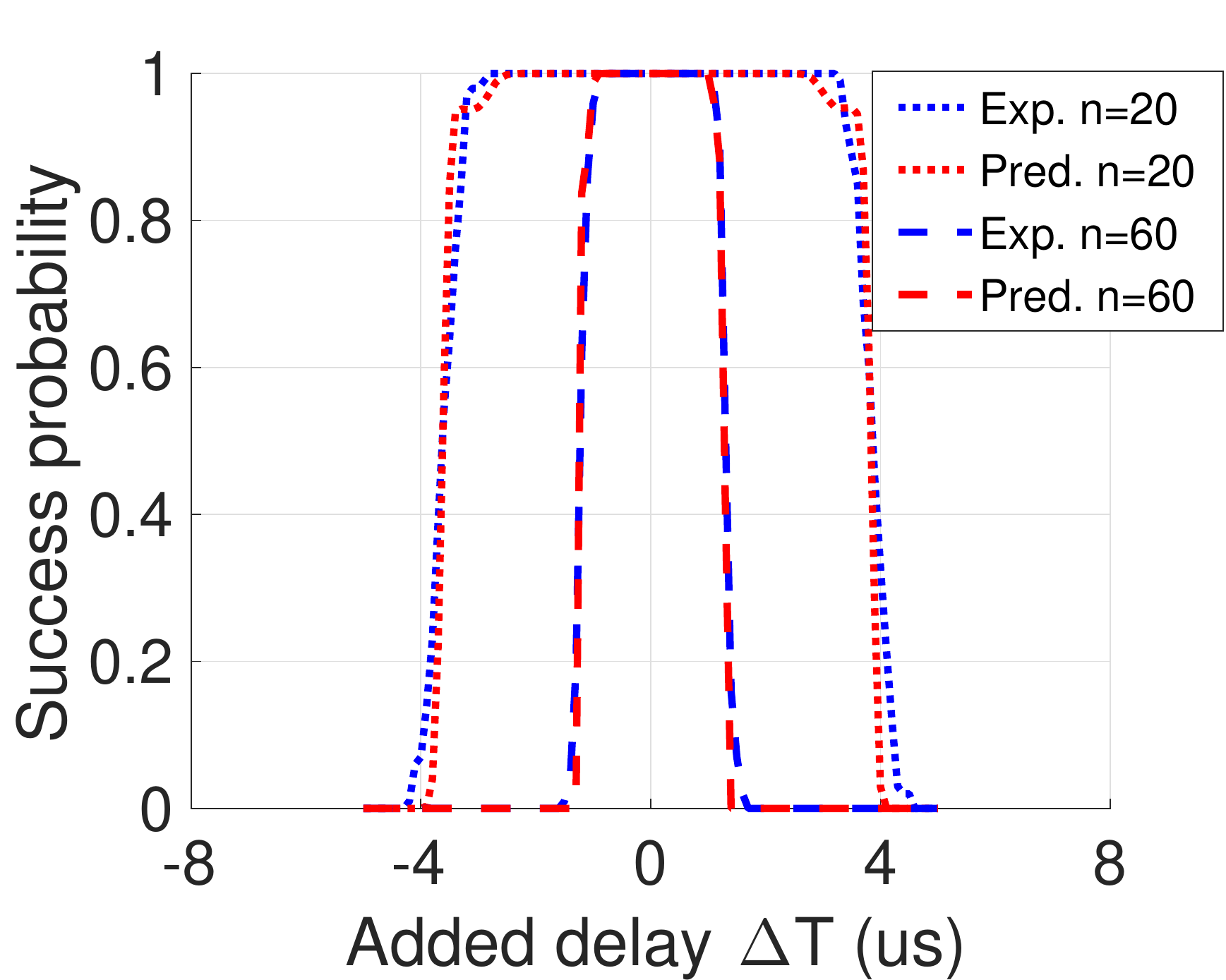}
		\caption{0x020 as normal data,\\NTP-based; $\text{ADE}=3.7\% $} %n=20,rho=3.7%; n=40,rho=3.3%; n=60,rho=3.7%
		\label{fig:ntp_toyota_normal_0x020_attack_0x0B2}
	\end{subfigure}
	\begin{subfigure}[h]{0.45\columnwidth}
		\captionsetup{justification=centering}
		\includegraphics[width=\columnwidth]{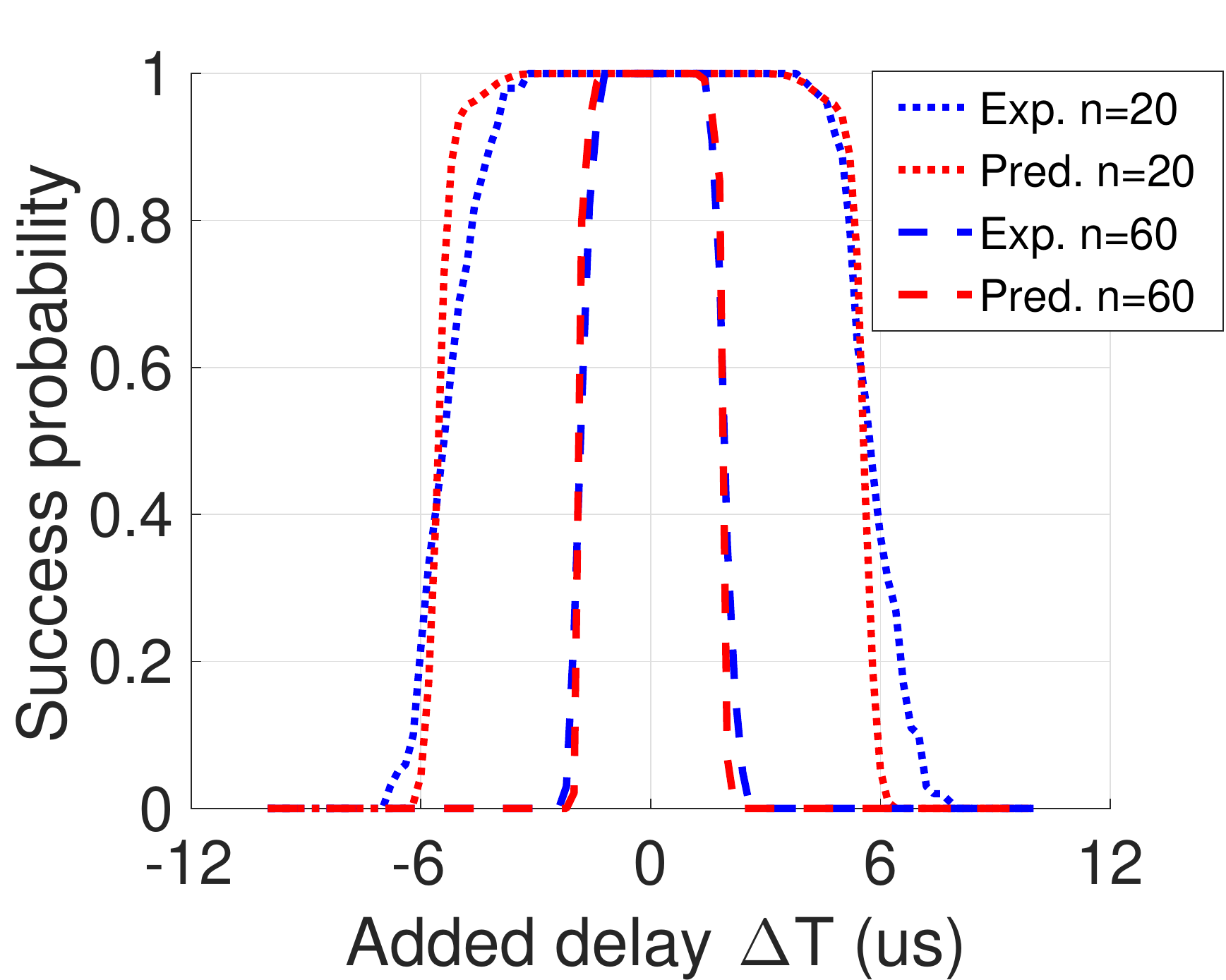}
		\caption{0x025 as normal data,\\NTP-based; $\text{ADE}=7.1\%$} %n=20,rho=7.3%; n=40,rho=6.3%; n=60,rho=7.1%
		\label{fig:ntp_toyota_normal_0x025_attack_0x0B2}
	\end{subfigure}
	\caption{Experimental versus predicted attack success probability curves for both SOTA and NTP-based IDSs using the Toyota data. In each case, the curves are closely matched in terms of shape.
		The proposed model for the SOTA IDS tends to overestimate the attack success probability for a CAN bus with less traffic, while the proposed model for the NTP-based IDS achieves comparable prediction accuracy as on our testbed. Note that the same attack data is used, and the ADE values are for $n=60$.
	}
	\label{fig:results_toyota}
	%\vspace{-0.3cm}
\end{figure}
}
\section{Conclusions}
\label{sec:conclusion}
In this paper, we proposed the cloaking attack and provided formal analyses of the attack for two clock skew-based IDSs, i.e., the SOTA IDS and the NTP-based IDS.
We incorporated parameters of the attacker, the detector, and the hardware platform and derived attack success probabilities for both IDSs.
We demonstrated the cloaking attack on hardware testbeds and validated the proposed models through extensive experiments using the data collected from the UW EcoCAR testbed \hl{and a publicly available dataset.}
\hl{Our results have demonstrated the predictive capability and the accuracy of our formal analyses for different messages, IDS settings, and vehicles.}
%The results have also demonstrated the predictive capability and the accuracy of our formal analysis across to different IDS settings and vehicles.
Our results also illustrate the feasibility of developing formal analysis and models for other variants of CAN used in vehicles and applications such as computer-integrated manufacturing.

%Our future work will include improving the modeling accuracy by accounting for other uncertainties such as the variations in the network delay due to arbitration and the processing delay due to the jitter at the IDS. We will also explore how to leverage our formal approach to design more effective IDSs for CAN buses.

\section*{Acknowledgment}
\label{sec:ack}
We would like to thank Drs. Sukarno Mertoguno and David Corman for discussions on this problem.
%, and Dr. Kang G. Shin, for his pioneering work that introduced the clock skew-based IDS for automotive CAN networks.  
This work was supported by NSF grants CNS-1446866 and CNS-1656981, ONR grants N00014-16-1-2710 and N00014-17-1-2946, and ARO grant W911NF-16-1-0485. Views and conclusions expressed are that of the authors and not be interpreted as that of the NSF, ONR or ARO.

%\input{Sections/acknowledgement}

%\newpage
\bibliographystyle{IEEEtran}
\bibliography{tifs-2018}

\begin{figure}[t!]
	\centering
	\begin{subfigure}[b]{0.49\columnwidth}
		\includegraphics[width=\columnwidth, trim={0 0 0 0.6cm}, clip]{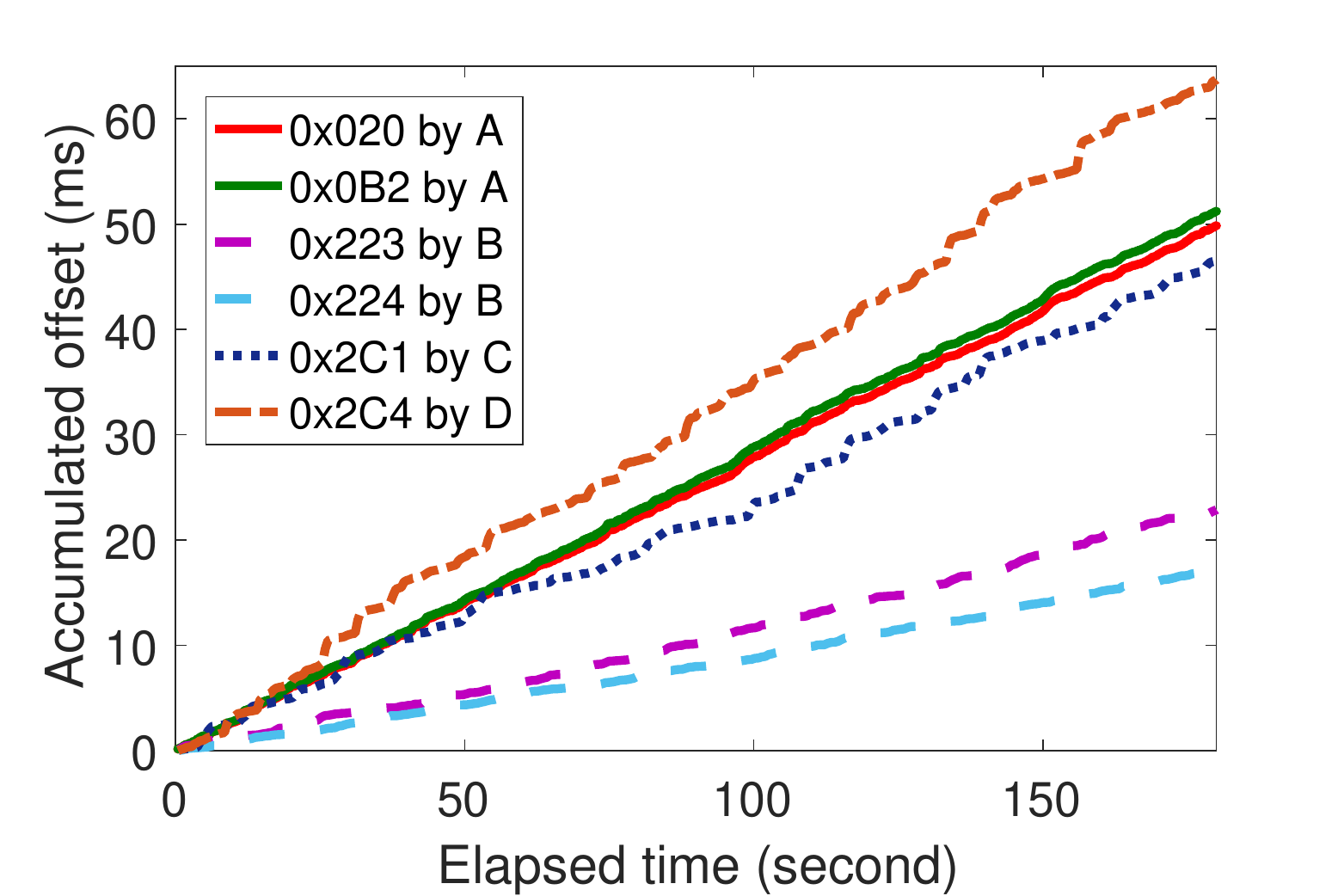}
		\caption{SOTA IDS, $N=20$}
	\end{subfigure}
	\begin{subfigure}[b]{0.49\columnwidth}
		\includegraphics[width=\columnwidth,trim={0 0 0 0.6cm}, clip]{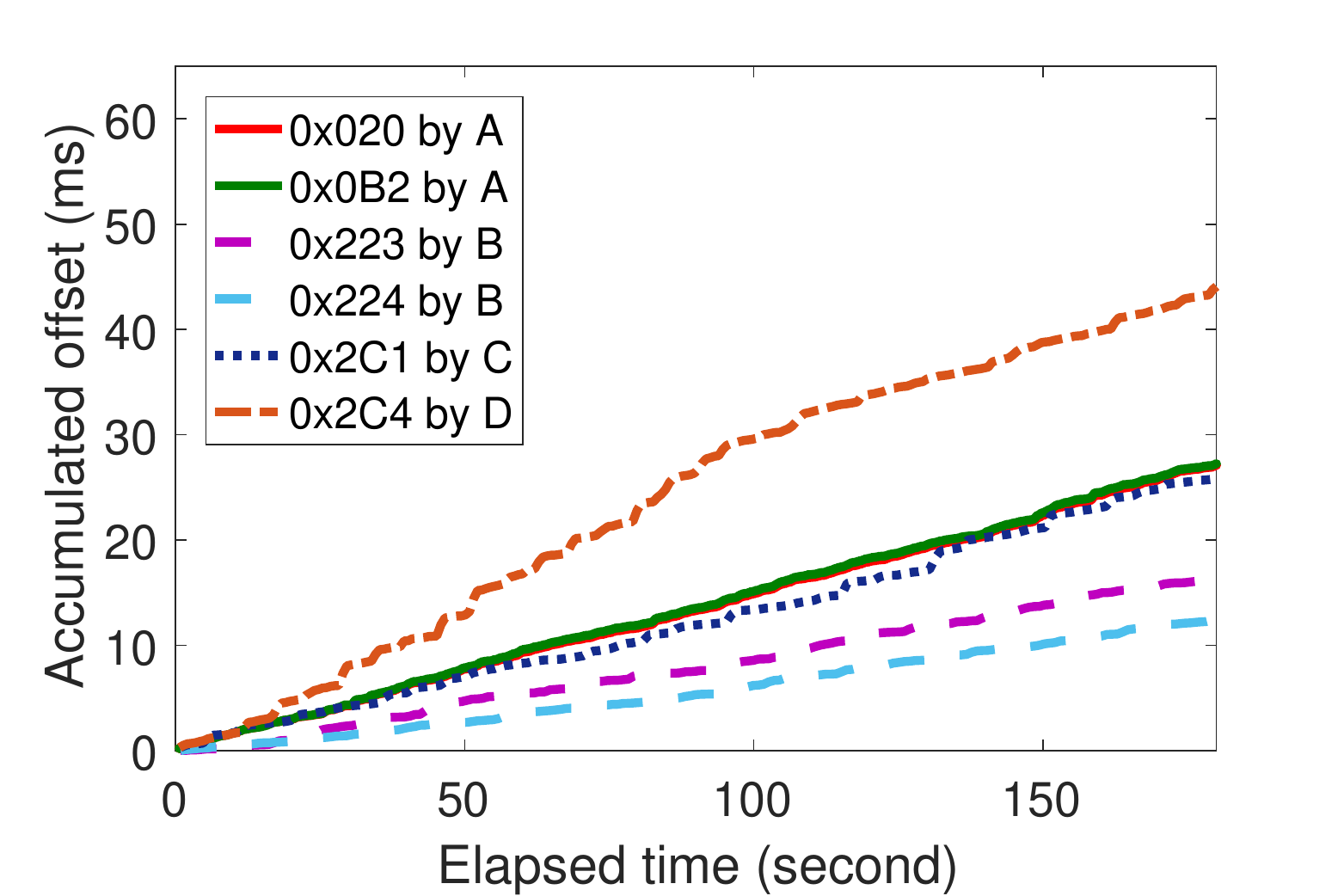}
		\caption{SOTA IDS, $N=30$}		
	\end{subfigure}
	\\
	\begin{subfigure}[b]{0.49\columnwidth}
		\includegraphics[width=\columnwidth]{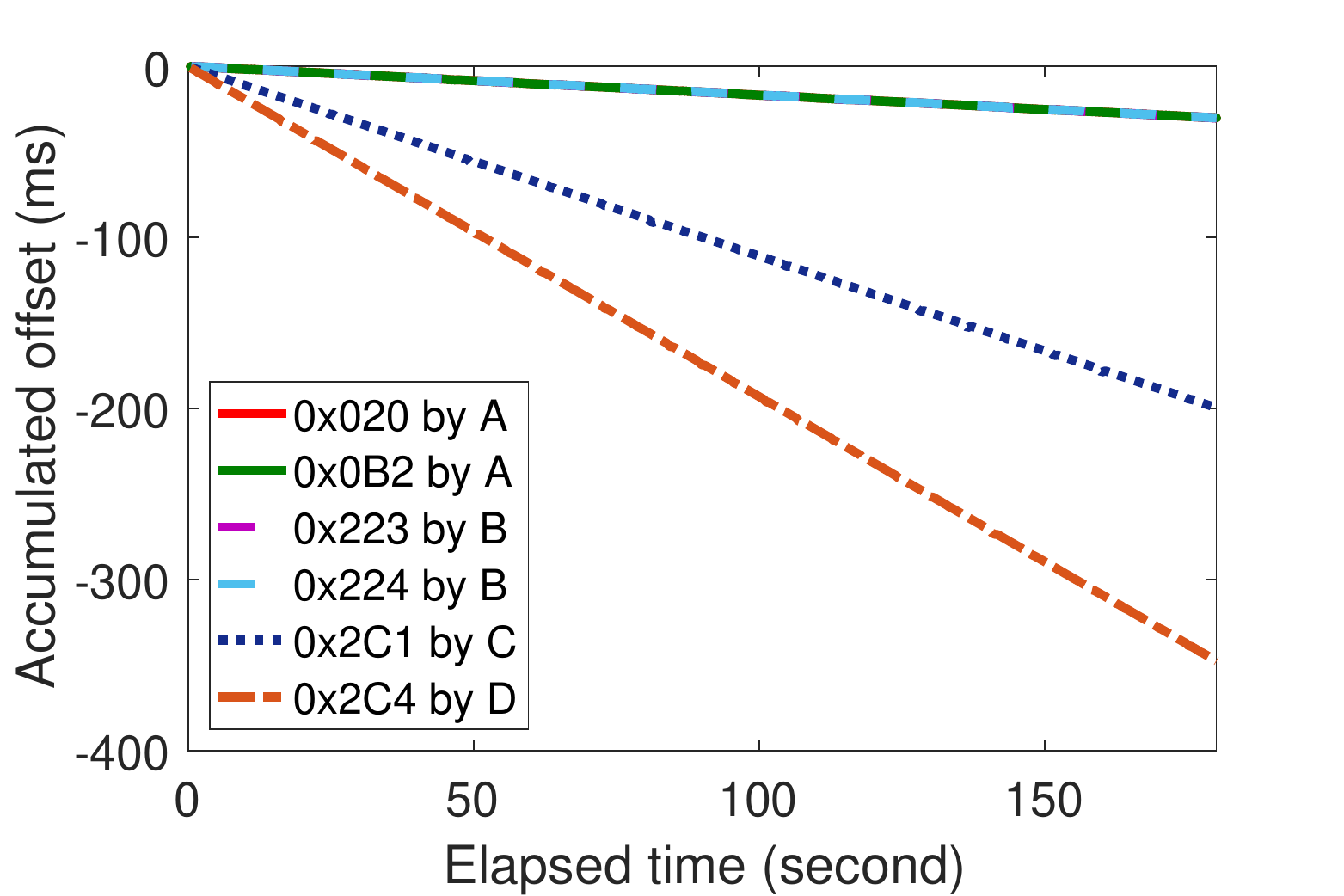}
		\caption{NTP-based IDS, $N=20$}
	\end{subfigure}
	\begin{subfigure}[b]{0.49\columnwidth}
		\includegraphics[width=\columnwidth]{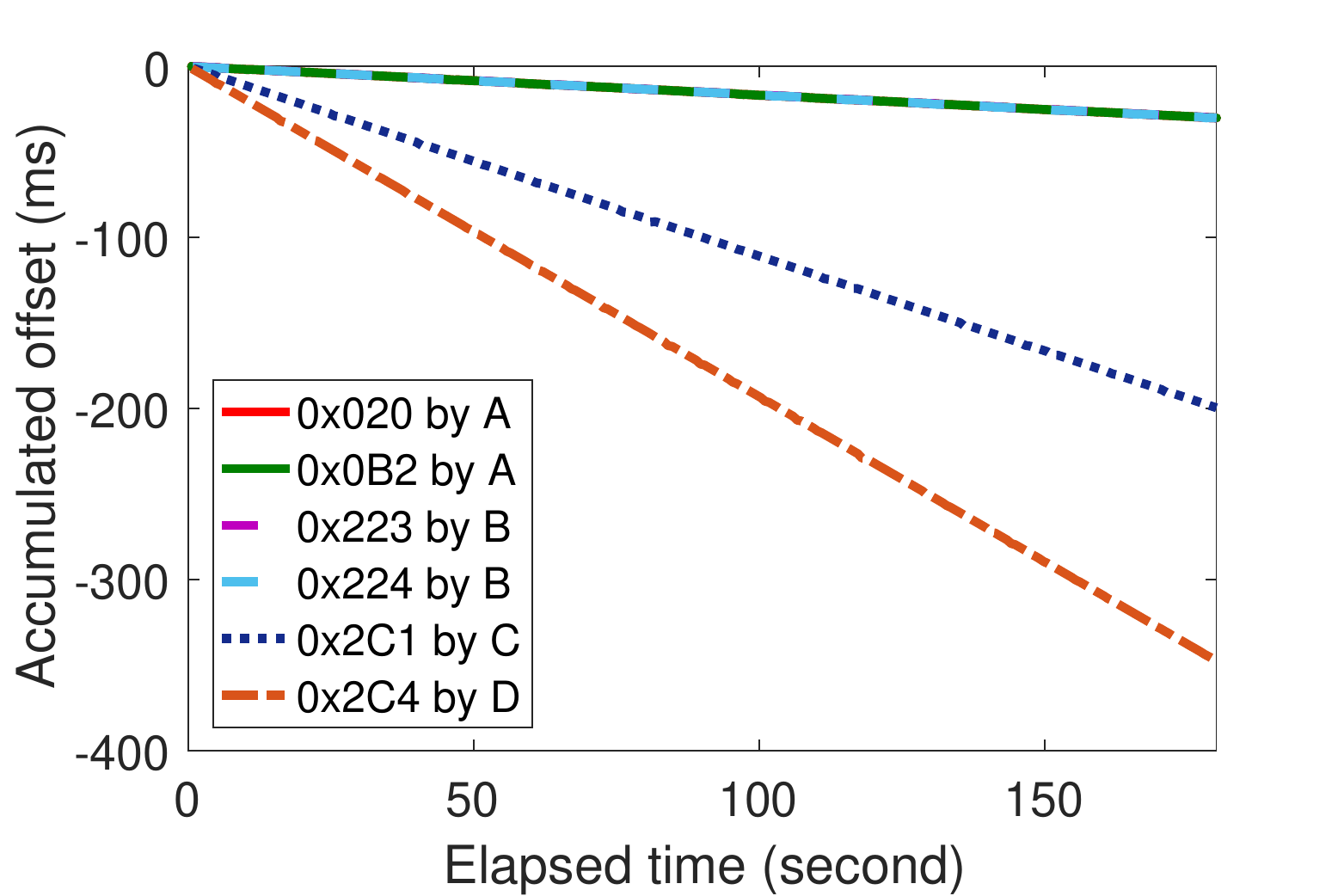}
		\caption{NTP-based IDS, $N=30$}		
	\end{subfigure}
	\caption{Accumulated offsets provided by the SOTA and NTP-based IDSs with batch sizes of $20$ and $30$.
	The same portion of the data trace (with ID=$25$) from the Toyota dataset is used.
	Significant differences in slopes (estimated clock skew) are observed for the same message using the SOTA IDS, whereas the clock skew estimated by the NTP-based IDS is almost identical with different batch sizes.
	}
	\label{fig:cho_shin_impact_of_batch_size}
\end{figure}

\appendix
\hl{\subsection{Estimation Consistency of SOTA and NTP-Based IDSs}
\label{appendix:estimation_consistency}
As a physical property of an ECU, clock skew is considered to be stable over time, and thus the estimated values should be  \textit{consistent}, across 1) different batch sizes used by an IDS, 2) different portions of the same trace, and 3) different traces of the same ECU.
Hence, we use the Toyota Camry dataset \cite{ruth2012accuracy} that was used in \cite{Shin:2016:finger} to compare the NTP-based IDS against the SOTA IDS in terms of estimation consistency.

Fig.~\ref{fig:cho_shin_impact_of_batch_size} illustrates the accumulated offsets estimated by the two IDSs with different batch sizes\footnote{Due to the lack of ground truth, the authors in \cite{Shin:2016:finger} empirically identified that 0x020, 0x0B2, 0x223 and 0x224 are transmitted by two different ECUs. However, based on our NTP-based clock skew estimation results, we believe that the four messages come from the same ECU.}. 
Significant differences in slopes for the same message are observed for the SOTA IDS. 
For example, the estimated clock skew (based on the end point) of message 0x020 is around $273$ ppm with $N=20$, but dropped to around $151$ ppm with $N=30$.
In contrast, the NTP-based IDS provides consistent estimation. 

To further quantify estimation consistency, we consider the following three cases: 1) use the same portion of the same trace, and vary $N$ from $20$ to $100$ with a step of $20$, 2) set $N=20$, and use different portions of the same trace by omitting the first $m$ messages, where $m$ is varied from $1$ to $19$, and 3) set $N=20$, and use $14$ different traces from the Toyota dataset. 
The standard deviation ($\sigma$) of estimated clock skews are adopted as the metric, and a smaller $\sigma$ value implies more consistent estimation.
As shown in Table~\ref{table:cho_shin_impact_of_batch_size}, the NTP-based IDS has a significantly smaller $\sigma$ than the SOTA IDS for all messages in all cases.}

\begin{table}[ht!]
	\caption {Standard deviations ($\sigma_1$, $\sigma_2$, $\sigma_3$) of clock skews estimated by IDSs in three different cases. The NTP-based IDS has a significantly smaller $\sigma$ than the SOTA IDS, which demonstrates its consistency in clock skew estimation.
	}
	%as the batch size changes from 20 to 100 by step size of 20 using Toyota data. Unit is in ppm.}
	\label{table:cho_shin_impact_of_batch_size}
	\begin{center}
		\resizebox{1\columnwidth}{!}
		{
			\begin{tabular}{|c|c|c|c|c|c|c|}
				\hline
				\multirow{2}{*}{Message ID} & \multicolumn{3}{c|}{State-of-the-art IDS} &
				\multicolumn{3}{c|}{NTP-based IDS} \\
				\cline{2-7}
				& $\sigma_1$ & $\sigma_2$ & $\sigma_3$ & $\sigma_1$ & $\sigma_2$ & $\sigma_3$  \\
				\hline
				0x020 & 92.3682 & 12.0589 & 20.1727 & 0.3706 & 0.2000 & 1.7716 \\
				\hline
				0x0B2 & 94.4480 & 11.7543 & 19.4549 & 0.4252 & 0.2045 & 1.7929 \\
				\hline
				0x223 & 41.6631 & 16.2060 & 25.4885 & 0.3083 & 0.4429 & 1.6437 \\
				\hline
				0x224 & 29.0736 & 17.0442 & 32.6059 & 0.8348 & 0.5491 & 2.3660\\
				\hline
				0x2C1 & 85.3753 & 7.9963 & 26.2618 & 0.0866 & 1.2191 & 3.2977\\
				\hline
				0x2C4 & 116.5630 & 13.0896 & 53.7820 & 1.0763 & 1.1599 & 3.3602\\
				\hline
			\end{tabular}
		}
	\end{center}
\end{table}
\normalsize
\subsection{Proof of Lemma~\ref{lemma:sota-initial-error}}\label{appendix:lemma1}
\begin{proof} 
To compute the mean of $e[m]$, the normalized identification error in the first attack batch (Eq.~(\ref{eq:sota-e})), 
we first consider the mean of $|O_{avg}[m]|$, the distribution of the absolute value of the average offset in the $m$-th batch. 
Under the assumption in Eq.~(\ref{eq:O_avg_approx}), we have
\begin{align}
\mathbb{E}(|O_{avg}[m]|)&= \frac{1}{N-1}\sum_{i=2}^{N}{\mathbb{E}[(i-1)(|\mu + \Delta T - \mu[m-1]|) } \nonumber\\
& ~~~~~~~~~~~~~~~~~~~~~~~~+ (\eta_{m,i}-\eta_{m,1})] \nonumber \\
&= \frac{N}{2}(|\mu + \Delta T - \mu[m-1]|). \label{eq:appendix_mean_abs_O_avg}
\end{align}
Now we compute the mean of the third term in Eq.~(\ref{eq:sota-e}). Since $\mathbb{E}[a_{m,N}]=\mathbb{E}[a_{m,1}]+(N-1)(\mu+\Delta T)$, we have
\begin{align}
&\mathbb{E}[S[m-1](t[m-1] + T_{m,0} + a_{m,N} - a_{m,1})] \nonumber \\
&~~~= S[m-1](t[m-1] + T_{m,0} + (N-1)(\mu + \Delta T). \label{eq:appendix_mean_2nd_term}
\end{align}
Combining Eq.~(\ref{eq:appendix_mean_abs_O_avg}) and (\ref{eq:appendix_mean_2nd_term}) with Eq.~(\ref{eq:sota-e}) yields (\ref{eq:sota-mu-e}).

The variance of $e[m]$ from Eq.~(\ref{eq:sota-e}) is
\begin{align*}
\sigma_e^2 &=  \text{Var}\left( \frac{1}{N-1} \sum_{i=2}^N (\eta_{m,i} - \eta_{m,1}) - S[m-1](\eta_{m,N} - \eta_{m,1}) \right)\\
%&= \left( \frac{N-2S[m-1]}{N-1} + 2 S^2[m-1] - 2 S[m-1] \right) \sigma_{\eta}^2 \\
&= \frac{1}{2}\left( \frac{N-2S[m-1]}{N-1} + 2 S^2[m-1] - 2 S[m-1] \right) \sigma^2,
\end{align*} 
which completes our proof. 
\end{proof}

\subsection{Proof of Theorem~\ref{theorem:sota-success}}\label{appendix:theory}
\begin{proof}
Let $e_n[k]$ be the normalized identification error in the $k$-th batch when the attack begins in the $m$-th batch, where $k\geq m$.  Let $\tau$ be the decreasing rate of $e_n[k]$. 
Then we have $e_n[k] \approx e_n[m] - \tau(k-m)$. 
For the upper control limit $L^+$, its maximum is reached at the $l$-th batch, where $l = \max \{k: e_n[k] \geq \kappa, k\geq m\} = \lceil (e_n[m] - \kappa)/\tau \rceil + (m-1)$. 
We then have 
\begin{align*}
L^+_{\max} &= \sum_{k=m}^l (e_n[k] - \kappa) = (l-m+1)\cdot e_n[m] \nonumber \\
&- \frac{(l-m+1)(l-m)}{2}\tau - (l-m+1) \cdot \kappa. 
\end{align*}
%For simplicity, we set $m=(e_n[1]-\kappa)/\tau$. Then we have
Simplifying the above equation yields
\begin{equation}
L^+_{\max} = \frac{(e_n[m] - \kappa)^2}{2\tau} + \frac{e_n[m]-\kappa}{2}.
\end{equation}
In order for the attack to be undetected, the condition that $L^{+}_{\max} \leq \Gamma$ needs to be met, or equivalently,
%Having $L^+_{\max} \leq \Gamma$ leads to 
\begin{equation}
(e_n[m]-\kappa)^2 + \tau (e_n[m]-\kappa) - 2 \tau \cdot \Gamma \leq 0.
\end{equation}
%Since $e[k] > \kappa$, we have
%\begin{equation}
%e_n[1] \leq \frac{-\tau + \sqrt{\tau^2 + 8 \tau \Gamma}}{2} + \kappa. 
%\end{equation}

Since $e_n[m]\sim N \left(\frac{\mu_e - \mu_{\text{CUSUM}}}{\sigma_{\text{CUSUM}}}, \frac{\sigma_e^2}{\sigma^2_{\text{CUSUM}}} \right)$, the probability of $L^+_{\max}\leq \Gamma$ is $Pr(e_n[m] \leq \frac{-\tau + \sqrt{\tau^2 + 8 \tau \Gamma}}{2} + \kappa)$. 

Similarly, for the lower control limit $L^-$, the probability of $L^+_{\min} \leq \Gamma$ is $Pr(e_n[m] \geq  - \frac{-\tau + \sqrt{\tau^2 + 8 \tau \Gamma}}{2} - \kappa)$. Combining these results yields (\ref{eq:sota-success}).
\end{proof}

\subsection{Proof of Lemma~\ref{lemma:NTP-CUSUM}}\label{appendix:lemma2}
\begin{proof}
The law of total probability implies 
\begin{align*}
&~~~~Pr(\alpha > n | L^{+}[k] = z^{+}, L^{-}[k] = z^{-})\\ 
&= \int_{-\infty}^{\infty}{Pr(\alpha > n | L^{+}[k] = z^{+}, L^{-}[k] = z^{-}, e_n[k] = r)f_{k}(r) \ dr},
\end{align*}
and we have
\begin{align*}
L^{+}[k+1] &= \left\{
\begin{array}{ll}
 0, & r < \kappa - z^{+} \\
z^{+} + r - \kappa, & r \geq \kappa - z^{+}
\end{array}
\right., \\
L^{-}[k+1] &= \left\{
\begin{array}{ll}
z^{-}-r-\kappa, & r < z^{-} - \kappa \\
0, & r \geq z^{-} - \kappa
\end{array}
\right..
\end{align*}
Now, first, suppose that $\kappa - z^{+} < z^{-} - \kappa$. Then $z^{+} + z^{-} > 2\kappa$, and hence by assumption $z^{+} + z^{-} \geq 2\Gamma$. Thus either $z^{+} \geq \Gamma$ or $z^{-} \geq \Gamma$, implying that $\tau = 0 < n$. 

We can therefore write
\footnotesize
\begin{align*}
&~~~~~ g_{n,k}(z^{+},z^{-}) \\
&=\int_{-\infty}^{z^{-}-\kappa}{Pr(\alpha > n | L^{+}[k] = z^{+}, L^{-}[k] = z^{-}, e_n[k] = r)f_{k}(r) \ dr} \\
& + \int_{z^{-}-\kappa}^{\kappa - z^{+}}{Pr(\alpha > n | L^{+}[k] = z^{+}, L^{-}[k] = z^{-}, e_n[k] = r)f_{k}(r) \ dr} \\
&+ \int_{\kappa - z^{+}}^{\infty}{Pr(\alpha > n | L^{+}[k] = z^{+}, L^{-}[k] = z^{-}, e_n[k] = r)f_{k}(r) \ dr} \\
%&=& \int_{-\infty}^{z^{-}-\kappa}{Pr(\tau > n | L_{k+1}^{+} = 0, L_{k+1}^{-} = z^{-}-r-\kappa)f_{k}(r) \ dr} \\
%&& + \int_{z^{-}-\kappa}^{\kappa - z^{+}}{Pr(\tau > n | L_{k+1}^{+} = 0, L_{k+1}^{-} = 0)f_{k}(r) \ dr} \\
%&& + \int_{\kappa - z^{+}}^{\infty}{Pr(\tau > n | L_{k+1}^{+} = z^{+} + r - \kappa, L_{k+1}^{-} = 0)f_{k}(r) \ dr} \\
&= \int_{z^{-}-\kappa-\Gamma}^{z^{-}-\kappa}{g_{n,k+1}(0,z^{-}-r-\kappa)f_{k}(r) \ dr} \\
&~~~~ + g_{n,k+1}(0,0)Pr(e_n[k] \in [z^{-} - \kappa, \kappa - z^{+}]) \\
&~~~~ + \int_{\kappa - z^{+}}^{\kappa-z^{+}+\Gamma}{g_{n,k+1}(z^{+} + r - \kappa, 0)f_{k}(r) \ dr}.
\end{align*}
\normalsize
This completes the proof.
\end{proof}

\hl{
\subsection{Impact of Mistiming on Cloaking Attack}\label{appendix:mistiming}
In a masquerade or cloaking attack, the strong attacker needs to start transmitting the spoofed message at the time constant at which the targeted message should have been transmitted, if it had not been suspended. 
It naturally raises the question whether mistiming affects the cloaking attack performance.
In this simulation, we introduce a mistiming delay (either positive or negative) between the last message of normal data and the first message of attack data, in addition to the message period.

% Successful Rates
\begin{figure}[t]
	\vspace{-0.3cm}
	\centering
	\begin{subfigure}[h]{0.49\columnwidth} % {0.48\columnwidth}
		\captionsetup{justification=centering}
		\includegraphics[width=\columnwidth]{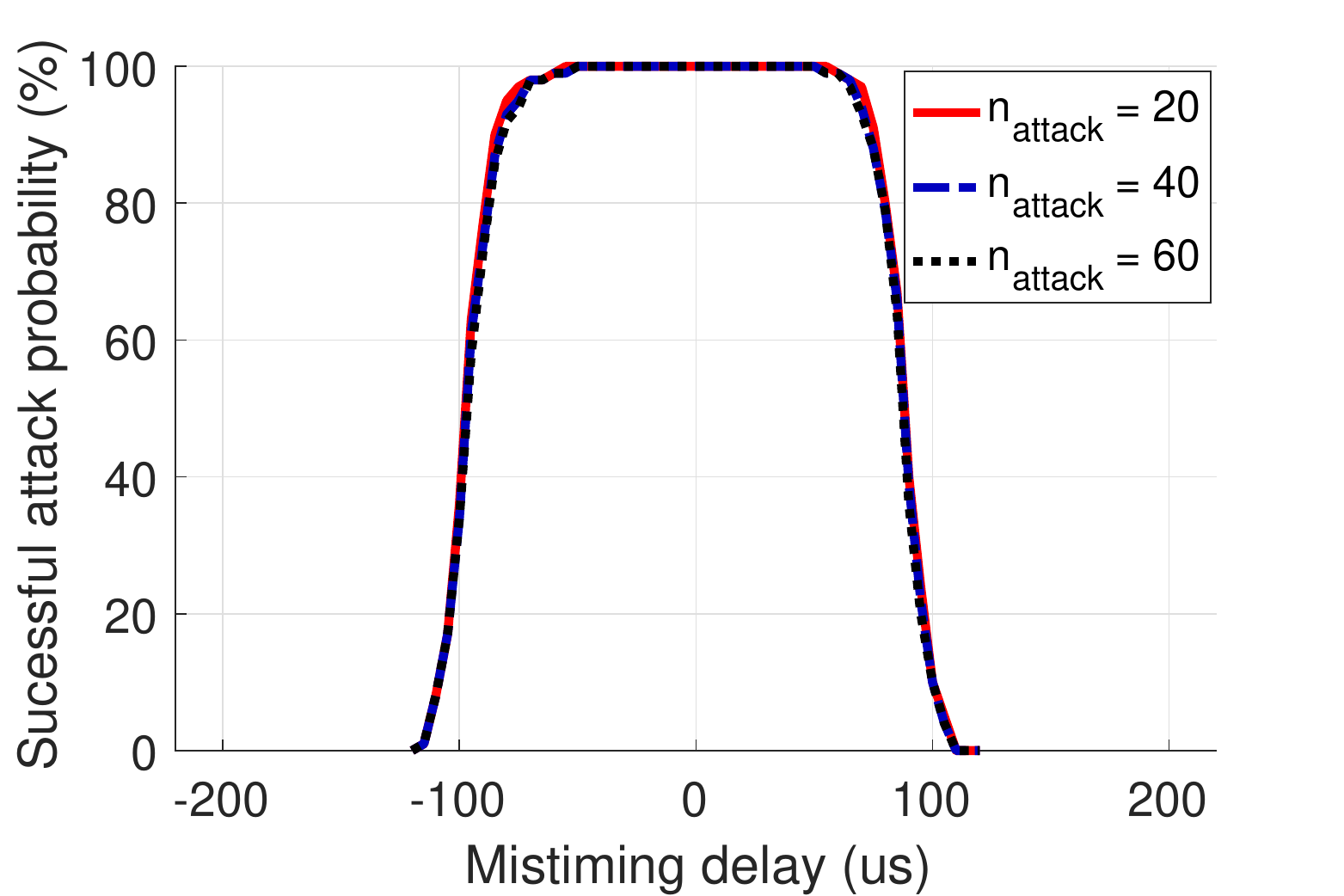}
		\caption{CAN prototype, SOTA}
		\label{fig:arduino_clock_skew_attack_success_rate_Cho_mistime}
	\end{subfigure}
	\begin{subfigure}[h]{0.49\columnwidth} % {0.48\columnwidth}
		\captionsetup{justification=centering}
		\includegraphics[width=\columnwidth]{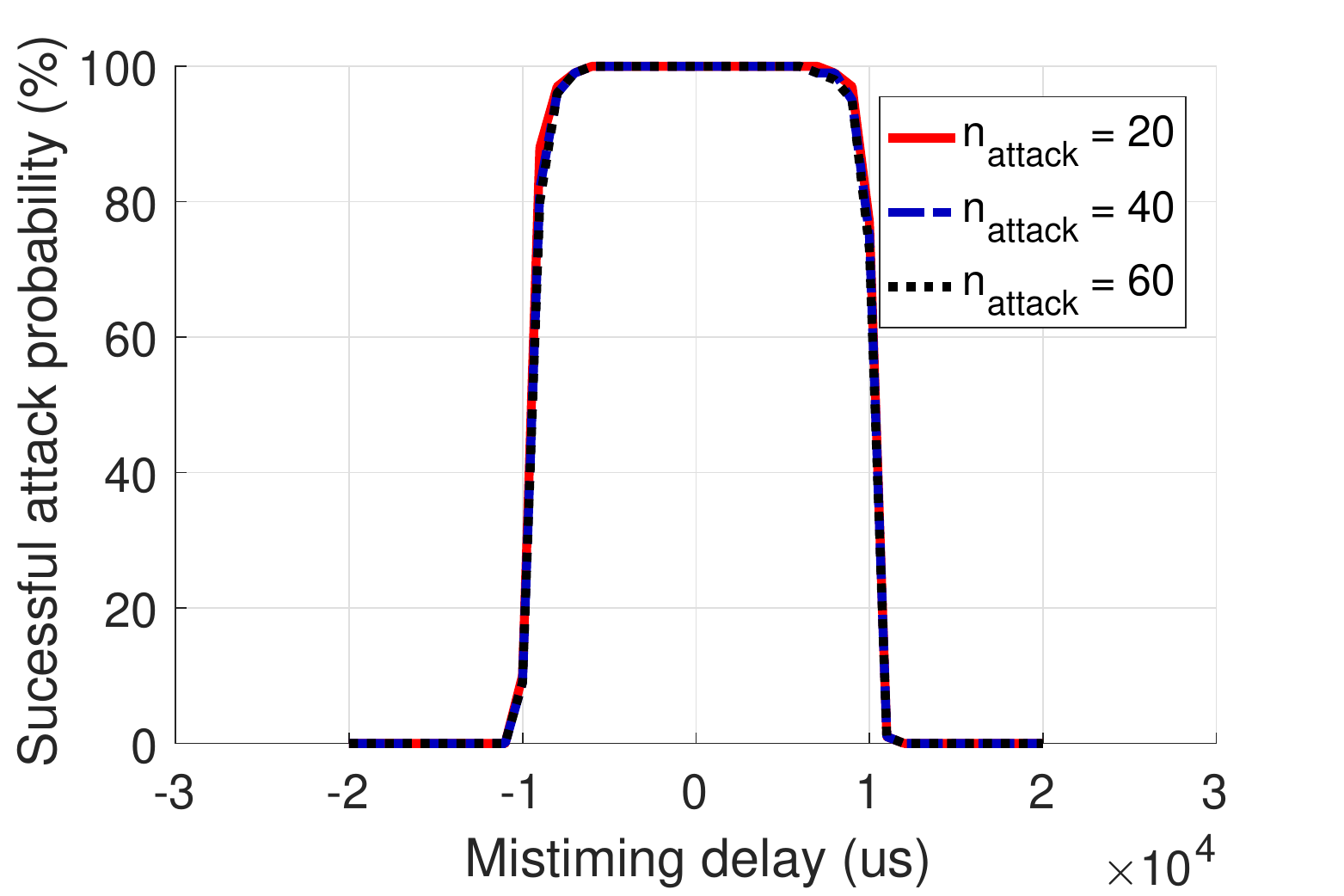}
		\caption{EcoCAR testbed, SOTA}
		\label{fig:ecocar_clock_skew_attack_success_rate_Cho_mistime}
	\end{subfigure}
	\\
	\begin{subfigure}[h]{0.49\columnwidth} % {0.48\columnwidth}
		\captionsetup{justification=centering}
		\includegraphics[width=\columnwidth]{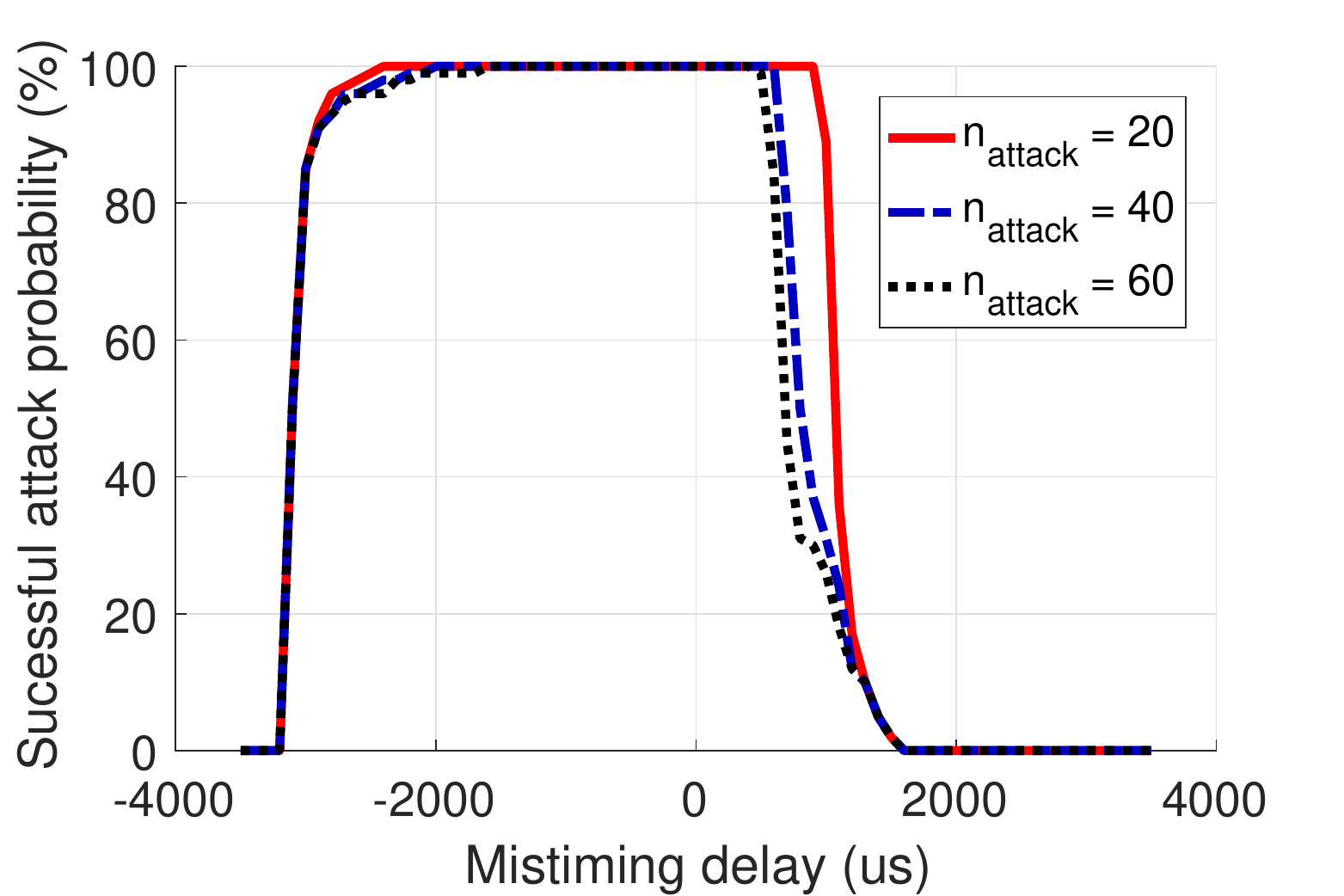}
		\caption{CAN prototype, NTP-based}
		\label{fig:arduino_clock_skew_attack_success_rate_ntp_mistime}
	\end{subfigure}
	\begin{subfigure}[h]{0.49\columnwidth} % {0.48\columnwidth}
		\captionsetup{justification=centering}
		\includegraphics[width=\columnwidth]{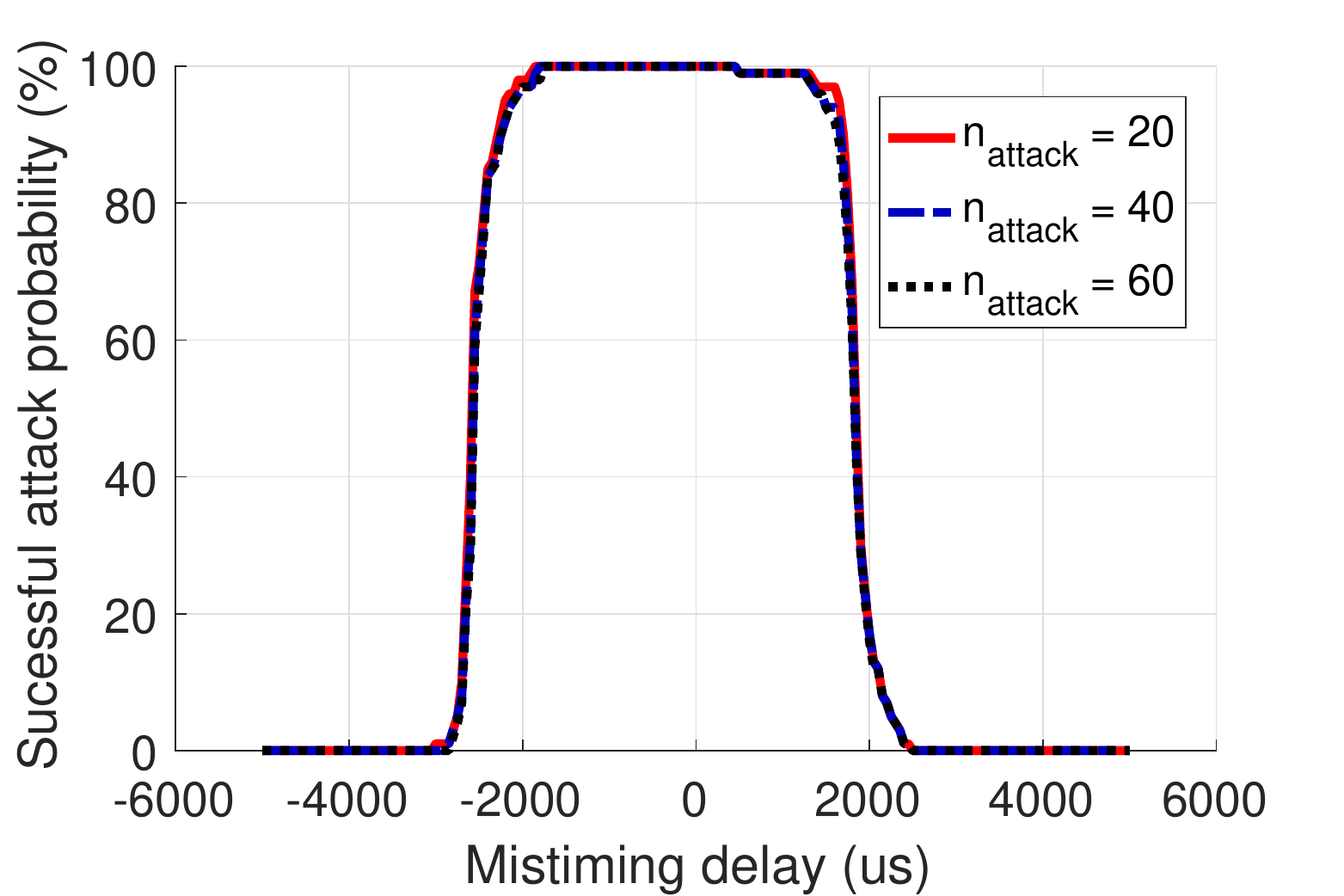}
		\caption{EcoCAR testbed, NTP-based}
		\label{fig:ecocar_clock_skew_attack_success_rate_ntp_mistime}
	\end{subfigure}
	\caption{Impact of the mistimed cloaking attack on the SOTA IDS and the NTP-based IDS. If the strong attacker can inject the fist attack message on the proper time, the cloaking attack can bypass both IDSs.
	}
	\label{fig:clock_skew_attack_success_rate_mistime}
	\vspace{-0.3cm}
\end{figure}

The IDS is fed with $1000$ batches of normal data, followed by $n$ batches of attack data with a batch size of $20$ in each experiment.
$\Gamma$ is $5$ for both IDSs, and $\gamma$ is set to $3$ and $4$ for SOTA and NTP-based IDSs, respectively.
Also, $\kappa$ is set to $5$ for the CAN bus prototype and to $8$ for the UW EcoCAR testbed, respectively.

Fig. \ref{fig:clock_skew_attack_success_rate_mistime} shows the impact of the mistiming of the cloaking attack on the SOTA and NTP-based IDSs.
In general, larger mistiming causes the attack performance to decrease.
On the CAN bus prototype, any amount of mistiming between $-55~\mu$s and $55~\mu$s does not affect the attack performance (i.e., $P_s$ is $100\%$ with $n=60$) against the SOTA IDS, whereas the allowed mistiming is much larger for the NTP-based IDS, mainly due to the difference in clock skew estimation. 
Since the clock skew of the Arduino-based ECU slowly decreases due to the temperature change in hardware as it warms up, the estimator tends to overestimate the clock skew, and thus is more sensitive to larger positive mistiming (that would further decrease the clock skew), which explains the skewness of the curves in Fig. \ref{fig:arduino_clock_skew_attack_success_rate_ntp_mistime}. 

On the UW EcoCAR tested, the allowed mistiming is much larger, which is between $-6$ ms and $7$ ms for the SOTA IDS, and between $-1.8$ ms and $0.5$ ms for the NTP-based IDS, due to much heavier CAN traffic in a real vehicle. 
The above observations imply that the timing is hardly a strict requirement for the adversary to launch a clocking attack in a real vehicle.
}

\end{document}